\documentclass[12pt,letterpaper]{article}

\usepackage[utf8]{inputenc}
\usepackage[T1]{fontenc}
\usepackage{lmodern}
\usepackage{amsmath,amssymb,amsthm,mathtools}
\usepackage{microtype}

\usepackage{geometry}
\usepackage{graphicx}
\DeclareGraphicsExtensions{.png,.pdf}
\usepackage{float,placeins,needspace,booktabs}
\usepackage[font=small,labelfont=bf,labelsep=period]{caption}
\usepackage{subcaption}
\usepackage[dvipsnames]{xcolor}
\definecolor{yaleblue}{RGB}{15,77,146}
\definecolor{hypblue}{HTML}{285B79}
\definecolor{hyplight}{HTML}{E7F0F5}
\definecolor{hyporange}{HTML}{AE662D}

\usepackage{tikz}
\usetikzlibrary{arrows.meta,calc,positioning,patterns}
\usepackage{pgfplots}
\pgfplotsset{compat=1.18}
\usepgfplotslibrary{groupplots,colormaps}
\usepackage{pgfplotstable}

\usepackage{enumitem}
\usepackage[multiple]{footmisc}
\usepackage{comment,soul,marginnote}
\usepackage{todonotes}

\usepackage[round,authoryear]{natbib}
\usepackage{hyperref}
\hypersetup{
    colorlinks=true,
    linkcolor=black,
    citecolor=black,
    urlcolor=Mulberry,
    pdftitle={Selling Training Data},
    pdfauthor={Jingmin Huang, Wei Zhao, Renjie Zhong},
    bookmarksnumbered=true
}
\usepackage[nameinlink,noabbrev]{cleveref}

\numberwithin{equation}{section}

\newtheorem{theorem}{Theorem}

\newtheorem{corollary}{Corollary}

\newtheorem{definition}{Definition}

\newtheorem{lemma}{Lemma}

\newtheorem{proposition}{Proposition}

\newcommand{\ubar}{\underline{u}}

\definecolor{hypgreen}{RGB}{34,139,34}

\title{Selling Training Data\thanks{We are deeply grateful to Laura Doval for her constant guidance and support, as well as to Yeon-Koo Che,  Jacopo Perego, Andreas Kleiner, Frank Yang, and Kai-Hao Yang for helpful discussions at various stages of this project. We also thank  Daniel Chen, Sulagna Dasgupta, Zhimin Feng, Nima Haghpanah, Nicole Immorlica, Shachar Kariv, Navin Kartik, Jan Knoepfle, Zihao Li, Xiao Lin, Elliot Lipnowski, Axel Niemeyer, Alessandro Pavan, Chris Shannon, Ali Shourideh, Alex Smolin, Tristan Tomala, Maren Vario, Cole Wittbrodt,  Kun Zhang,  Jidong Zhou, and Weijie Zhong as well as seminar and conference participants at the 2026 NBER/NSF/CEME Conference on Mathematical Economics, the 2026 Econometric Society IF-AIML, the Columbia Theory Lunch, 2025 PKU-NUS conference on Quantitative Finance and Economics, 2025 CES Annual conference, the Berkeley Theory Lunch, the 7th World Congress of the Game Theory Society (GAMES 2024), the 2024 Stony Brook Game Theory Conference, and the 2024 GAIMSS for helpful comments.  We thank ChatGPT for research assistance. Jingmin Huang thanks Yale University and Monash University for hosting his visits. Renjie Zhong thanks UC Berkeley and the Hausdorff Institute for Mathematics at the University of Bonn (DFG, German Research Foundation under Germany\textquoteright{}s Excellence Strategy EXC-2047/2\textemdash{}390685813) for hosting his visits.  All errors are our own. \href{https://www.china-ces.org/Files/3058abstract/202501050708364538.pdf}{An early version of this paper, compiled on Nov 26, 2024, can be found on Google Scholar. It is posted  by the CES society on Jan 5, 2025.} }}
\author{Jingmin Huang\thanks{School of Economics, Renmin University of China, China. \textit{\ Email}:
			jingmin.huang@ruc.edu.cn } \and Wei Zhao\thanks{
		School of Economics and Management, Tsinghua University, China. \textit{\ Email}:
		wei.zhao@outlook.fr } \and Renjie Zhong\thanks{
			Department of Economics, Columbia University, USA. \textit{\ Email}:
			renjiezhongecon@gmail.com } }

\date{This version: Nov 22, 2026\\ \href{https://www.china-ces.org/Files/3058abstract/202501050708364538.pdf}{First posted version: Nov 26, 2024}}

\begin{document}

% Local layout for the title page
\begingroup
\linespread{1.16}\selectfont
\setlength{\parskip}{2pt}
\setlength{\footnotesep}{.18cm}
\setlength{\skip\footins}{8pt plus 2pt minus 1pt}

\makeatletter
\def\@maketitle{%
  \newpage
  \null
  \vskip .65em%
  \begin{center}%
    \let\footnote\thanks
    {\LARGE\bfseries \@title\par}%
    \vskip .85em%
    {\large
      \lineskip .5em%
      \begin{tabular}[t]{c}\@author\end{tabular}\par}%
    \vskip .65em%
    {\small \@date\par}%
  \end{center}%
  \par
}
\makeatother

\maketitle

\begin{abstract}

This paper studies the sale of supplementary training dataset to improve predictive model. The buyer not only has private knowledge of its own dataset, but also can choose the way to merge these two datasets. This flexibility imposes a new constraint on the menu, requiring the buyer\textquoteright{}s total net utility be monotonic in the Blackwell order on private datasets. Given that, we show that a single posted price for a fully informative dataset is optimal in hypothesis testing, under differential virtual-value condition. The result extends to multi-class classification in which private datasets share a clean component but differ in how their labels are contaminated.

\end{abstract}

\clearpage
\endgroup

%\tableofcontents

\newpage
%\tableofcontents

\section{Introduction}
\label{sec:introduction}
Artificial intelligence has been delegated to make a growing number of decisions, including medical diagnosis, AI content safety moderation, credit assessment, and autonomous driving. AI differs from automation in its capacity to learn. This capability is grounded in scaling laws whereby large volumes of high-quality data act as fuel for the model. This has in turn
incubated a large upstream training-data industry, spanning data collection, labeling, and decontamination.\footnote{In practice, firms can supplement their own training data by licensing existing datasets, obtaining data through partnerships, or commissioning suppliers to collect or label examples.} For instance, Wayve supplements driving data from its own fleet with datasets from fleet partners and automakers.\footnote{See \href{https://wayve.ai/thinking/multi-country-generalization/}{Wayve's website account} (https://wayve.ai/thinking/multi-country-generalization/).} In AI content safety, OpenAI hired Sama to label text for a classifier that detects harmful content in ChatGPT.\footnote{See \href{https://time.com/6247678/openai-chatgpt-kenya-workers/}{Billy Perrigo} \emph{Time}, January 18, 2023.}

Training data has two distinguishing properties as an economic good. On the one hand, datasets are inherently multidimensional. They differ not only in the number of mistakes but also in the types of mistakes that arise in prediction. On the other hand, buyers can choose how to merge a supplementary dataset with their private data.\footnote{When adapting its model to U.S.\ roads, Wayve fine-tuned it on additional U.S.\ driving data to handle situations such as four-way stops (see \href{https://wayve.ai/thinking/multi-country-generalization/}{Wayve's website account}). In one content moderation, OpenAI used an existing classifier to select examples from their own production traffic on which the classifier was uncertain. They obtained labels for those examples and retrained the classifier on the expanded dataset \citep{MarkovEtAl2023}. In both settings, what the existing model has learned shapes which additional data are worth obtaining and how the firm uses them.} These two properties have \textit{conflicting} implications for data sales. The first enlarges, while the second limits, the seller's ability to discriminate among buyers. This paper tries to examine these two implications in the design and pricing of supplementary datasets.

A buyer (he) privately knows the statistical performance of the model trained on his private data. He can also purchase supplementary training data from a monopolist (she) to improve his predictive model. Each dataset is summarized by the prediction it generates, and we assume that the predictions produced by the two datasets are conditionally independent given the state. The seller designs a menu of supplementary datasets, each with an associated price. The buyer first selects a dataset from the menu and pays the corresponding price. He then chooses how to combine the supplementary dataset with his private data when training the model. In other words, once the model is deployed, the buyer must decide, in each scenario, how to aggregate the two predictions drawn from the two datasets into a final prediction.

%a monopolist selling supplementary training data to a buyer to improve his predictive model. This model is previously trained based on who already has a dataset. At the time of purchase, the buyer privately knows the statistical performance of the model trained on its own data. The contract is chosen before the predictions for a future case arrive. We represent each dataset by the prediction machine it produces and assume that the two predictions are independent conditional on the state. The seller can designs the menu of supplement training dataset and its associated price. The buyer decides how to combine the datasets to train the prediction machine and can always rely on its own model alone. The buyer's private dataset thus determines both its outside option and what it can gain from each supplement.

In the first step, we focus on direct mechanisms without loss of generality, in which, upon learning the buyer's reported private dataset, the seller recommends both a supplementary dataset and a rule for aggregating predictions. A challenge here is that the buyer can make a double deviation, i.e. he can misreport his private dataset and choose a different aggregation rule at the same time. When aggregating conflicting predictions, if the buyer puts less weight on the prediction from his private dataset, he behaves as if his private dataset were less informative. To rule out such deviations, we impose a new constraint other than truth-telling and obedience, requiring that  total net utility be monotone in the informativeness of the private dataset. Under this additional constraint, termed as \textit{Blackwell monotonicity}, we establish a revelation principle. This characterization recasts data sales as a bundling problem. In particular, hypothesis testing can be framed as the sale of two goods, namely reductions in false positives ($\alpha$, Type I error) and false negatives ($\beta$, Type II error).

We proceed to examine the implication of Blackwell monotonicity in optimal menu. One classical conclusion is that a posted price for the grand bundle is no longer optimal in multi-good sales, even under the condition (termed as regularity condition) of a differential virtual-value (See \citet{McAfeeMcMillan1988}).\footnote{This condition is a multidimensional analogue of a nondecreasing density-weighted virtual value \citep{Myerson1981}. \citet{McAfeeMcMillan1988} argued that it suffices for a deterministic mechanism to be optimal in two-good monopoly pricing. \citet{Thanassoulis2004} gave a counterexample in which lotteries strictly increase revenue.} The sub-optimality is due to its failure to incorporate horizontal differences in price discrimination. In data sales, a fully informative dataset is the analogue of the grand bundle because it eliminates both false positives and false negatives. However, we show that, under the same regularity condition, such simple mechanism becomes optimal in hypothesis testing. The argument exactly centers on Blackwell monotonicity. First, for buyers with completely uninformative private dataset, their predictions can have different mixes of false positives and false negatives. However, each prediction can reproduce the others by randomly switching their default model. Blackwell monotonicity requires that these buyers achieve the same net utility, and their common outside option gives them the same rent. Second, for buyers with more informative dataset, the combination of incentive and participation constraint then give a lower bound on the rent. A single price for full information can attain all these bounds at once. Third, the regularity condition then implies that reducing these rents weakly raises revenue. The same argument can be generalized to multiclass classification under our generalized contamination type space, in which private datasets share a clean component and differ in how their labels are contaminated. It nests many important real applications for data training.

Beyond regularity, we establish strong duality for the hypothesis-testing problem and derive a certification condition for a posted price to be optimal. Blackwell monotonicity also coarsens the canonical menu structures studied by \citet{Pavlov2011,DDT2017} to posted pricing. We trace Blackwell monotonicity to buyers' ability to choose  a set of feasible prediction models trained by dataset. We then discuss how Blackwell monotonicity constrains allocations, rent extraction, and dual measure design. Finally, we compare our ex-ante data-sales model with the interim information-sales model of \citet{BergemannBonattiSmolin2018}.

\interfootnotelinepenalty=10000
\subsection*{An Illustrative Example: AI Content Safety.}
\label{sec:illustrative-example}

An AI provider is considering buying additional labeled data to improve the safety classifier it uses to classify responses before release.\footnote{For an example of a safety classifier trained on labeled prompt--response pairs, see \href{https://arxiv.org/abs/2312.06674}{Inan et al. (2023)}.} We use this example to show why providers with equally accurate classifiers can value the same dataset differently, and how their ability to combine new data with their existing data shapes the seller’s menu. To facilitate interpretation, we only provide intuition and abstract away the details, when reformulating the selling data into selling multiple goods. 

%the details of reformulating 

Take a potential provider's classifier $A$ as an example. Trained alone on private data, $A$ flags $20$ of $100$ safe responses and allows $40$ of $100$ harmful responses to pass. Its accuracy is $70\%$. Another provider's classifier $B$ has the same accuracy but flags $16$ of $100$ safe responses and allows $44$ of $100$ harmful responses to pass. We can represent a provider's private data by the unconditional probabilities of flagging a safe response ($\alpha$, false positives) and allowing a harmful response to pass ($\beta$, false negatives). The AI provider can further decrease these two kinds of statistical errors, after acquiring supplementary data. Let $q_\alpha$ ($q_\beta$) be the proportion of false positives (negatives) that can be reduced through the supplementary dataset, if the buyer follows the seller's recommended way of dataset merging.

However, the AI provider may choose a different way of dataset merging. When there exists a conflict in predictions drawn from these two datasets, if the AI provider places less weight on that drawn from private dataset, then he acts \emph{as if} he has less informative private dataset and follows the recommended way of merging dataset. We term this approach as model retune. For instance, in Figure \ref{fig:example-garbling-region}, the AI provider \(A\) can retune to any point in \(\mathcal{R}_A\). Consider supplementary training data \(q=(0.55,1)\), which correct 55\% of false positives and all false negatives. Under \(A\)'s original rule, these data raise its correct classification rate from 70\% to 95.5\%. Now suppose \(A\) retunes to \((0,0.50)\) by initially allowing every response. Its private classifier's accuracy falls from 70\% to 50\%, but all its errors are now false negatives. The supplementary data correct every one of them, raising the final correct classification rate to 100\%.

%AI providers may use the data differently from what the training data seller intended. They can adjust how they use their private classifiers’ predictions. This adjustment can lower the accuracy of the private classifier on its own while increasing the value of the supplementary data. For example, provider \(A\) can retune from \((0.10,0.20)\) to \(C=(0.20,0.15)\) by randomly flagging one-quarter of the responses its classifier would otherwise allow, as shown in Figure \ref{fig:retuning-experiment}.

\begin{figure}[htbp]%
\centering%
\begingroup%
\definecolor{menuBlue}{HTML}{255C78}%
\definecolor{menuShade}{HTML}{E8F0F5}%
\definecolor{menuRust}{HTML}{A74E23}%
\definecolor{menuTeal}{HTML}{346967}%
\definecolor{menuViolet}{HTML}{655190}%
\definecolor{menuGray}{HTML}{616C75}%
\definecolor{menuDashed}{HTML}{737D83}%
\definecolor{menuInk}{HTML}{20252A}%
%
% The drawing uses printer's points (bp). Text sizes and strokes scale together.
\newcommand{\menuFace}[2]{\usefont{T1}{stix}{#2}{n}\fontsize{#1bp}{#1bp}\selectfont}%
% Isolate the three mathematical glyphs from the manuscript's math font.
\newcommand{\menuGlyph}[3]{{\font\menulocalglyph=#1 at #2bp\relax%
                             \menulocalglyph\char#3\relax}}%
\newcommand{\menuSub}[1]{\raisebox{-2.1bp}{\menuFace{9.1}{m}#1}}%
\resizebox{\linewidth}{!}{%
\begin{tikzpicture}[x=1bp,y=1bp,
    every node/.style={inner sep=0pt,outer sep=0pt,text=menuInk},
    line cap=round,line join=round]
  \path[use as bounding box] (0,0) rectangle (787,335.5);

  % Left panel. Points are given in probability units, with equal axis scales.
  \begin{scope}[shift={(54,51.1)},x=514bp,y=514bp]
    \coordinate (A) at (.10,.20);
    \coordinate (B) at (.08,.22);
    \coordinate (C) at (.20,.15);
    \coordinate (N) at (0,.50);
    \coordinate (P) at (.50,0);

    \fill[menuShade] (N)--(A)--(P)--cycle;
    \draw[menuBlue,line width=1.25bp] (N)--(P)--(A);
    \draw[menuDashed,line width=1bp,dash pattern=on 4bp off 3bp]
      (0,.30)--(.30,0);
    \draw[menuGray,line width=.9bp] (-.012,0)--(.523,0);
    \draw[menuGray,line width=.9bp] (0,-.012)--(0,.535);
    \foreach \x in {0,0.1,0.2,0.3,0.4,0.5}{
      \draw[menuGray,line width=.8bp] (\x,0)--(\x,-.006);
      \node[anchor=base,font=\menuFace{11}{m},text=menuGray]
        at (\x,-.0302) {\x};
    }
    \foreach \y in {0.1,0.2,0.3,0.4,0.5}{
      \draw[menuGray,line width=.8bp] (0,\y)--(-.006,\y);
      \node[anchor=base east,yshift=-3.75bp,font=\menuFace{11}{m},text=menuGray]
        at (-.0156,\y) {\y};
    }

    \draw[menuRust,line width=2.2bp,-{Triangle[length=10.1bp,width=8.9bp]},
          shorten <=4bp,shorten >=4.5bp] (A)--(C);
    \draw[menuViolet,line width=1.65bp,dash pattern=on 6.77bp off 3.47bp,
          -{Triangle[length=8.9bp,width=8bp]},shorten <=5bp,shorten >=5bp]
          (A)--(N);
    \foreach \name/\col in {A/menuRust,B/menuTeal,C/menuRust}{
      \filldraw[fill=\col,draw=white,line width=.7bp] (\name) circle[radius=3.35bp];
    }
    \foreach \name in {N,P}{\fill[menuBlue] (\name) circle[radius=2.75bp];}

    \node[anchor=base west,font=\menuFace{12}{m},text=menuGray]
      at (.0097,.5117) {Never flag N};
    \node[anchor=base west,font=\menuFace{12}{m},text=menuGray]
      at (.5058,.0175) {Always flag};
    \node[anchor=base west,font=\menuFace{18}{m}\itshape,text=menuRust]
      at (.0686,.1723) {A};
    \node[anchor=base west,font=\menuFace{18}{m}\itshape,text=menuTeal]
      at (.0427,.2162) {B};
    \node[anchor=base west,font=\menuFace{18}{m}\itshape,text=menuRust]
      at (.1960,.1652) {C};

    \node[anchor=base west,rotate=-45,font=\menuFace{12}{m},text=menuGray]
      at (.1082,.1568) {70\% accuracy};
    \node[anchor=base west,rotate=-71.565,font=\menuFace{12}{m},text=menuViolet]
      at (.0428,.4235) {Uses B's contract};
    \node[anchor=base west,font=\menuFace{19}{m},text=menuBlue]
      at (.187,.2268) {\menuGlyph{stix-mathcal}{19}{82}%
                       \raisebox{-2.6bp}{\menuFace{13.3}{m}\itshape A}};
  \end{scope}

  \node[anchor=base west,font=\menuFace{13}{m}] at (140,12.3)
    {False positives, \menuGlyph{stix-mathit}{13}{11}};
  \node[anchor=base west,rotate=90,font=\menuFace{13}{m}] at (18.3,138.4)
    {False negatives, \menuGlyph{stix-mathit}{13}{12}};

  % Right panel. All entries in one row use a common baseline.
  \begin{scope}[shift={(401.6,335.5)},y=-1bp]
    \newcommand{\menuTitle}[3]{%
      \node[anchor=base west,font=\menuFace{16}{b},text=#2] at (0,#1) {#3};}
    \newcommand{\menuRevenue}[2]{%
      \node[anchor=base east,font=\menuFace{14}{m}] at (375.6,#1) {Revenue\hspace{.5em}#2};}
    \newcommand{\menuRow}[6]{%
      \begin{scope}[every node/.append style={font=\menuFace{#2}{m}}]
        \node[anchor=base west] at (0,#1) {#3};
        \node[anchor=base] at (195.5,#1) {#4};
        \node[anchor=base] at (282.0,#1) {#5};
        \node[anchor=base] at (345.9,#1) {#6};
      \end{scope}}
    \newcommand{\menuHeader}[1]{%
      \menuRow{#1}{13}{Dataset}{(\textit{q}\menuSub{$\alpha$},\,\textit{q}\menuSub{$\beta$})}{Price}{Types}}
    \newcommand{\menuRule}[3]{%
      \draw[#3,line width=#2bp,line cap=rect] (0,#1)--(376,#1);}

    \node[anchor=base west,font=\menuFace{17}{b}] at (0,29.55) {Optimal menus under truthtelling and participation};
    \menuTitle{65.14}{menuBlue}{Without Retuning}
    \menuRevenue{64.63}{0.31}
    \menuRule{73.24}{1.25}{menuInk}
    \menuHeader{87.86}
    \menuRule{95.86}{.60}{menuGray}
    \menuRow{112.97}{14}{A-targeted}{(0.20,\,0.40)}{0.10}{\textit{A}}
    \menuRow{135.87}{14}{B-targeted}{(0.55,\,1)}{0.26}{\textit{B}}
    \menuRow{158.76}{14}{Full information}{(1,\,1)}{0.35}{\textit{C}}
    \menuRule{167.09}{1.10}{menuInk}

    \menuTitle{206.87}{menuRust}{With Retuning}
    \menuRevenue{206.25}{0.30}
    \menuRule{214.86}{1.25}{menuInk}
    \menuHeader{229.48}
    \menuRule{237.48}{.60}{menuGray}
    \menuRow{254.59}{14}{Full information}{(1,\,1)}{0.30}{\textit{A, B, C}}
    \menuRule{262.92}{1.10}{menuInk}
  \end{scope}
\end{tikzpicture}%
}%
\endgroup
\setlength{\abovecaptionskip}{4pt}
\caption{Retuning and optimal menus.}
\label{fig:example-garbling-region}
\end{figure}

Consider a market with three types of providers, \(A\), \(B\), and \(C\), with \((\Pr(A),\Pr(B),\Pr(C))=(0.025,0.375,0.600)\). When providers cannot retune, the seller tailors the two kinds of error reduction to offer each type a personalized contract.\footnote{In this example, the benchmark imposes only truth-telling and
individual rationality. Imposing obedience yields the
same maximum revenue of $0.30$ as allowing retuning. In the hypothesis testing with full type space, obedience and Blackwell monotonicity
are equivalent given truth-telling (see Appendix \ref{reduce}).} Although \(A\) and \(B\) have the same accuracy and their error rates differ only slightly, their menus differ significantly. Without retuning, \(A\)'s payoff, measured by its correct classification rate net of price, is \(0.700\) under its own contract and \(0.695\) under \(B\)'s contract. Thus, \(A\) has no incentive to report \(B\) if it keeps its way of using private data. But \(A\) can first report \(B\) and then retune its classifier to never flag \(N=(0,0.50)\). This double deviation raises \(A\)'s net payoff to \(0.740\). Once the seller accounts for retuning, it maximizes revenue in turn by offering a single contract to all three types, selling full information at a price of \(0.30\).

\subsection*{Related Literature}
Our paper relates to several strands of literature. It is primarily related to the literature on information markets,\footnote{It is large literature dating back to \cite{AdmatiPfleiderer1986} and \cite{AdmatiPfleiderer1990}. See \citet{BergemannBonatti2019} for an overview and recent developments in \citet{AliEtAl2022,DovalSkreta2026, DworczakPeregoVeldkamp2026, HaghpanahLinSiegel2026}.} especially work on monopoly pricing of information structures for decision makers. The closest paper is \citet{BergemannBonattiSmolin2018}, who study interim sales of information to buyers with private posterior beliefs. We develop this comparison in Section \ref{sec:bbs-discussion}. \citet{Yang2022} studies the sale of consumer information to a producer who uses it to price discriminate. \citet{SeguraRodriguez2022} studies the sale of statistics to firms with private forecasting targets and valuations for information. \citet{EliazSpiegler2026} study the storage and sale of current and historical data to buyers with different prediction horizons. To our knowledge, we are the first to study the optimal sale of information structures when the buyer's private type is itself an information structure. The buyer chooses a contract before observing either signal realization, and the signals are independent conditional on state. This ex ante timing captures a firm buying training data to improve a classifier before applying it to future cases.

Our results also deliver new insights into multidimensional mechanism
design and optimal bundling, a literature dating back to
\citet{AdamsYellen1976} and \citet{McAfeeMcMillanWhinston1989}.\footnote{See \citet{LahrNiemeyer2025} for an overview.}
In the multiple-good monopoly problem, \citet{LahrNiemeyer2025} show
that mechanisms that can be rationalized as uniquely revenue maximizing
for some type distribution are dense among incentive-compatible and
individually rational mechanisms. Some of the existing literature focus on imposing structures to guarantee the optimality of simple menu.\footnote{Robustness over ambiguity set can justify separate sales or categorical bundling
\citep{Carroll2017,CheZhong2025}. Conditions on the distribution of
values or on relative valuations can justify pure bundling
\citep{DDT2017,HaghpanahHartline2021}. With one-dimensional consumer
heterogeneity, conditions on bundle demand can justify nested menus
\citep{Yang2025}. Sufficiently strong complementarities around a core
of goods can justify selling the core with optional add-ons
\citep{Zhong2026}.} An early attempt to extend the
one-dimensional ``no-haggling'' result of \citet{RileyZeckhauser1983}
to multiple goods was the claim by \citet{McAfeeMcMillan1988} that
a differential virtual value condition suffices for an optimal menu
of posted prices for grand bundles. This differential virtual value condition is a multi-dimensional version of Myersonian nondecreasing virtual value. However, \citet{Thanassoulis2004}
showed that lotteries can be necessary and \citet{Pavlov2011}  restricts the form of an optimal two-good menu to partial allocations under this condition. Training data are a particular kind
of multidimensional good that buyers can combine with their own data
and use flexibly in model training. This flexibility imposes Blackwell
monotonicity and coarsens the optimal menu. With this additional
constraint, the differential virtual value condition yields
``no-haggling'' result. 

The ex ante timing also connects our model to sequential screening.
In \citet{CourtyLi2000}, buyers initially know the distribution of
their valuations and learn their realized valuations later.
\citet{EsoSzentes2007} and \citet{LiShi2017} allow the seller to design
the disclosure of information about those valuations. Here the
buyer's initial type specifies how he learns about an uncertain state
and is multidimensional even in binary state.

Methodologically, we follow \citet{Rochet1987} and
\citet{RochetChone1998} by formulating the seller's problem in terms of
the buyer's indirect rent and optimizing directly over it.  However, Pricing information structures for decision makers raises three
challenges. \footnote{Since we have of type-dependent outside option, we use indirect rent instead of indirect utility to avoid mixing it with indirect total net utility.} First, both the seller's allocation and the buyer's private
type are information structures, giving rise to a multidimensional
screening problem over information structures. Second, the buyer's private information determines
what he can achieve without a purchase, so the individual rationality
constraint is type dependent. Third, the value of supplementary
information depends on the buyer's action after observing the signals.
In a direct recommendation mechanism, a buyer can therefore jointly
deviate by misreporting his type and departing from the recommended
action \citep{Myerson1982}. 

We establish a revelation principle
shows that, the buyer can reproduce
the payoff from his optimal continuation after a misreport by
garbling his private signal and then following the recommendation.
This argument parallels the compensating second report in
\citet{EsoSzentes2007} and imitating feasible technology in
\citet{BullWatson2007} and \citet{GanLi2026}. Under the differential virtual value condition,
we obtain the posted-price result.
Beyond regularity, we establish strong duality for hypothesis testing,
building on the duality approach to monopoly pricing
\citep{DDT2017,KleinerManelli2019,Zhong2026}. Blackwell monotonicity
changes the stochastic order in the dual by allowing revenue weights
to be reshuffled among Blackwell-equivalent types. Blackwell monotonicity thus reduces the resulting optimization over measures from a coupling problem \citep{DDT2017,KleinerManelli2019,Zhong2026} to a generalized persuasion problem \citep{YangYang2026}.

Section~\ref{sec:model} presents the model and
Section~\ref{sec:BM} establishes the revelation principle. Section~\ref{sec:main-result}  establishes the optimality of posted price, and Section~\ref{GCS} further extends the optimality of posted price to general type space. Section~\ref{sec:final-discussion} develops the
duality framework, compares with interim information sales, and
discusses the implications of Blackwell monotonicity.

\section{Model}
\label{sec:model}

\paragraph{Decision problem and private datasets.}
There are finitely many states $\Omega=\{1,\ldots,m\}$ and actions
$A=\{1,\ldots,J\}$. The prior $\pi\in\operatorname{int}\Delta(\Omega)$
and the decision payoff $\upsilon:A\times\Omega\to\mathbb R$ are common
knowledge and the same for all buyers. A dataset is an information structure
about the state. The buyer's private dataset has signal set
$S=\{1,\ldots,n\}$ and is represented by a joint distribution
\begin{equation}
 P\in\Theta_{m,n}:=
 \left\{P\in\Delta(\Omega\times S):
 \sum_{s\in S}P(\omega,s)=\pi_\omega
 \text{ for every }\omega\in\Omega\right\}.
 \label{eq:model-types}
\end{equation}
Its conditional likelihoods are $P(s\mid\omega)=P(\omega,s)/\pi_\omega$.
The type domain is a Borel set $\Theta\subseteq\Theta_{m,n}$. The
seller knows its distribution $F$, but observes neither the buyer's
dataset $P$ nor his private signal. In prediction applications, $s$
is the output of a prediction machine trained on the private dataset.
The joint law $P$ describes the machine's statistical performance.

At contracting, the buyer knows $P$ but has not observed $s$.
Every type has the same prior over the state: $\pi\in\Delta(\Omega)$.
The buyer's private information therefore describes how he will learn
about the state. The distinction between a privately known distribution and its
subsequent realization connects our model to sequential screening
\citep{CourtyLi2000}.
In our setting, the private type $\{P(\cdot\mid\omega)\}_\omega$  is 
multidimensional.
We return to this connection in Section~\ref{sec:discussion}.

\paragraph{Example: multiclass classification.}\label{example} A buyer classifies targets into $m$ groups, $\Omega=A=\{1,2,...,m\}$. Each group is equally likely, $\pi_\omega=1/m$ and the buyer receives one for a correct classification and zero otherwise, $\upsilon(a,\omega)=\mathbf1\{a=\omega\}$. The buyers' model trained by private dataset generates $n$ labels, $S=\{1,2,...,n\}$. Hypothesis testing is a binary classification with $n=2$.\footnote{Notice that any binary state-decision problem $\Omega=A=\{1,2\}$ with general utility $\upsilon(a,\omega)$, undominated actions, and balanced prior $\pi$, can be normalized to binary classification.}

\paragraph{Dataset contracts.}
The seller commits to a menu before the buyer chooses a contract. A contract specifies a supplementary
dataset $E:\Omega\to\Delta(Y)$ with a finite signal set $Y$, and a
payment $t\geq0$. The seller can supply any such dataset at zero cost. The private and supplementary signals are
independent conditional on the state. Consequently, if type $P$ purchases $E$, the joint law is
\begin{equation}
 \Pr(\omega,s,y\mid P,E)=P(\omega,s)E(y\mid\omega).
 \label{eq:model-joint-law}
\end{equation}
The buyer pays before observing $(s,y)$ and then chooses an action.
Neither the state, two signals, or the action are contractible.
A buyer who declines the contract observes only $s$ and chooses an action
using his private dataset. So the outside option is type-dependent.

\paragraph{Direct mechanisms and recommendations.}

Following \citet{Myerson1982}, a direct recommendation mechanism assigns a contract and a decision rule
to each report $r\in\Theta$:
\begin{equation}
 M=\{(E_r,\sigma_r,t_r)\}_{r\in\Theta},
 \qquad \sigma_r:S\times Y\to A, \qquad t_r\ge0.
 \label{eq:model-mechanism}
\end{equation}
The rule $\sigma_r$ specifies the obedient use of both datasets. The buyer remains free to choose any action after observing
both signals.\footnote{In \citet[Section~II.B]{BergemannBonattiSmolin2018}, a supplemental
dataset recommends actions to a buyer who already knows his private
belief. Here the private signal arrives after contracting, so the
recommendation specifies a strategy for each realization of that signal pairs $(s,y)$.}
Equivalently, the mechanism may request a second report $\widehat s$
after the signals arrive and recommend $\sigma_r(\widehat s,y)$, while leaving $E_r$ and $t_r$ fixed.

The timing is summarized in Figure~\ref{fig:model-timeline}.

\begin{figure}[htbp]
\centering
\begin{tikzpicture}[x=1cm,y=1cm,>=Latex,font=\small,
                     line cap=round,line join=round]
 \draw[->,black!65,line width=.7pt] (.4,0)--(14.1,0);
 \foreach \x in {1.3,5.2,9.1,13.0}
    \fill[hypblue] (\x,0) circle (2.2pt);
 \node[above=8pt,font=\small\bfseries] at (1.3,0) {Commitment};
 \node[above=8pt,font=\small\bfseries] at (5.2,0) {Contract choice};
 \node[above=8pt,font=\small\bfseries] at (9.1,0) {Signals};
 \node[above=8pt,font=\small\bfseries] at (13.0,0) {Action};
 \node[anchor=north,align=center,text width=2.8cm,font=\footnotesize]
    at (1.3,-.29) {Menu committed.\\Buyer knows $P$.};
 \node[anchor=north,align=center,text width=2.8cm,font=\footnotesize]
    at (5.2,-.29) {Buyer reports $r$\\and pays $t_r$.};
 \node[anchor=north,align=center,text width=2.8cm,font=\footnotesize]
    at (9.1,-.29) {Buyer observes\\signals $s$ and $y$.};
 \node[anchor=north,align=center,text width=2.8cm,font=\footnotesize]
    at (13.0,-.29) {Buyer chooses $a$\\using both signals.};
 \draw[black!35] (.35,.86)--(6.15,.86);
 \draw[black!35] (.35,.80)--(.35,.92) (6.15,.80)--(6.15,.92);
 \node[above=3pt,font=\footnotesize,text=black!65] at (3.25,.86)
    {Neither signal is observed};
\end{tikzpicture}
\caption{Timing. }
\label{fig:model-timeline}
\end{figure}

\paragraph{Payoffs and allocations.}
For a true type $P$ and a report $r$, let
\begin{align}
 W^o(P,r)&=\sum_{\omega,s,y}P(\omega,s)E_r(y\mid\omega)
                   \upsilon(\sigma_r(s,y),\omega),
                   \label{eq:model-prescribed}\\
 W(P,r)&=\sum_{s,y}\max_{a\in A}
           \sum_\omega P(\omega,s)E_r(y\mid\omega)\upsilon(a,\omega).
                   \label{eq:model-optimized}
\end{align}
The obedient payoff $W^o$ evaluates the recommended decision rule.
The optimized payoff $W$ lets the buyer choose his preferred action at
every signal pair. Thus $W^o(P,r)\leq W(P,r)$.

\paragraph{Designer's problem.}
The type-dependent outside option is the value of private dataset
\begin{equation}
 \ubar(P)=\sum_s\max_{a\in A}\sum_\omega P(\omega,s)\upsilon(a,\omega).
 \label{eq:model-outside}
\end{equation}
The revenue-maximization problem is
\begin{equation}
\begin{aligned}
 \sup_M\quad &\operatorname{Rev}_F(M)=\int_\Theta t_P\,dF(P)\\[4pt]
 \text{subject to}\quad
 &W^o(P,P)-t_P
   \geq W(P,r)-t_r  & (P,r\in\Theta),\qquad\mathrm{(IC)}\\[4pt]
 &W^o(P,P)-t_P\geq\ubar(P)
       & (P\in\Theta),\qquad\mathrm{(IR)}
\end{aligned}
\label{eq:model-full-ic}
\end{equation}
Selling training data presents three challenges. First, both the
buyer's private type and the seller's allocation are datasets, making
the problem multidimensional screening over information structures.
Second, the outside option is type dependent. Third, IC must rule out
joint deviations in reporting and subsequent data use. In Section \ref{sec:BM}, we establish a revelation principle to handle the joint deviations. In Section \ref{sec:main-result}, we focus on the full type space $\Theta=\Theta_{2,2}$ in hypothesis testing. Then in Section \ref{GCS}, we extend the optimality of posted price to \textit{generalized contamination type space}, which preserves all the three challenges and has wide applications in real life training data settings. 
 
% \sum_{\omega,s,y}P(\omega,s)E_r(y\mid\omega)
%                \upsilon(\delta(s,y),\omega)-t_r,
%                \qquad\mathrm{(IC)}

\section{A revelation principle for data sales}\label{sec:BM}
In this section, we first introduce Blackwell monotonicity. We then establish a revelation principle, characterizing when this restriction,
together with truthtelling and obedience, rules out all joint
deviations. In Section \ref{sec:hypothesis}, we will apply the revelation principle to  reformulate the designer's problem in hypothesis testing as monopolistic selling two goods. We discuss the sources and implications of Blackwell monotonicity in Section \ref{BMI}.

\begin{figure}[htbp]
\centering
\begin{tikzpicture}[x=1cm,y=1cm,>=Latex,font=\small,
                     line cap=round,line join=round]
 \node (state) at (0,0) {$\omega$};
 \node[draw=black!45,rounded corners=2pt,minimum width=.85cm,
       minimum height=.65cm] (signal) at (2.7,1.05) {$s$};
 \node[draw=hypblue,fill=hyplight,rounded corners=2pt,
       minimum width=.85cm,minimum height=.65cm] (retuned) at (5.65,1.05) {$s'$};
 \node[draw=black!45,rounded corners=2pt,minimum width=.85cm,
       minimum height=.65cm] (supplement) at (4.15,-1.05) {$y$};
 \node[draw=hypblue,fill=hyplight,rounded corners=2pt,
       minimum width=2.0cm,minimum height=.85cm]
       (rule) at (8.55,0) {$\sigma_r(s',y)$};
 \node (action) at (10.65,0) {$a$};
 \draw[->,black!65] (state)--(signal.west)
    node[midway,above=5pt,font=\footnotesize] {$P(s\mid\omega)$};
 \draw[->,hypblue,line width=1pt] (signal.east)--(retuned.west)
    node[midway,above=5pt] {$G(s'\mid s)$};
 \draw[->,black!65] (state)--(supplement.west)
    node[midway,below=5pt,font=\footnotesize] {$E_r(y\mid\omega)$};
 \draw[->,hypblue,line width=.8pt] (retuned.east)--(rule.north west);
 \draw[->,black!65] (supplement.east)--(rule.south west);
 \draw[->,hypblue,line width=.8pt] (rule.east)--(action);
\end{tikzpicture}
\caption{Imitation through garbling private predictions.}
\label{fig:bm-imitation}
\end{figure}
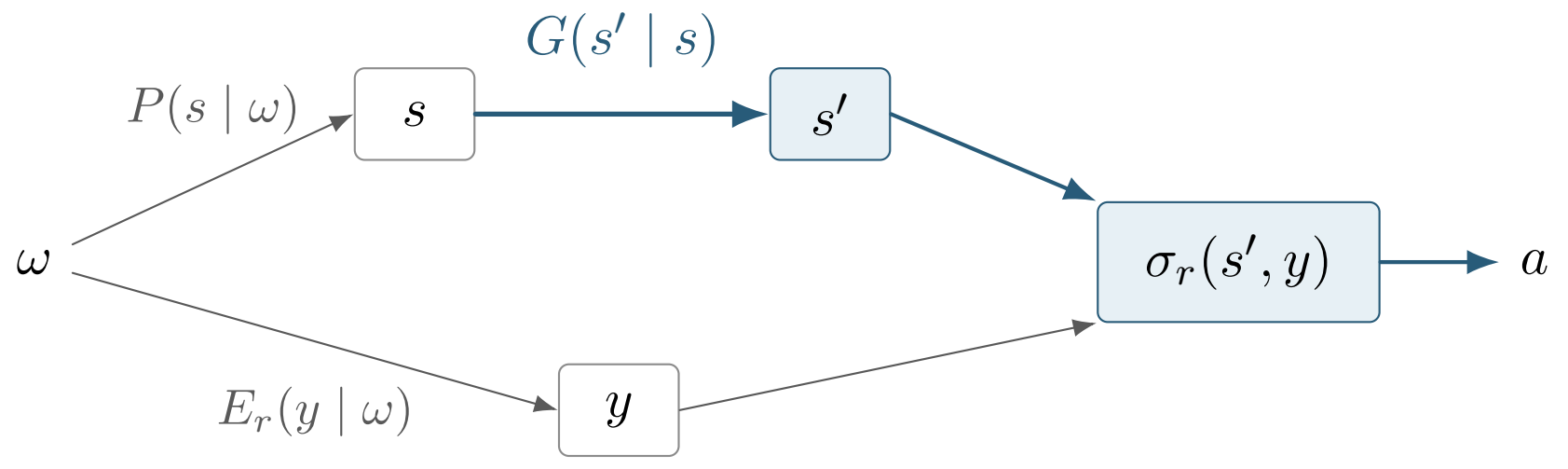

Write $\mathcal G_n$ for the set of stochastic matrices  $G:S\to\Delta(S)$. For a given $G$, define
\begin{equation}
     (PG)(\omega,s')=\sum_sP(\omega,s)G(s'\mid s).\label{PG}
\end{equation}
Write $P\succeq_B Q$ 
if $Q=PG$ for some $G\in\mathcal G_n$ \citep{Blackwell1953}. At any contract $r$, type
$P$ can garble to $Q$  through $G$ and follow
$\sigma_r(s',y)$. Conditional independence gives the joint law
$Q(\omega,s')E_r(y\mid\omega)$, so this reproduces type $Q$'s
obedient payoff at contract $r$, as in
Figure \ref{fig:bm-imitation}. Thus (IC) implies Blackwell monotonicity (BM):
\begin{align}
 U(P)&\geq U(Q)
        &&(P,Q\in\Theta,\ P\succeq_B Q).
        \tag{BM}\label{eq:model-bm}
\end{align}
where $U(P):=W^o(P,P)-t_P$ is the truthful and obedient total utility.

\label{sec:revelation}
\label{sec:discussion}

Notice that (IC) also implies  truth telling  (\ref{eq:model-tt}) and  Obedience (\ref{eq:model-ob}),
\begin{align}
 U(P)&\geq W^o(P,r)-t_r &&(P,r\in\Theta),
        \tag{TT}\label{eq:model-tt}\\
 W^o(P,P)&=W(P,P) &&(P\in\Theta),
        \tag{OB}\label{eq:model-ob}
\end{align}
Truth telling (TT) compares reports under obedient behavior;
obedience (OB) requires optimal use after a truthful report.
Blackwell monotonicity (BM) requires $U(P)\geq U(Q)$ whenever
$P\succeq_B Q$. 

Write $\mathcal M_{\mathcal C}$ for the mechanisms satisfying
constraints $\mathcal C$ and nonnegative payments.
Assume that every action is uniquely optimal at some full-support
belief.\footnote{Formally, for every $a\in A$, there exists
$\mu^a\in\operatorname{int}\Delta(\Omega)$ such that
$\sum_\omega\mu^a_\omega\upsilon(a,\omega)>
\sum_\omega\mu^a_\omega\upsilon(b,\omega)$ for every $b\ne a$.} 
We consider \emph{binary decisions}, with $\Omega=\{\omega_1,\omega_2\}$, $A=\{a_1,a_2\}$, relabeling $a_i\in\arg\max_{a} v(a,\omega_i)$. Let type space $\Theta=\Theta_{2,n}$.

For binary decisions, consider a direct mechanism
$\{(E_r,\sigma_r,t_r)\}_{r\in\Theta}$ with \emph{normalized contracts}.
The supplementary dataset $E_r$ has signal labels $y_0,\ldots,y_n$.
Whenever they occur, $y_0$ reveals state $1$ and $y_n$ reveals
state $2$. Each intermediate label $y_j$, $1\leq j\leq n-1$, is
either active, with positive probability in both states, or null,
with zero probability in both states. The likelihood ratios
$\ell_j=E_r(y_j\mid 2)/E_r(y_j\mid 1)$ are weakly increasing in $j$
across the active intermediate signals.

For each type $r$, order the nonnull private signals by increasing
posterior probability of state $2$ under $r$, and place the null
signals last. Write $s_1^r,\ldots,s_n^r$ for the resulting list.
For a fixed private signal, the posterior probability of state $2$
is weakly increasing in $j$ over supplementary signals that can occur
with it. An optimal continuation can therefore be chosen to switch
at most once from $a_1$ to $a_2$ as $j$ increases. The normalized
recommendation rule assigns one cutoff to each private signal:
\[
\sigma_r(s_i^r,y_j)=
\begin{cases}
a_1 & \text{if } j\leq n-i,\\
a_2 & \text{if } j>n-i.
\end{cases}
\]
Appendix~\ref{app:normalization} gives the details of the normalization. Table~\ref{fig:normalized-contract} illustrates a normalized contract
for $n=3$. Its three rows contain all three cutoffs on the two
imperfect signals, while every row recommends the matching action
at the revealing signals. In general, the $n$ rows contain all
$n$ cutoffs. Obedience requires the row
assigned to each private signal to be optimal for the intended type. Write
$\mathcal N_n$ for the mechanisms with normalized
contracts.
\begin{table}[!htbp]
\centering
\begin{tikzpicture}[x=1cm,y=1cm,font=\small,
  endpoint/.style={draw=black!20,fill=black!4,
    minimum width=1.70cm,minimum height=.50cm},
  firstaction/.style={draw=white,fill=hyplight,
    minimum width=1.70cm,minimum height=.50cm},
  secondaction/.style={draw=white,fill=hyporange!12,
    minimum width=1.70cm,minimum height=.50cm}]
 \node[font=\footnotesize] at (1.30,1.50) {Reveals state $1$};
 \node[font=\footnotesize] at (4.30,1.50)
    {Imperfect: $\ell_1\leq\ell_2$};
 \node[font=\footnotesize] at (7.30,1.50) {Reveals state $2$};
 \node[font=\footnotesize] at (-.30,.95) {Private signal};
 \foreach \xx/\jj in {1.30/0,3.30/1,5.30/2,7.30/3}
   \node at (\xx,.95) {$y_{\jj}$};
 \foreach \yy/\ii in {.30/1,-.30/2,-.90/3}{
   \node at (-.30,\yy) {$s^r_{\ii}$};
   \node[endpoint] at (1.30,\yy) {$a_1$};
   \node[endpoint] at (7.30,\yy) {$a_2$};}
 \node[firstaction] at (3.30,.30) {$a_1$};
 \node[firstaction] at (5.30,.30) {$a_1$};
 \node[firstaction] at (3.30,-.30) {$a_1$};
 \node[secondaction] at (5.30,-.30) {$a_2$};
 \node[secondaction] at (3.30,-.90) {$a_2$};
 \node[secondaction] at (5.30,-.90) {$a_2$};
\end{tikzpicture}
\caption{Recommended strategies in a normalized contract.
Each entry is the $\sigma_r(s_i^r,y_j)$.}
\label{fig:normalized-contract}
\end{table}

\begin{theorem}[Revelation principle]
\label{thm:dataset-revelation}\label{thm:binary}\label{thm:converse}
For binary decisions, restricting attention to
$\mathcal N_n\cap\mathcal M_{\mathrm{IC,IR}}$ is without loss
of revenue.\footnote{Every $M\in\mathcal{M}_{IC,IR}$ admits a $M'\in\mathcal{M}_{IC,IR}\cap \mathcal N_n$ with unchanged truthful utilities
and weakly higher payments.} Moreover,
\begin{equation}
 \mathcal N_n\cap\mathcal M_{\mathrm{IC,IR}}
 =\mathcal N_n\cap\mathcal M_{\mathrm{TT,OB,BM,IR}}.
 \label{eq:encoding-characterization}
\end{equation}
\end{theorem}

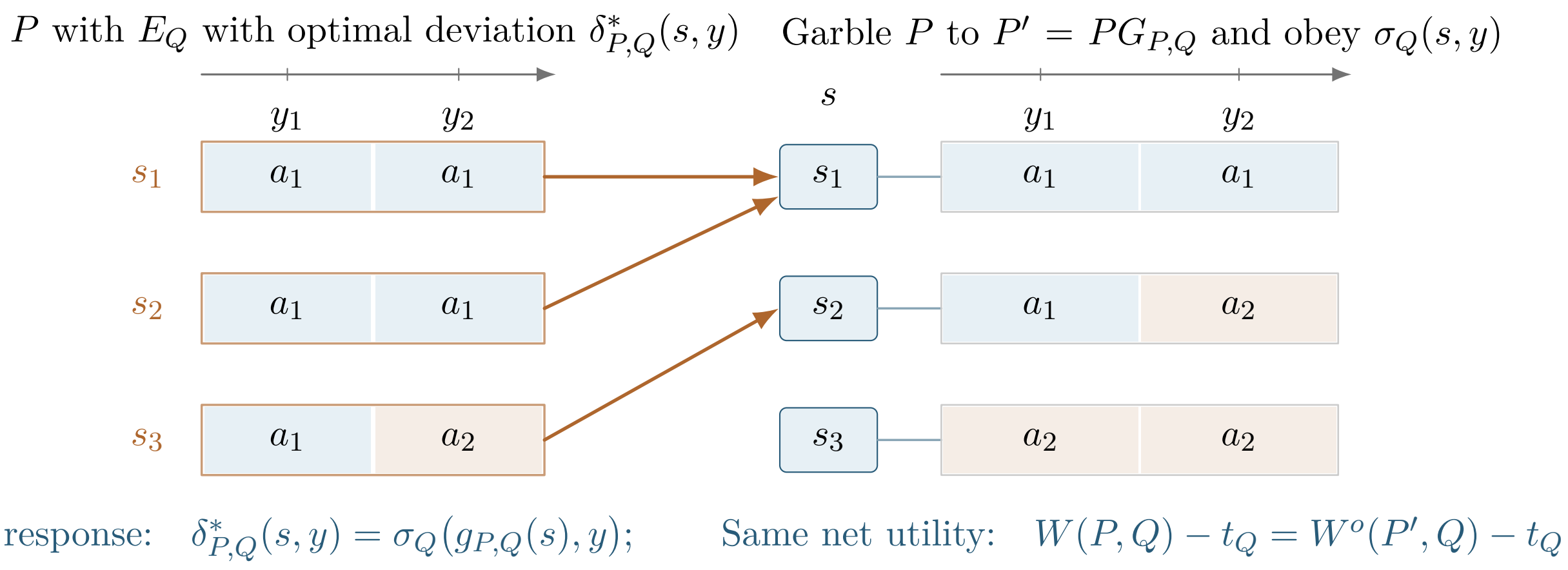
\begin{figure}[!htbp]
\centering
\begin{tikzpicture}[x=.925cm,y=1cm,>=Latex,font=\small,
  line cap=round,line join=round,
  firstaction/.style={fill=hyplight,minimum width=1.70cm,
    minimum height=.68cm},
  secondaction/.style={fill=hyporange!12,minimum width=1.70cm,
    minimum height=.68cm},
  rowlabel/.style={draw=hypblue,fill=hyplight,rounded corners=2pt,
    minimum width=1.00cm,minimum height=.66cm}]
 \node[align=center,text width=8.85cm] at (1.95,5.75)
    {$P$ with $E_Q$ with
     optimal deviation $\delta^*_{P,Q}(s,y)$};
 \draw[->,black!55,line width=.6pt] (.05,5.35)--(3.98,5.35);
 \foreach \xx/\jj in {1.00/1,2.90/2}{
   \draw[black!55] (\xx,5.29)--(\xx,5.41);
   \node[below=4pt] at (\xx,5.29) {$y_{\jj}$};}
 \foreach \yy/\ii in {4.30/1,2.95/2,1.60/3}{
   \node[text=hyporange] at (-.55,\yy) {$s_{\ii}$};
   \draw[hyporange!65,line width=.65pt]
      (.05,\yy-.36) rectangle (3.85,\yy+.36);}
 \node[firstaction] at (1.00,4.30) {$a_1$};
 \node[firstaction] at (2.90,4.30) {$a_1$};
 \node[firstaction] at (1.00,2.95) {$a_1$};
 \node[firstaction] at (2.90,2.95) {$a_1$};
 \node[firstaction] at (1.00,1.60) {$a_1$};
 \node[secondaction] at (2.90,1.60) {$a_2$};
 \node at (7.00,5.12) {$s$};
 \node[rowlabel] (r1) at (7.00,4.30) {$s_1$};
 \node[rowlabel] (r2) at (7.00,2.95) {$s_2$};
 \node[rowlabel] (r3) at (7.00,1.60) {$s_3$};
 \draw[->,hyporange,line width=1pt] (3.85,4.30)--(r1.west);
 \draw[->,hyporange,line width=1pt] (3.85,2.95)--([yshift=-.20cm]r1.west);
 \draw[->,hyporange,line width=1pt] (3.85,1.60)--(r2.west);

 % Right: all recommended rows of contract Q.
 \node[align=center,text width=7.85cm] at (10.45,5.75)
    {Garble $P$ to $P'=PG_{P,Q}$ and obey $\sigma_Q(s,y)$};
 \draw[->,black!55,line width=.6pt] (8.25,5.35)--(12.80,5.35);
 \foreach \xx/\jj in {9.35/1,11.55/2}{
   \draw[black!55] (\xx,5.29)--(\xx,5.41);
   \node[below=4pt] at (\xx,5.29) {$y_{\jj}$};}
 \foreach \rr/\yy in {r1/4.30,r2/2.95,r3/1.60}{
   \draw[hypblue!55,line width=.65pt] (\rr.east)--(8.25,\yy);
   \draw[black!20,line width=.5pt]
      (8.25,\yy-.36) rectangle (12.65,\yy+.36);}
 \node[firstaction,minimum width=2.00cm] at (9.35,4.30) {$a_1$};
 \node[firstaction,minimum width=2.00cm] at (11.55,4.30) {$a_1$};
 \node[firstaction,minimum width=2.00cm] at (9.35,2.95) {$a_1$};
 \node[secondaction,minimum width=2.00cm] at (11.55,2.95) {$a_2$};
 \node[secondaction,minimum width=2.00cm] at (9.35,1.60) {$a_2$};
 \node[secondaction,minimum width=2.00cm] at (11.55,1.60) {$a_2$};
 \node[text=hypblue] at (5.95,0.6)
    {Same response:\quad
     $\delta^*_{P,Q}(s,y)=\sigma_Q\bigl(g_{P,Q}(s),y\bigr)$; \qquad Same net utility:\quad $W(P,Q)-t_Q=W^o(P',Q)-t_Q$};
\end{tikzpicture}
\caption{Double deviation and response encoding.}
\label{fig:binary-codebook}
\end{figure}

We use BM to separate the joint deviations in reporting and obeying in a class of optimal menus. The similar decomposition is also in the sequential screening literature
\citep{CourtyLi2000,EsoSzentes2007}.\footnote{Sequential move reduces or simplifies double deviation. In
\citet{CourtyLi2000}, initial private information describes the
distribution of future valuations. So the second-stage private information has the same preference order. In \citet[Lemma~4]{EsoSzentes2007}, 
the buyer will adjust his second report so that the misreported pair of type
implies his true realized valuation.}
Here it comes from a strategic deviation \emph{already
available} to the buyer: retuning his model by garbling 
predictions.\footnote{This
imitation relation resembles mechanism design with evidence
\citep{BullWatson2007}. Here garbling can perfectly reproduce another type's obedient
payoff at any fixed contract.} Appendix \ref{app:binary-nested-bundling} uses the revelation principle to give a bundling representation of the problem. In hypothesis testing setting as in Section \ref{sec:hypothesis}, the representation reduces to a standard selling two goods problem. Appendix~\ref{app:binary} prove
Theorem~\ref{thm:dataset-revelation}.
\paragraph{Why the revelation principle holds.}
\label{sec:binary-logic}
Consider a normalized contract for type $Q$ for $n=3$, as in Figure~\ref{fig:binary-codebook}. We preserve only the two likelihood-ratio ordered imperfect $y_1,y_2$ for illustration.

We can encode optimal continuation for any $s$ from misreporting by garbling the private signals, since the optimal response can be chosen from $(a_1,a_1)$, $(a_1,a_2)$, and $(a_2,a_2)$, all of which appear among the rows in $\sigma_Q(s,y)$. We suppress the report superscript on the signal labels in $s^r$.
For example, suppose type $P$ misreports $Q$ and chooses optimal continuation $\delta^*_{P,Q}(s,y)$, where $\delta^*_{P,Q}(s_1,y)=\delta^*_{P,Q}(s_2,y)=\sigma_{Q}(s_1,y)=(a_1,a_1)$, $\delta^*_{P,Q}(s_3,y)=\sigma_{Q}(s_2,y)=(a_1,a_2)$. Let $g_{P,Q}$ maps $s_1,s_2$ to $s'_1$
and $s_3$ to $s'_2$. Write $G_{P,Q}(s'|s)=1_{s'=g_{P,Q}(s)}$ for the matrix of this map. We have 
\begin{equation}
 W(P,Q)=W^o(PG_{P,Q},Q).
 \label{eq:response-encoding}
\end{equation}

Thus type $P$ attains the same response and payoff at contract $Q$
by garbling his private dataset to $P'=PG_{P,Q}$ and following the
obedient rule. The garbled dataset belongs to the type space.
TT at $(P',Q)$ and BM at $(P,P')$ imply
\begin{equation}
 W(P,Q)-t_Q=W^o(P',Q)-t_Q
 \overset{\mathrm{TT}}{\leq}U(P')
 \overset{\mathrm{BM}}{\leq}U(P).
 \label{eq:encoding-chain}
\end{equation}
Every optimal continuation in the joint deviation can be encoded in this form. TT and BM therefore rule them all out. With more actions or
payoff-relevant states, an optimal continuation may be absent from
the recommended rows, and this decomposition can fail.

\section{Hypothesis testing}\label{sec:hypothesis}
\label{sec:main-result}
\begin{figure}[htbp]
\centering
\begin{minipage}[c]{.49\linewidth}
\centering
{\small\bfseries Private dataset $P_x$}\par\medskip
\[
\renewcommand{\arraystretch}{1.3}
\begin{array}{c|cc}
 &s=1&s=2\\ \hline
 \omega=1&\tfrac12-\alpha&\alpha\\
 \omega=2&\beta&\tfrac12-\beta
\end{array}
\]
{\footnotesize Rows: states. Columns: predictions.}
\end{minipage}\hfill
\begin{minipage}[c]{.47\linewidth}
\centering
\begin{tikzpicture}[x=3.4cm,y=3.4cm,>=Latex,font=\small,
                     line cap=round,line join=round]
 \fill[hyplight] (0,0)--(1,0)--(0,1)--cycle;
 \draw[->,black!70] (0,0)--(1.13,0) node[right] {$\alpha$};
 \draw[->,black!70] (0,0)--(0,1.12) node[above] {$\beta$};
 \draw[hyporange,line width=1.1pt] (0,1)--(1,0)
   node[pos=.52,sloped,above=4pt] {$H:\ \alpha+\beta=\tfrac12$};
 \draw[hypblue,densely dotted] (.28,0)--(.28,.33)--(0,.33);
 \fill[hypblue] (.28,.33) circle (1.6pt) node[above right] {$x$};
 \node[hypblue] at (.22,.62) {$D$};
 \node[below left] at (0,0) {$0$};
 \node[below] at (1,0) {$\tfrac12$};
 \node[left] at (0,1) {$\tfrac12$};
 \node[below] at (.28,0) {$\alpha$};
 \node[left] at (0,.33) {$\beta$};
\end{tikzpicture}
\end{minipage}
\caption{Private datasets and the type space.}
\label{fig:hyp-private-types}
\end{figure}
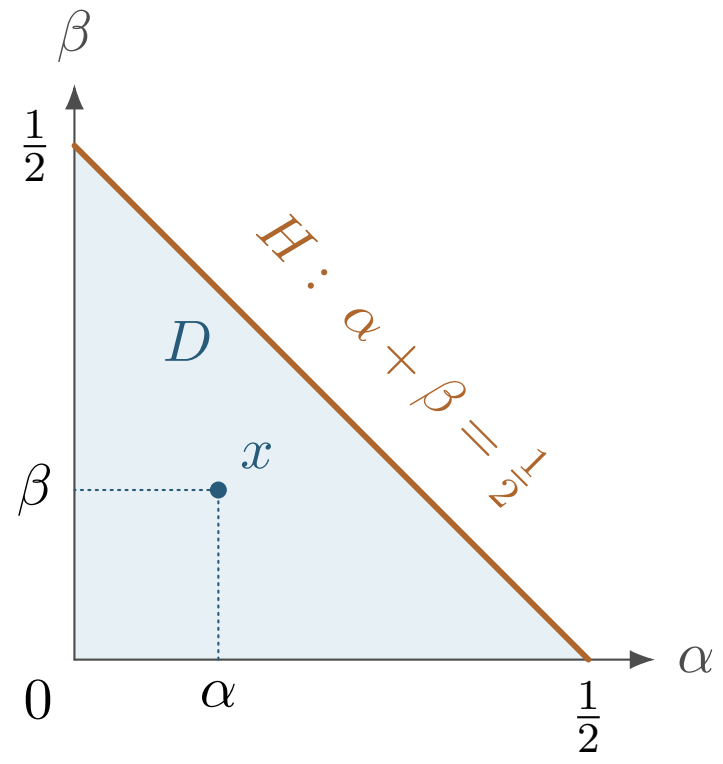

In this section, we focus on the optimal menu when the data buyer uses training datasets to improve prediction performance in hypothesis testing, as introduced in Section \ref{example}. State $1$ is the negative case
and state $2$ the positive case; private signals $1$ and $2$ are
negative and positive predictions. Define
$\alpha=\Pr(\omega=1,s=2)$ and $\beta=\Pr(\omega=2,s=1)$, the joint
probabilities of a false positive and a false negative.\footnote{These
are Type I and Type II error events. Their conventional conditional
rates are $2\alpha$ and $2\beta$.}
These probabilities determine the private dataset $P_x$, where
$x=(\alpha,\beta)$ (Figure~\ref{fig:hyp-private-types}). Label the
signals so that each positive-probability prediction is weakly
optimal. This requires $\alpha+\beta\leq \tfrac12$, giving the type domain
\begin{equation}
 D=\{x\in\mathbb R_+^2:s(x):=\alpha+\beta\leq \tfrac12\},
 \qquad H=\{x\in D:s(x)=\tfrac12\}.
 \label{eq:binary-error-representation}
\end{equation}

The outside value is the probability of a correct prediction,
$\ubar(x)=1-s(x)$. On $H$, the private dataset is uninformative and buyer just switch between default errors $(\alpha,\beta)$ of the model. Full information eliminates both errors, so its incremental
value is $s(x)$. 

Let \(f\) be a strictly positive \(C^1\) density of the error pair \(X=(\alpha,\beta)\) on \(D\). Define regularity 
condition \citep{McAfeeMcMillan1988} as
\begin{equation}
 g(x):=3f(x)+x\cdot\nabla f(x)\geq0
 \qquad (x\in D).
 \label{eq:gc1-density}
\end{equation}

\begin{theorem}
\label{thm:generalized-contamination1} Under \eqref{eq:gc1-density}, a posted price for fully informative dataset is optimal.
\end{theorem}

By Theorem 1, we can restrict attention to normalized contracts without loss of revenue. Within this class, we bound revenue under TT, BM, and IR and show that a single posted price for full information attains the bound. In fact, in hypothesis testing on $D$, BM is equivalent to IC or Obedience given TT and normalized contract. We will return to the duality framework in Section \ref{sec:hyp-beyond}.

\paragraph{The regularity condition.}
Condition~\eqref{eq:gc1-density} is differential virtual value
condition in multidimensional monopoly pricing
\citep{McAfeeMcMillan1988,Pavlov2011,Zhong2026}.  
Its connection to \citet{Myerson1981} is through the
\emph{density-weighted} virtual value, as explained by
\citet{Pavlov2011}. This condition alone does not imply a single grand-bundle
price in multidimensional screening. The proposed extension of
one-dimensional ``no-haggling'' to deterministic menus need not hold
\citep{McAfeeMcMillan1988,Thanassoulis2004}. We discuss how (BM) coarsens those canonical menus \citep{Pavlov2011,DDT2017} to the no-haggling
in Section~\ref{coarsen}.

\subsection{Proof sketch}

\begin{proof}[Proof of Theorem~\ref{thm:generalized-contamination1}]
We first reformulate the problem as a selling two goods problem: bundling the elimination ratios of statistical errors $q=(q_1,q_2)$ to a buyer with private error pairs $(\alpha,\beta)$. We then identify a \textit{no-distortion-at-the-frontier} constraint imposed by BM. That value determines a candidate price for the posted-price menu. The utility induced by this menu gives the upper bound of revenue, and it is feasible under IC. Proofs are in
Appendix~\ref{app:hyp-representation}.

\paragraph{Step 1: Represent a dataset as reductions in the two errors.} By revelation principle, without loss of revenue, we can always represent supplementary dataset $E_x$ as $q(x)=(q_1(x),q_2(x))$ and incremental rent $u(x):=U(P_x)-\underline{u}(P_x)$ in this form
\begin{equation}
 E_x=\begin{pmatrix}q_1(x)&1-q_1(x)&0\\0&1-q_2(x)&q_2(x)\end{pmatrix},
 \qquad  u(x)=x\cdot q(x)-t(x),\qquad q\in[0,1]^2.
 \label{eq:hyp-q}
\end{equation}
The vector $q=(q_1,q_2)$ records the fractions of false-positive
and false-negative errors corrected under the recommended decision
rule. The first and third outcomes reveal states 1 and 2,
respectively; at the middle outcome, the recommendation is to
retain the decision at private prediction. Conditional on state 1, a revealing
outcome arrives with probability $q_1$, correcting that fraction
of the buyer's false-positive errors. Similarly, $q_2$ is the
fraction of false-negative errors corrected. Conditional
independence therefore implies that the two error probabilities
fall from $\alpha$ and $\beta$ to $\alpha(1-q_1)$ and
$\beta(1-q_2)$. The resulting improvement in the decision payoff is
\( \alpha q_1+\beta q_2=x\cdot q.
\)

This gives a selling-two-goods interpretation: the two goods are
reductions in the two statistical errors, and the buyer's initial
error probabilities $(\alpha,\beta)$ determine his values for
these reductions. His incremental payoff from the recommended
use of contract $(q,t)$ is $x\cdot q-t$. Under TT and IR for these
recommended payoffs, Rochet's (1987) characterization applies:
the indirect rent $u$ is convex, coordinatewise nondecreasing,
1-Lipschitz for the $\ell^1$ norm, and satisfies $u(0)=0$.
At almost every type, the allocation is $q(x)=\nabla u(x)$.
\paragraph{Step 2: Replace BM by frontier restrictions.}
The types obtained by garbling $P_x$ form
\begin{equation}
 \{z:P_x\succeq_B P_z\}
 =\operatorname{co}\{x,(\tfrac12,0),(0,\tfrac12)\}.
 \label{eq:hyp-reachable}
\end{equation}
Figure~\ref{fig:hyp-bm} illustrates this set. In particular, all frontier
types are Blackwell equivalent, and every private dataset dominates
them. Since their outside values coincide, BM requires
\begin{equation}
 u|_H=r,\qquad 0\leq r-u(x)\leq d(x):=\tfrac12-s(x).
 \label{eq:hyp-boundary}
\end{equation}
The first inequality follows from coordinate
monotonicity between $x$ and types on $H$. The second follows from BM between $x$ and types on  $H$. 

Condition \eqref{eq:hyp-boundary} is a \textit{no-distortion-on-the-frontier condition.} $u(x)\le r$ implies that frontier types have the highest indirect rent in any feasible mechanisms. Since the designer always adds fully informative dataset for revenue-maximization in her optimal menu, $x\in H$ will choose this option in the menu. Thus their allocation is never distorted, which shares the ``no-distortion-at-the-top'' in one-dimensional screening problem.\footnote{
On a product domain $[0,1]^n$, the usual no-distortion-at-the-top
condition concerns the maximal type $\mathbf1$; here BM imposes a
common value on an entire frontier.} This condition is sufficient for BM, $l_1$ 1-Lipschitz continuity, and Obedience, we will return the discussion in Section \ref{sec:hyp-beyond}.

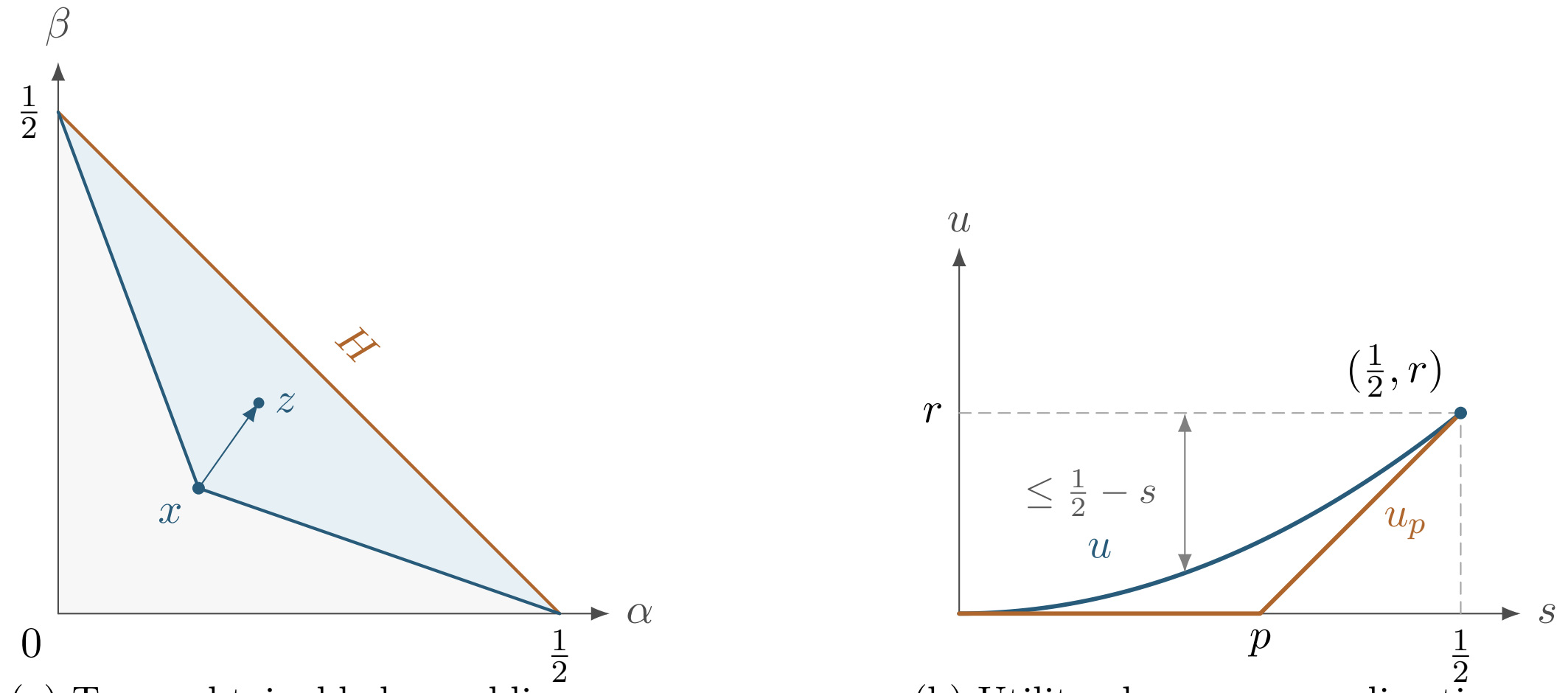
\begin{figure}[tbp]
\centering
\begin{minipage}[t]{.48\linewidth}
\centering
\begin{tikzpicture}[x=4.5cm,y=4.5cm,>=Latex,font=\small,
                     line cap=round,line join=round]
 \fill[black!3] (0,0)--(1,0)--(0,1)--cycle;
 \fill[hyplight] (.28,.25)--(1,0)--(0,1)--cycle;
 \draw[->,black!70] (0,0)--(1.10,0) node[right] {$\alpha$};
 \draw[->,black!70] (0,0)--(0,1.10) node[above] {$\beta$};
 \draw[hyporange,thick] (0,1)--(1,0)
    node[pos=.53,sloped,above=4pt] {$H$};
 \draw[hypblue,thick] (.28,.25)--(1,0) (.28,.25)--(0,1);
 \draw[->,hypblue] (.28,.25)--(.40,.42);
 \fill[hypblue] (.28,.25) circle (1.6pt) node[below left] {$x$};
 \fill[hypblue] (.40,.42) circle (1.4pt) node[right] {$z$};
 \node[below left] at (0,0) {$0$};
 \node[below] at (1,0) {$\tfrac12$};
 \node[left] at (0,1) {$\tfrac12$};
 \node[align=center,font=\footnotesize] at (.50,-.18)
   {(a) Types obtainable by garbling $x$};
\end{tikzpicture}
\end{minipage}\hfill
\begin{minipage}[t]{.48\linewidth}
\centering
\begin{tikzpicture}[x=4.5cm,y=4.5cm,>=Latex,font=\small,
                     line cap=round,line join=round]
 \draw[->,black!70] (0,0)--(1.12,0) node[right] {$s$};
 \draw[->,black!70] (0,0)--(0,.73) node[above] {$u$};
 \draw[black!35,densely dashed] (0,.4)--(1,.4);
 \draw[black!35,densely dashed] (1,0)--(1,.4);
 \draw[hypblue,line width=1.1pt,domain=0:1,samples=60]
   plot (\x,{.4*\x*\x});
 \draw[hyporange,line width=1.1pt] (0,0)--(.6,0)--(1,.4);
 \fill[hypblue] (1,.4) circle (1.6pt);
 \node[above left] at (1,.4) {$(\tfrac12,r)$};
 \node[left] at (0,.4) {$r$};
 \node[below] at (.6,0) {$p$};
 \node[below] at (1,0) {$\tfrac12$};
 \node[hypblue] at (.28,.13) {$u$};
 \node[hyporange] at (.89,.18) {$u_p$};
 \draw[<->,gray] (.45,{.4*.45*.45})--(.45,.4)
   node[midway,left=3pt,font=\footnotesize,text=black!65] {$\leq \tfrac12-s$};
 \node[align=center,font=\footnotesize] at (.50,-.18)
   {(b) Utility along an error direction};
\end{tikzpicture}
\end{minipage}
\caption{BM and the frontier. In (a), the shaded triangle is the set of types obtainable by garbling $x$.
In (b), the frontier rent $r$ and the bound $r-u\leq \tfrac12-s$ place
utility above $u_p=(s-p)_+$, where $p=\tfrac12-r$. }
\label{fig:hyp-bm}
\end{figure}

\paragraph{Step 3: Fix frontier utility and minimize interior rents.}
For a feasible utility, divergence theorem gives expected revenue
\begin{equation}
 \begin{split}
 R_f(u)&:=\int_D[x\cdot\nabla u(x)-u(x)]f(x)\,dx\\
       &=Br-\int_Du(x)g(x)\,dx,\qquad
 B:=\int_0^{1/2} \tfrac12 f(a,\tfrac12-a)\,da.
 \end{split}
 \label{eq:hyp-revenue}
\end{equation}
This is the revenue representation used in
multidimensional screening \citep{Rochet1987,DDT2017}.
For a fixed frontier utility $r$, interior rents enter revenue with
weight $-g$. When $g\geq0$, lowering them weakly raises revenue.

Set $p=\tfrac12-r$. IR and the vertical bound imply
\begin{equation}
 u(x)\geq\max\{0,r-d(x)\}=(s(x)-p)_+=u_p(x).
 \label{eq:hyp-envelope}
\end{equation}
The lower envelope $u_p$ is itself feasible and has the same frontier
value. It follows that
\begin{equation}
 R_f(u_p)-R_f(u)
 =\int_Dg(x)[u(x)-u_p(x)]\,dx\geq0.
 \label{eq:hyp-dominance}
\end{equation}
Every feasible menu is therefore revenue-dominated by the
full-information posted price with the same frontier utility. That
menu is IC and attains the bound obtained from TT, BM, and IR.
\end{proof}

\section{Multiclass classification}\label{GCS}
\begin{figure}[htbp]
\centering
\begin{tikzpicture}[
    x=1cm,y=1cm,>=Latex,
    font=\sffamily\small,
    line cap=butt,line join=miter
]
\definecolor{contamBlue}{HTML}{3577A6}
\definecolor{contamBlueFill}{HTML}{EAF3F9}
\definecolor{contamBlueBorder}{HTML}{98BBD4}
\definecolor{contamBlueText}{HTML}{285573}
\definecolor{contamOrange}{HTML}{B87535}
\definecolor{contamOrangeFill}{HTML}{FBF0E3}
\definecolor{contamOrangeBorder}{HTML}{DDB992}
\definecolor{contamOrangeText}{HTML}{875120}
\definecolor{contamSlateFill}{HTML}{EFF2F5}
\definecolor{contamSlateBorder}{HTML}{C9D1DA}
\definecolor{contamSlateText}{HTML}{414B58}
\definecolor{contamSlateLine}{HTML}{919CAA}

\tikzset{
    card/.style={
        rounded corners=8pt,line width=.7pt,
        align=center,inner sep=6pt
    },
    source/.style={
        card,draw=contamSlateBorder,fill=contamSlateFill,
        text=contamSlateText
    },
    animal/.style={
        card,draw=contamBlueBorder,fill=contamBlueFill,
        text=contamBlueText,
        minimum width=2.75cm,minimum height=.90cm
    },
    vehicle/.style={
        card,draw=contamOrangeBorder,fill=contamOrangeFill,
        text=contamOrangeText,
        minimum width=2.75cm,minimum height=.90cm
    },
    animalleaf/.style={
        animal,minimum width=1.72cm,
        minimum height=1.08cm,inner sep=0pt
    },
    vehicleleaf/.style={
        vehicle,minimum width=1.72cm,
        minimum height=1.08cm,inner sep=0pt
    },
    bluepath/.style={draw=contamBlue,line width=.9pt},
    orangepath/.style={draw=contamOrange,line width=.9pt},
    pill/.style={
        fill=white,rounded corners=6pt,line width=.5pt,
        font=\sffamily\footnotesize,
        inner xsep=5pt,inner ysep=3pt
    },
    bluepill/.style={
        pill,draw=contamBlueBorder,text=contamBlueText
    },
    orangepill/.style={
        pill,draw=contamOrangeBorder,text=contamOrangeText
    }
}

\node[font=\sffamily\footnotesize,text=contamSlateText!75]
    at (0,5.64) {Conditional on contamination};

% Original label
\node[source,minimum width=3.95cm,
      minimum height=1.08cm,inner sep=0pt]
    (original) at (0,4.83) {};

\node[inner sep=0pt]
    at ($(original.center)+(-1.43,0)$)
    {\includegraphics[height=.58cm]{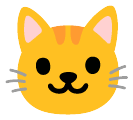}};

\node[align=center,inner sep=0pt,text=contamSlateText]
    at ($(original.center)+(.38,0)$)
    {Original label: cat\\[-1pt]
     {\footnotesize Animal group S\textsubscript{1}}};

% Target groups
\node[animal] (animal) at (-3.3,2.75)
    {Animal group S\textsubscript{1}\\[-1pt]
     {\footnotesize label drawn from h\textsubscript{1}}};

\node[vehicle] (vehicle) at (3.3,2.75)
    {Vehicle group S\textsubscript{2}\\[-1pt]
     {\footnotesize label drawn from h\textsubscript{2}}};

% Final labels
\node[animalleaf]  (cat) at (-5.2,.24) {};
\node[animalleaf]  (dog) at (-1.4,.24) {};
\node[vehicleleaf] (car) at (1.4,.24) {};
\node[vehicleleaf] (bus) at (5.2,.24) {};

\node[inner sep=0pt] at ($(cat.center)+(0,.13)$)
    {\includegraphics[height=.62cm]{contamination_icons/cat.png}};
\node[font=\sffamily\footnotesize,text=contamBlueText]
    at ($(cat.center)+(0,-.35)$) {cat};

\node[inner sep=0pt] at ($(dog.center)+(0,.13)$)
    {\includegraphics[height=.62cm]{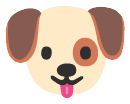}};
\node[font=\sffamily\footnotesize,text=contamBlueText]
    at ($(dog.center)+(0,-.35)$) {dog};

\node[inner sep=0pt] at ($(car.center)+(0,.13)$)
    {\includegraphics[height=.62cm]{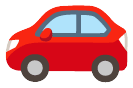}};
\node[font=\sffamily\footnotesize,text=contamOrangeText]
    at ($(car.center)+(0,-.35)$) {car};

\node[inner sep=0pt] at ($(bus.center)+(0,.13)$)
    {\includegraphics[height=.62cm]{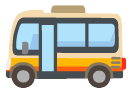}};
\node[font=\sffamily\footnotesize,text=contamOrangeText]
    at ($(bus.center)+(0,-.35)$) {bus};

% Group transitions
\draw[draw=contamSlateLine,line width=.9pt]
    (original.south)--(0,3.77);
\draw[bluepath,->]
    (0,3.77)--(-3.3,3.77)--(animal.north);
\draw[orangepath,->]
    (0,3.77)--(3.3,3.77)--(vehicle.north);

\node[bluepill] at (-1.65,3.77)
    {R\textsubscript{11} = 0.9};
\node[orangepill] at (1.65,3.77)
    {R\textsubscript{12} = 0.1};

% Within-group draws
\draw[bluepath] (animal.south)--(-3.3,1.54);
\draw[bluepath,->]
    (-3.3,1.54)--(-5.2,1.54)--(cat.north);
\draw[bluepath,->]
    (-3.3,1.54)--(-1.4,1.54)--(dog.north);

\draw[orangepath] (vehicle.south)--(3.3,1.54);
\draw[orangepath,->]
    (3.3,1.54)--(1.4,1.54)--(car.north);
\draw[orangepath,->]
    (3.3,1.54)--(5.2,1.54)--(bus.north);

\node[bluepill]   at (-4.25,1.54) {$h_{11}$=0.75};
\node[bluepill]   at (-2.35,1.54) {$h_{12}$=0.25};
\node[orangepill] at (2.35,1.54)  {$h_{21}$=0.4};
\node[orangepill] at (4.25,1.54)  {$h_{22}$=0.6};
\end{tikzpicture}
\caption{Two-stage contamination: $R$
selects a target group, $h_r$ selects a label within it. }
\label{fig:gc-contamination}
\end{figure}

In this section, we extend the optimality of posted price to multiclass classification problems, as introduced in Section \ref{example}, in generalized contamination space for private datasets. Buyers train classifiers for the same task using datasets that differ
in label quality. We take a common clean-label benchmark $C$ and model
the differences in prediction noise through a two-stage confusion process: when a
prediction is subject to noise, the classifier first selects a
possibly incorrect group and then draws a label within that group.\footnote{$C\in\Theta_{m,n}$ is the joint distribution of the
true class and the clean label. 
$\{\operatorname{Supp}C(\cdot\mid\omega)\}_\omega$ is a partition of $S$ , so a clean label
identifies the true class. A class may have several clean labels.} Figure~\ref{fig:gc-contamination} illustrates the
two stages.

Partition the label set $S$ into nonempty correct groups $S_1,\ldots,S_k$,\footnote{Each true class belongs to a single group, and several classes may share a group. Let $\ell(\omega)$ denote the group of class $\omega$. All clean labels for class $\omega$ belong to that group:
\(
 \operatorname{supp} C(\cdot\mid\omega)
 \subseteq S_{\ell(\omega)}
 \,\text{for every }\omega\in\Omega.
\)}
and let $R:Z\rightarrow\Delta(Z)$ be a fixed regular confusion matrix on the group set
$Z=\{1,\ldots,k\}$.\footnote{$R$ is nonsingular and
$(R^{-1})_{\ell r}\leq 0$ for $\ell\neq r$.}
The buyer privately knows the contamination rate
$\varepsilon\in[0,1]$ and the within-group distributions
$h=(h_1,\ldots,h_k)$, where $h_r\in\Delta(S_r)$.

Given the true class $\omega$, draw a clean label $j$ from
$C(\cdot\mid\omega)$. With probability $1-\varepsilon$, the classifier
outputs $j$. With probability $\varepsilon$, the output is drawn in
two stages. First, if $j$ belongs to group $S_\ell$, draw a target group
$r$ with probability $R_{\ell r}$. Second, draw the output label $j'$
from $h_r$. Thus all clean labels in the same group face the same
confusion process. 
Conditional on contamination, the transition from the clean label
$j$ to the output label $j'$ is
\begin{equation}
 \Gamma_h(j'\mid j)=R_{\ell r}h_r(j'),
 \qquad j\in S_\ell,\quad j'\in S_r.
 \label{eq:gc-channel}
\end{equation}
The buyer observes only the classifier's output $j$. Its joint distribution with the true class $\omega$ is
\begin{equation}
 P_{\varepsilon,h}
 =C\big[(1-\varepsilon)I_n+\varepsilon\Gamma_h\big]\in\Delta(\Omega\times S).
 \label{eq:gc-family}
\end{equation}
Set $x_j=\varepsilon h_r(j)$ for $j\in S_r$, and let $\mathcal D$
denote the domain of $x$. Each group has the same total mass
$\varepsilon=\sum_{j\in S_r}x_j\in[0,1]$, so the domain has
dimension $d=n-k+1$. Let \(f\) be a strictly positive \(C^1\) density of \(X\) on \(\mathcal D\). Now regularity 
condition becomes
\begin{equation}
 g(x):=(d+1)f(x)+x\cdot\nabla f(x)\geq0
 \qquad (x\in\mathcal D).
 \label{eq:gc-density}
\end{equation}

\begin{theorem}
\label{thm:generalized-contamination} Under \eqref{eq:gc-density}, a posted price for fully informative dataset is optimal.
\end{theorem}

\paragraph{Applications.} The generalized contamination space covers the following environments.

\textit{Hypothesis testing/binary classification.}
A platform purchases verified safety labels to improve a proprietary
content classifier with privately known false-positive and false-negative
errors.
For $m=n=2$, take $C=I_2/2$, $k=1$, and
$h_1=h\in\Delta(S)$. Then $\Gamma_h(j'\mid j)=h(j')$.
After relabeling the signals, every binary dataset has the form
\begin{equation}
P_{\alpha,\beta}
=
\begin{pmatrix}
\frac12-\alpha & \alpha\\
\beta & \frac12-\beta
\end{pmatrix},
\qquad
\alpha,\beta\geq0,
\quad
\alpha+\beta\leq\frac12.
\label{eq:gc-binary}
\end{equation}
It equals $(1-\varepsilon)C+\varepsilon\mathbf1_2h/2$ with
$\varepsilon=2(\alpha+\beta)$ and
$h=(\beta,\alpha)/(\alpha+\beta)$ when $\alpha+\beta>0$.
The theorem therefore covers the $\Theta_{2,2}$ environment and theorem \ref{thm:generalized-contamination1}.\footnote{Allowing the clean component to vary, the posted-price
result also extends to the full $\Theta_{2,n}$ domain under
a conditional price-revenue concavity condition. See Appendix \ref{sec:binary-private-datasets}.}
This binary label-noise formulation is standard in
\citet{NatarajanEtAl2013}.

\textit{Contamination.}
An image-recognition developer purchases expert annotations to supplement
a private corpus containing labels supplied by uninformed annotators.
Take $k=1$, $R=(1)$, and $h_1=h\in\Delta(S)$.
Then $\Gamma_h(j'\mid j)=h(j')$, independently of $j$. Writing
$a^\omega_j=mC(\omega,j)$ gives
\begin{equation}
P_{\varepsilon,h}(j\mid\omega)
=
(1-\varepsilon)a^\omega_j+\varepsilon h_j.
\label{eq:gc-standard}
\end{equation}
A record is clean or replaced by a label independent of its true class.
Here $d=n$ and $g=(n+1)f+x\cdot\nabla f\geq0$.
The replacement component is the spammer model of
\citet{RaykarYu2012}.

\textit{Class-dependent label noise.}
A buyer purchases verified image labels to supplement data partly labeled
by an automated tool with systematic class-confusion errors.
Take $n=m=k$, $C=I_m/m$, and singleton groups $S_r=\{r\}$.
Each $h_r$ then puts probability one on $r$, so
\begin{equation}
 \Gamma_h=R,\qquad
 P_{\varepsilon,h}
 =\frac1m\big[(1-\varepsilon)I_m+\varepsilon R\big].
 \label{eq:gc-class-noise}
\end{equation}
The relative confusion pattern is fixed, while $\varepsilon$ measures
reliance on the automated labels. The transition-matrix formulation
of class-dependent noise follows \citet{PatriniEtAl2017}.

\textit{Coarse and noisy coarse labels.}
A buyer purchases fine-grained labels, such as dog breeds, to supplement
data with coarse or imprecise annotations.
Take $k<m$, so the grouping pools some true classes. When $R=I_k$,
contamination preserves the coarse class and redraws the private
label within it:
\begin{equation}
 \Gamma_h(j'\mid j)=h_\ell(j')
 \quad\text{for }j,j'\in S_\ell,
 \qquad \Gamma_h(j'\mid j)=0\text{ otherwise}.
 \label{eq:gc-coarse}
\end{equation}
The contaminated record carries exactly the information in the coarse
class, such as ``dog'' rather than the breed. Off-diagonal entries of $R$
allow confusion between coarse classes. The case $R=I_k$ nests a
fixed-partition coarse-label model of the kind studied by
\citet{FotakisEtAl2021}.

\textit{Multiple noise sources.}
A buyer pools training records from several annotation providers and
purchases a curated supplement.
Suppose source $i$ has within-group distributions $h^{(i)}$, with
the same grouping and confusion matrix $R$. For private weights
$\varepsilon_i\geq0$ with $\varepsilon=\sum_i\varepsilon_i\leq1$,
\begin{equation}
\begin{aligned}
 P
 &=C\left[\left(1-\sum_i\varepsilon_i\right)I_n
                    +\sum_i\varepsilon_i\Gamma_{h^{(i)}}\right]\\
 &=C\big[(1-\varepsilon)I_n+\varepsilon\Gamma_{\bar h}\big],
 \qquad
 \bar h_r=\sum_i\frac{\varepsilon_i}{\varepsilon}h^{(i)}_r
 \quad(\varepsilon>0).
\end{aligned}
\label{eq:gc-multiple}
\end{equation}
For $k=1$, the sources can be several annotation pools with
different state-independent labeling propensities. For $k>1$,
they may differ in how their records are distributed within
the same coarse categories. Such a learning from heterogeneous
annotators is studied by \citet{RaykarEtAl2010}.
\section{Discussion}
\label{sec:final-discussion}

\subsection{Beyond regularity}
\label{sec:hyp-beyond}
\subsubsection{A duality framework}
We now drop the differential virtual value condition and develop the duality framework in the hypothesis-testing
model. The change of variables $w(x)=r-u(x)$ gives
$R_f(u)=\int_D w\,d\nu$, where $\nu=g(x)\,dx-\delta_0$, and replaces the Lipschitz restriction by
$w\leq d$. Define
\begin{align*}
    &\mathcal K=\{w\in C(D):w\text{ is concave and coordinatewise
 nonincreasing},\ w|_H=0\}， \\ &\lambda\succeq_{\mathcal K}\nu \text{ iff }
\int_Dw\,d\lambda\geq\int_Dw\,d\nu  \text{ for every } w\in\mathcal K.
\end{align*}

\begin{theorem}[Strong duality]
\label{thm:hyp-strong-duality}
\begin{equation}
 \sup_{M\in\mathcal M_{\mathrm{IC,IR}}}\operatorname{Rev}_F(M)
 =\max_{\substack{w\in\mathcal K\\w\leq d}}\int_D w\,d\nu
 =\min_{\substack{\lambda\geq0\\\lambda\succeq_{\mathcal K}\nu}}
       \int_D d\,d\lambda,
 \label{eq:hyp-obstacle-dual}
\end{equation}
\end{theorem}
For convex, coordinatewise nondecreasing rents $u$ with \(u|_H=r\),
the global \(\ell^1\) \(1\)-Lipschitz condition is equivalent to the
vertical IC bound \(r-u(x)\leq d(x)\) for every
\(x\in D\). This bound only compares each type $x$ with the frontier $H$. Its dual
multiplier is a measure on \(D\), yielding a generalized persuasion
problem \citep{YangYang2026}. This is different from standard transport formulations in multi-dimensional screening, where the pairwise Lipschitz constraints is associated with a coupling on
\(D\times D\) \citep{DDT2017,KleinerManelli2019,Zhong2026}.\footnote{We thank Andreas Kleiner for pointing us to this point.}
Appendix~\ref{app:hyp-obstacle-dual} establishes the dual representation.

\paragraph{Certifying posted price.}
Let $u_p=(s-p)_+$,
$w_p=\min\{d,\tfrac12-p\}$, and $W_p=\{s\geq p\}$.
\begin{proposition}[Posted price certificate]
\label{prop:hyp-grand-bundling-certificate}
Posted price for a fully informative dataset at $p$ is optimal if and only if a nonnegative
measure $\lambda$ on $D$ satisfies
\begin{equation}
 \lambda\succeq_{\mathcal K}\nu,\qquad
 \operatorname{supp}\lambda\subseteq W_p,\qquad
 \int_Dw_p\,d(\lambda-\nu)=0.
 \label{eq:hyp-grand-bundling-certificate}
\end{equation}
\end{proposition}
These conditions make both dual inequalities tight at $w_p$.
Conversely, strong duality gives an attaining $\lambda$ for which
both inequalities are tight whenever $w_p$ is optimal.
Under $g\geq0$, let $p^*$ maximize $p\Pr\{s(X)\geq p\}$.
Its first-order condition gives
$\int_{W_{p^*}}g\,dx=B$ and $\int_{D\setminus W_{p^*}}g\,dx=1$.
Then $\lambda^*=\mathbf1_{W_{p^*}}g\,dx$ is feasible because
$\int_Dw\,d(\lambda^*-\nu)
=w(0)-\int_{D\setminus W_{p^*}}wg\,dx\geq0$, and
\[
 \int_Dd\,d\lambda^*
 =\int_{W_{p^*}}d(x)g(x)\,dx
 =R_f(u_{p^*}).
\]
Equivalently, reshuffle the frontier measure to
$(S_H)_\#\lambda^*$, where $S_H(x)=(\tfrac12-\beta,\beta)$;
the coupling $(S_H,\mathrm{Id})_\#\lambda^*$ has cost
$\int d\,d\lambda^*$ and certifies grand bundling. In the equivalent two-good bundling problem with linear-additive utility,\footnote{Given $q=(q_1,q_2)$, let $\lambda_q$ be the lottery that
includes good $i$ independently with probability $q_i$. Its probabilities
for the bundles $\varnothing$, $\{1\}$, $\{2\}$, and $\{1,2\}$ are
$(1-q_1)(1-q_2)$, $q_1(1-q_2)$, $(1-q_1)q_2$, and $q_1q_2$,
respectively. The buyer pays $t$ and receives the drawn bundle, whose
expected net payoff is $\alpha q_1+\beta q_2-t$.}
the certificate generalizes the sufficient condition of
\citet{HaghpanahHartline2021}, which  requires distributions on each ray $r_k=\{(\alpha,\beta)| \alpha=k\beta\}$ to coincide.

\subsubsection{BM coarsens the menu.}\label{coarsen}
BM  restricts which options can be active, coarsening complex menus to simple ones.

\begin{proposition}[Replacing boundary contracts]
\label{prop:hyp-boundary-collapse}
Replacing all options with $q_1=1$ or $q_2=1$ can be replaced by $q=(1,1)$ at price $p=\tfrac12-r$ weakly increases the profits.
\end{proposition}
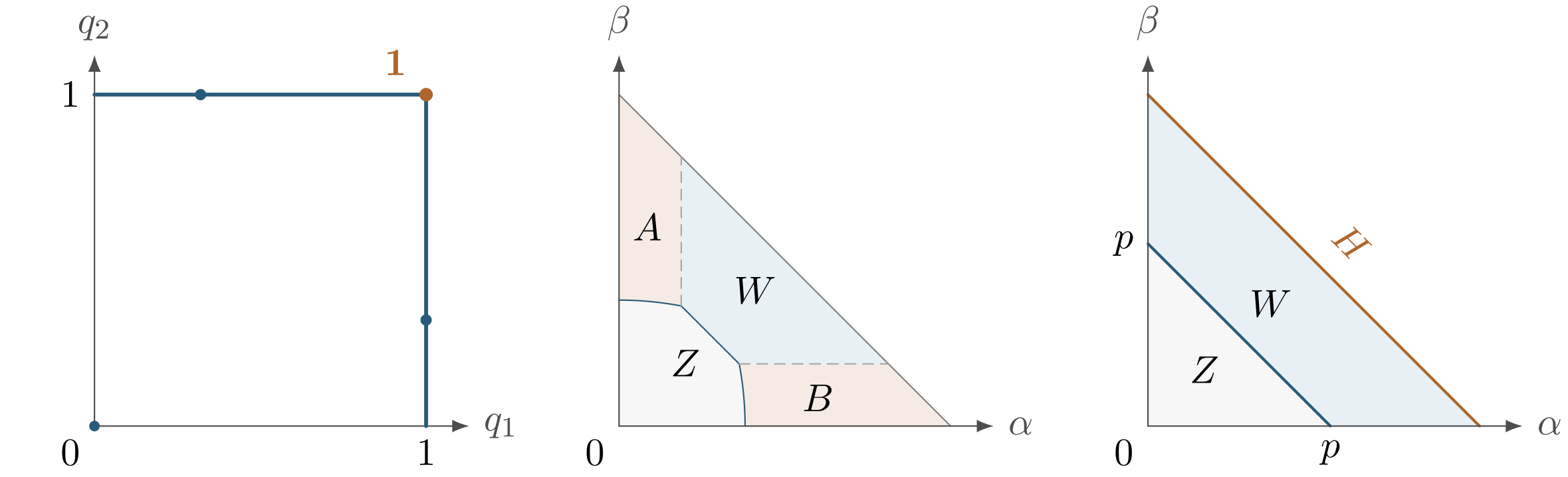
\begin{figure}[htbp]
\centering
\begin{minipage}[t]{.31\linewidth}
\centering
\begin{tikzpicture}[x=3.30cm,y=3.30cm,>=Latex,font=\small,
                     line cap=round,line join=round]
 \draw[->,black!70] (0,0)--(1.13,0) node[right] {$q_1$};
 \draw[->,black!70] (0,0)--(0,1.12) node[above] {$q_2$};
 \draw[hypblue,line width=1.1pt] (0,1)--(1,1)--(1,0);
 \fill[hypblue] (.32,1) circle (1.6pt);
 \fill[hypblue] (1,.32) circle (1.6pt);
 \fill[hyporange] (1,1) circle (1.9pt)
   node[above left=2pt] {$\mathbf1$};
 \fill[hypblue] (0,0) circle (1.5pt);
 \node[below left] at (0,0) {$0$};
 \node[below] at (1,0) {$1$};
 \node[left] at (0,1) {$1$};
 \node[font=\scriptsize] at (.50,-.23)
    {(a) Boundary allocation $(a,1)$ or $(1,a)$};
\end{tikzpicture}
\end{minipage}\hfill
\begin{minipage}[t]{.31\linewidth}
\centering
\begin{tikzpicture}[x=3.30cm,y=3.30cm,>=Latex,font=\small,
                     line cap=round,line join=round]
 \def\k{.187596}
 \fill[black!3] (0,0)--(1,0)--(0,1)--cycle;
 \fill[hyporange!13] (0,.38)--(0,1)--(\k,{1-\k})
   --(\k,{.55-\k})
   --plot[domain=\k:0,samples=35] (\x,{.38-.5*\x*\x})--cycle;
 \fill[hyporange!13] (.38,0)--(1,0)--({1-\k},\k)
   --({.55-\k},\k)
   --plot[domain=\k:0,samples=35] ({.38-.5*\x*\x},\x)--cycle;
 \fill[hyplight] (\k,{.55-\k})--({.55-\k},\k)
   --({1-\k},\k)--(\k,{1-\k})--cycle;
 \draw[gray] (0,1)--(1,0);
 \draw[hypblue] plot[domain=0:\k,samples=35] (\x,{.38-.5*\x*\x});
 \draw[hypblue] plot[domain=0:\k,samples=35] ({.38-.5*\x*\x},\x);
 \draw[hypblue] (\k,{.55-\k})--({.55-\k},\k);
 \draw[black!35,densely dashed] (\k,{.55-\k})--(\k,{1-\k})
   ({.55-\k},\k)--({1-\k},\k);
 \draw[->,black!70] (0,0)--(1.13,0) node[right] {$\alpha$};
 \draw[->,black!70] (0,0)--(0,1.12) node[above] {$\beta$};
 \node at (.085,.60) {$A$};
 \node at (.60,.085) {$B$};
 \node at (.20,.19) {$Z$};
 \node at (.41,.41) {$W$};
 \node[below left] at (0,0) {$0$};
 \node[font=\scriptsize] at (.50,-.23)
    {(b) Canonical partition$\{Z,A,B,W\}$};
\end{tikzpicture}
\end{minipage}\hfill
\begin{minipage}[t]{.31\linewidth}
\centering
\begin{tikzpicture}[x=3.30cm,y=3.30cm,>=Latex,font=\small,
                     line cap=round,line join=round]
 \fill[hyplight] (0,0)--(1,0)--(0,1)--cycle;
 \fill[black!3] (0,0)--(.55,0)--(0,.55)--cycle;
 \draw[hyporange,thick] (0,1)--(1,0)
   node[pos=.53,sloped,above=3pt] {$H$};
 \draw[hypblue,thick] (0,.55)--(.55,0);
 \draw[->,black!70] (0,0)--(1.13,0) node[right] {$\alpha$};
 \draw[->,black!70] (0,0)--(0,1.12) node[above] {$\beta$};
 \node at (.17,.17) {$Z$};
 \node at (.37,.37) {$W$};
 \node[below] at (.55,0) {$p$};
 \node[left] at (0,.55) {$p$};
 \node[below left] at (0,0) {$0$};
 \node[font=\scriptsize] at (.50,-.23)
    {(c) BM coarsens menu (a) and (b)};
\end{tikzpicture}
\end{minipage}
\caption{Menu coarsening. }
\label{fig:hyp-menu-collapse}
\end{figure}
 \citet{Pavlov2011} gives conditions
under which an optimal menu can be chosen to use only exclusion $q=(0,0)$ or
allocations of the form $(1,a)$ or $(a,1)$, with $a\in[0,1]$.
\citet{DDT2017} give sufficient conditions for the optimality of a
canonical partition $\{Z,A,B,W\}$. Let $b_1(\alpha)$ be the height
of the upper boundary of the exclusion region $Z$ at $\alpha$, and $b_2(\beta)$ be the width of its right boundary
at $\beta$. On $A$ and $B$, the buyer's rent is the vertical distance
$\beta-b_1(\alpha)$ and the horizontal distance $\alpha-b_2(\beta)$,
respectively. The allocations on $Z$, $A$, $B$, and $W$ are 
$(0,0)$, $(-b_1'(\alpha),1)$, $(1,-b_2'(\beta))$, and $(1,1)$,
respectively, almost everywhere. Under BM, a single full-information
contract can replace all these contracts. Figure~\ref{fig:hyp-menu-collapse} illustrates the resulting coarsening.

\begin{corollary}[Menu collapse]
\label{cor:hyp-canonical-bundling}
If an optimal dataset menu has Pavlov's boundary allocation or DDT's
canonical partition, a single posted price for full information is optimal.
\end{corollary}

\subsection{Analysis of Blackwell monotonicity}
\label{BMI}
We first reframe this question to explain the source of Blackwell monotonicity: designer screen agents' feasible data processing technology, related to \citet{GanLi2026}. We then compare Blackwell monotonicity with three characterizations of implementable mechanisms in multi-dimensional screening \citep{Rochet1987,DDT2017}: (i) monotonicity in indirect rent, (ii) cyclic monotonicity in allocation, (iii)  admissible movements of measures in duality form. The comparisons give an answer to why Blackwell monotonicity coarsens the menu. 

\subsubsection{BM and screening for choice sets of statistical models}
\label{sec:choice-sets}

A private dataset can be viewed as a privately known set of feasible
prediction models. Let
\begin{equation}
 \mathcal R_P:=\{Q\in\Theta:P\succeq_B Q\}
 \label{eq:choice-set-type}
\end{equation}
be the models a buyer with dataset $P$ can obtain by garbling his private
predictions. By transitivity of garbling and the fact that $Q\in\mathcal R_Q$,
\begin{equation}
 P\succeq_B Q
 \quad\Longleftrightarrow\quad
 \mathcal R_Q\subseteq\mathcal R_P.
 \label{eq:choice-set-inclusion}
\end{equation}
If $\mathcal R_Q\subseteq\mathcal R_P$, type $P$ can choose $Q$'s contract,
garble his private predictions to reproduce $Q$, and follow $Q$'s
prescribed rule. This yields $U(Q)$, so incentive compatibility requires
$U(P)\geq U(Q)$. Thus (BM) is monotonicity of total utility under inclusion
of feasible model sets.

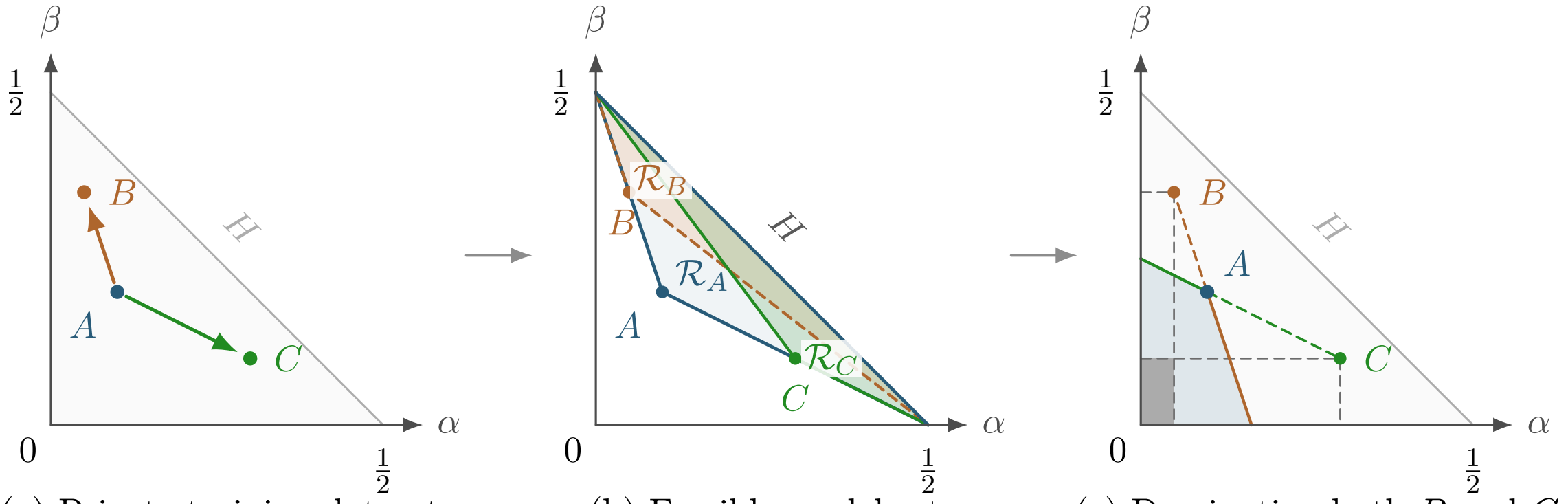
\begin{figure}[htbp]
\centering
\resizebox{\linewidth}{!}{%
\begin{tikzpicture}[x=7cm,y=7cm,>=Latex,font=\small,
  line cap=round,line join=round]
\begin{scope}
  \coordinate (A) at (.10,.20);
  \coordinate (B) at (.05,.35);
  \coordinate (C) at (.30,.10);
  \fill[black!2] (0,0)--(0,.5)--(.5,0)--cycle;
  \draw[black!32,line width=.6pt] (0,.5)--(.5,0)
    node[pos=.49,sloped,above=4pt] {$H$};
  \draw[->,black!70,line width=.65pt] (0,0)--(.56,0) node[right] {$\alpha$};
  \draw[->,black!70,line width=.65pt] (0,0)--(0,.56) node[above] {$\beta$};
  \node[below left] at (0,0) {$0$};
  \node[below=3pt] at (.5,0) {$\tfrac12$};
  \node[left=3pt] at (0,.5) {$\tfrac12$};
  \draw[->,hyporange,line width=1.05pt,shorten <=3pt,shorten >=4pt] (A)--(B);
  \draw[->,hypgreen,line width=1.05pt,shorten <=3pt,shorten >=4pt] (A)--(C);
  \fill[hypblue] (A) circle (2.1pt);
  \fill[hyporange] (B) circle (2.1pt);
  \fill[hypgreen] (C) circle (2.1pt);
  \node[below left=3pt,text=hypblue] at (A) {$A$};
  \node[right=3pt,text=hyporange] at (B) {$B$};
  \node[right=3pt,text=hypgreen] at (C) {$C$};
  \node at (.265,-.135) {(a) Private training datasets};
  \node at (.265,-.193) {$\alpha_C>\alpha_A>\alpha_B,\quad\beta_C<\beta_A<\beta_B$};
  \node at (.265,-.25) {$A\succeq_B B,\quad A\succeq_B C$};
\end{scope}

\draw[->,black!45,line width=.8pt] (.625,.255)--(.725,.255);

\begin{scope}[shift={(.82,0)}]
  \coordinate (AA) at (.10,.20);
  \coordinate (BB) at (.05,.35);
  \coordinate (CC) at (.30,.10);
  \coordinate (N) at (0,.5);
  \coordinate (F) at (.5,0);
  \fill[hypblue!7] (AA)--(N)--(F)--cycle;
  \fill[hyporange!18] (BB)--(N)--(F)--cycle;
  \fill[hypgreen,fill opacity=.17] (CC)--(N)--(F)--cycle;
  \draw[hypblue,line width=1pt] (N)--(AA)--(F);
  \draw[hyporange,line width=.9pt,dash pattern=on 3pt off 2pt] (N)--(BB)--(F);
  \draw[hypgreen,line width=.9pt] (N)--(CC)--(F);
  \draw[hypblue,line width=1pt] (N)--(F)
    node[pos=.49,sloped,above=4pt,text=black!65] {$H$};
  \draw[->,black!70,line width=.65pt] (0,0)--(.56,0) node[right] {$\alpha$};
  \draw[->,black!70,line width=.65pt] (0,0)--(0,.56) node[above] {$\beta$};
  \node[below left] at (0,0) {$0$};
  \node[below=3pt] at (.5,0) {$\tfrac12$};
  \node[left=3pt] at (0,.5) {$\tfrac12$};
  \fill[hypblue] (AA) circle (1.9pt);
  \fill[hyporange] (BB) circle (1.9pt);
  \fill[hypgreen] (CC) circle (1.9pt);
  \node[below left=3pt,text=hypblue] at (AA) {$A$};
  \node[text=hyporange] at (.039,.305) {$B$};
  \node[below=4pt,text=hypgreen] at (CC) {$C$};
  \node[text=hypblue] at (.16,.225) {$\mathcal R_A$};
  \node[text=hyporange,fill=white,fill opacity=.85,text opacity=1,inner sep=.9pt]
    at (.098,.368) {$\mathcal R_B$};
  \node[text=hypgreen,fill=white,fill opacity=.85,text opacity=1,inner sep=.9pt]
    at (.355,.099) {$\mathcal R_C$};
  \node at (.265,-.135) {(b) Feasible model sets};
  \node at (.265,-.193) {$\mathcal R_B,\mathcal R_C\subsetneq\mathcal R_A$};
  \node at (.265,-.25) {$\mathcal R_B\nsubseteq\mathcal R_C,\quad
    \mathcal R_C\nsubseteq\mathcal R_B$};
\end{scope}

\draw[->,black!45,line width=.8pt] (1.445,.255)--(1.545,.255);

\begin{scope}[shift={(1.64,0)}]
  \coordinate (Ocommon) at (0,0);
  \coordinate (Acommon) at (.10,.20);
  \coordinate (Bcommon) at (.05,.35);
  \coordinate (Ccommon) at (.30,.10);
  \coordinate (Bintercept) at ({1/6},0);
  \coordinate (Cintercept) at (0,.25);
  \fill[black!2] (0,0)--(0,.5)--(.5,0)--cycle;
  \draw[black!32,line width=.6pt] (0,.5)--(.5,0)
    node[pos=.49,sloped,above=4pt] {$H$};
  % The lower edges of R_P give the two garbling conditions for a target Q:
  % alpha/(1/2-beta) <= alpha_Q/(1/2-beta_Q) and
  % beta/(1/2-alpha) <= beta_Q/(1/2-alpha_Q), for P off H.
  % The first bound is tighter for B, and the second for C.
  \fill[hypblue!15] (0,0)--(Bintercept)--(Acommon)--(Cintercept)--cycle;
  \fill[black!33] (Ocommon) rectangle (Bcommon |- Ccommon);
  \draw[black!58,densely dashed,line width=.55pt]
    (Ocommon |- Bcommon)--(Bcommon)--(Bcommon |- Ocommon);
  \draw[black!58,densely dashed,line width=.55pt]
    (Ocommon |- Ccommon)--(Ccommon)--(Ccommon |- Ocommon);
  \draw[hyporange,line width=.9pt] (Bintercept)--(Acommon);
  \draw[hypgreen,line width=.9pt] (Cintercept)--(Acommon);
  \draw[hyporange,line width=.75pt,densely dashed] (Acommon)--(Bcommon);
  \draw[hypgreen,line width=.75pt,densely dashed] (Acommon)--(Ccommon);
  \draw[->,black!70,line width=.65pt] (0,0)--(.56,0) node[right] {$\alpha$};
  \draw[->,black!70,line width=.65pt] (0,0)--(0,.56) node[above] {$\beta$};
  \node[below left] at (0,0) {$0$};
  \node[below=3pt] at (.5,0) {$\tfrac12$};
  \node[left=3pt] at (0,.5) {$\tfrac12$};
  \fill[hypblue] (Acommon) circle (2.1pt);
  \fill[hyporange] (Bcommon) circle (1.9pt);
  \fill[hypgreen] (Ccommon) circle (1.9pt);
  \node[above right=1pt,text=hypblue] at (Acommon) {$A$};
  \node[right=3pt,text=hyporange] at (Bcommon) {$B$};
  \node[right=3pt,text=hypgreen] at (Ccommon) {$C$};
  \node at (.265,-.135) {(c) Dominating both $B$ and $C$};
  \node at (.265,-.193) {\textcolor{black!33}{\rule{.8ex}{.8ex}}\enspace
    Coordinatewise};
  \node at (.265,-.25) {\textcolor{hypblue!22}{\rule{.8ex}{.8ex}}\enspace
    Blackwell};
\end{scope}
\end{tikzpicture}%
}
\caption{Feasible model sets and Blackwell comparisons.}
\label{fig:bm-choice-sets}
\end{figure}

The Blackwell order compares strictly more pairs of models than the
coordinatewise order on error rates does in hypothesis testing. In Figure~\ref{fig:bm-choice-sets}, $A$ has different error compositions from $B$ and $C$, yet $A$ can reproduce
them. Panel~(c) shows the set of models that can reproduce both $B$ and $C$.
This region strictly contains the rectangle of
models with weakly lower errors than $B$ and $C$ in both coordinates. The feasible model sets can be incomparable under inclusion, as
$\mathcal R_B$ and $\mathcal R_C$ illustrate, so screening remains
multidimensional. 

This perspective connects our problem to \citet{GanLi2026}, who
study screening over privately known technology sets. The main difference is in \citet{GanLi2026}, the payoff only depends on the reported type given allocation. Their IC
constraint consequently takes the form of inclusion-monotone
promised utility, corresponding to BM here. Our seller must
additionally satisfy TT: a buyer can choose another type's
contract while retaining his own private dataset, and the payoff
from that contract depends on the dataset used.

\subsubsection{BM restricts the implementability}
\label{sec:bm-cm}

Although Blackwell order is a coarse order, garbling can change error composition in a variety of ways. BM therefore restricts screening on both the quality and the error composition, limiting the designer’s ability to extract rents from horizontal differences.

\paragraph{BM in indirect rent.} Consider two data types $P$ and $Q$, $P\succeq_B Q$. The buyer with less informative private data $Q$ has a lower
outside payoff $\ubar(Q)$ and more to gain from full information. Therefore $P$ is the Low type for screening while $Q$ is the High type. 
BM 
caps the rent received by High type $Q$ cannot below the gap in the outside payoff:
\begin{equation}
 u(Q)-u(P)\leq\ubar(P)-\ubar(Q)
 \qquad(P\succeq_B Q).
 \label{eq:bm-incremental}
\end{equation}
where $u(P)=U(P)-\ubar(P)$ is the incremental rent for $P$. Define $u(Q)$ for $Q$ similarly.

In the
\hyperref[sec:illustrative-example]{illustrative example},  $u(N)-u(A)>\ubar(A)-\ubar(N)$, so
$U(N)>U(A)$ despite $A\succeq_B N$.\footnote{Define $
U_N:=W^o(P_N,P_B)-t_{P_B}$ since $B$'s option is optimal for $N$.} Although $N$ even is not in the type space, it is the garbling of $A$.
The owner of $A$ can therefore adopt the composition that benefits
more from a targeted supplement. 

\paragraph{BM and Cyclic monotonicity.} Define the induced allocation of action recommendations by
\begin{equation}
 D_r(a\mid\omega,s)=\sum_y E_r(y\mid\omega)
                           \mathbf1\{\sigma_r(s,y)=a\}.
 \label{eq:model-allocation}
\end{equation}
Therefore the obedient payoff can admit a linear form
\begin{equation}
 \langle P,D_r\rangle_\upsilon :=W^o(P,r)=\sum_{\omega,s,a}P(\omega,s)
                       D_r(a\mid\omega,s)\upsilon(a,\omega).
 \label{eq:model-linear}
\end{equation}Garbling through $G$ gives
\begin{equation}
 D_r^G(a\mid\omega,s)=\sum_{s'}G(s'\mid s)D_r(a\mid\omega,s'),
 \qquad
 \langle P,D_r^G\rangle_\upsilon=\langle PG,D_r\rangle_\upsilon.
 \label{eq:retuned-allocation}
\end{equation}
IC requires
$U(P)\geq\langle P,D_r^G\rangle_\upsilon-t_r$ for every $P,r,G$.
Summing around a cycle:
\begin{equation}
 \sum_{i=1}^k\left[
 \langle P_i,D_{P_i}\rangle_\upsilon
 -\langle P_i,D_{P_{i+1}}^{G_i}\rangle_\upsilon
 \right]\geq0,\qquad P_{k+1}=P_1,\quad k\geq1.
 \label{eq:garbling-cycles}
\end{equation}
Cyclic monotonicity (CM) \citep{Rochet1987} sets every $G_i=I$ while $G_i$ can be any stochastic matrix in \eqref{eq:garbling-cycles}. The
\hyperref[sec:illustrative-example]{illustrative example} satisfies CM.
Consider the cycle $A\to B\to A$. On the first edge, type $A$
garbles his private prediction to the never-flag rule, $P_AG=P_N$,
and follows $B$'s recommendation. On the second edge, type $B$
follows $A$'s recommendation. It violates \eqref{eq:garbling-cycles},
\begin{equation}
\begin{aligned}
 &\bigl[W^o(P_A,P_A)-W^o(P_N,P_B)\bigr]
 +\bigl[W^o(P_B,P_B)-W^o(P_B,P_A)\bigr]\\
 &\qquad=0.800-1+0.964-0.804=-0.040<0.
\end{aligned}
\label{eq:cm-retuning-example}
\end{equation}

\paragraph{BM and Frontier Reshuffling in Duality.}
\label{sec:bm-frontier-duality}

BM also changes the admissible movements of revenue weights in the
dual. In hypothesis testing, all types on $H$ are Blackwell
equivalent and have the same outside payoff, so their incremental
rent is constant: $u|_H=r$. Recall equation \eqref{eq:hyp-revenue} 
\begin{equation}
 R_f(u)=r\lambda_H(H) -\int_Du(x)g(x)\,dx.
 \label{eq:bm-frontier-invariance}
\end{equation}
where $\lambda_H(H)=\int_0^{1/2} \tfrac12 f(a,\tfrac12-a)\,da$ is the frontier
revenue measure after the integration by parts. Thus frontier weights can be redistributed across error
compositions without changing revenue given $u|_H=r$.\footnote{In general, Blackwell monotonicity permits measure reshuffling within types equivalent in Blackwell order.}
Without BM, frontier rents can differ, so redistribution
may change the objective. 

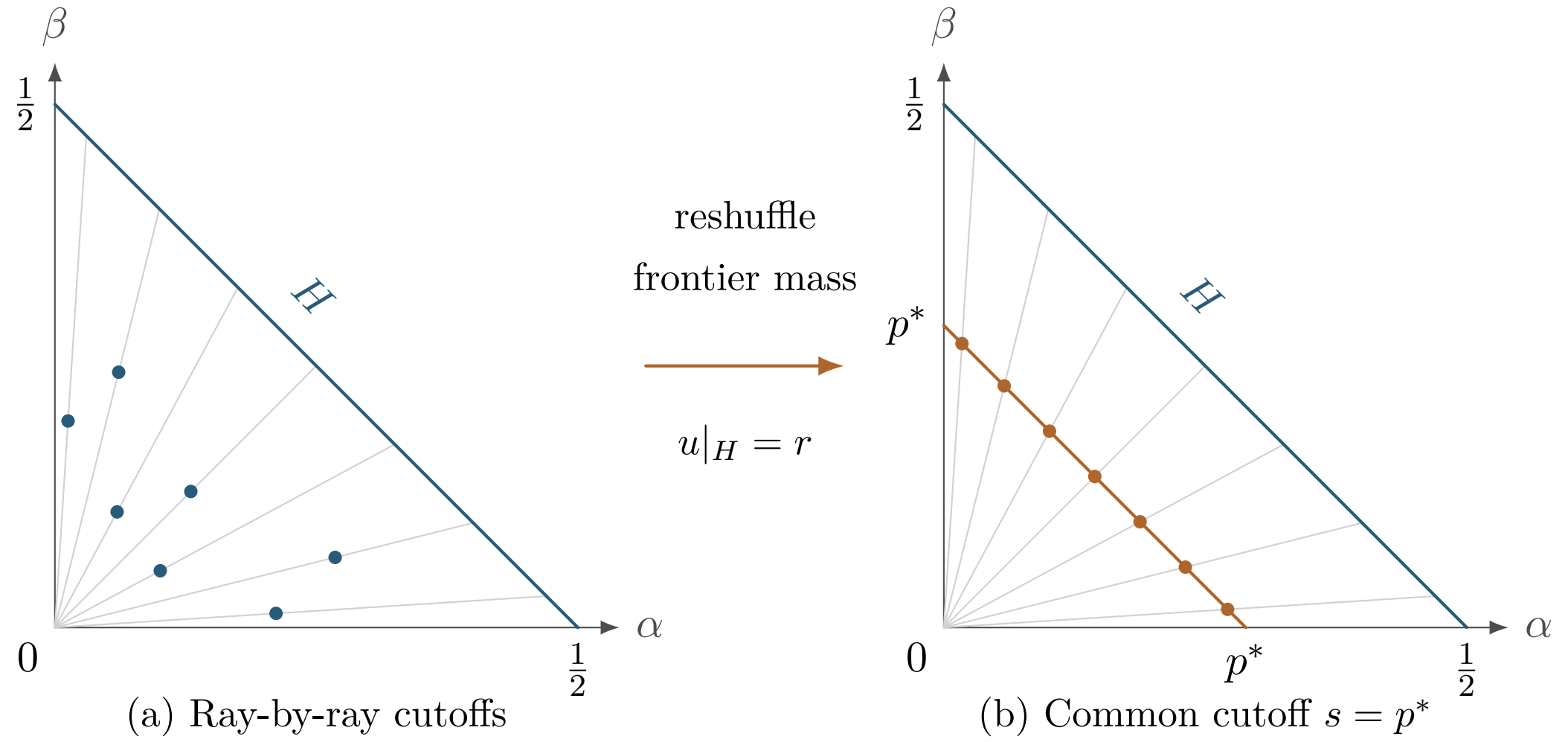
\begin{figure}[tbp]
\centering
\begin{tikzpicture}[x=4.65cm,y=4.65cm,>=Latex,font=\small,
                     line cap=round,line join=round]
 \begin{scope}
  \draw[->,black!70] (0,0)--(1.08,0) node[right] {$\alpha$};
  \draw[->,black!70] (0,0)--(0,1.08) node[above] {$\beta$};
  \foreach \a/\b/\t in {
   .06/.94/.42,.20/.80/.61,.35/.65/.34,.50/.50/.52,
   .65/.35/.31,.80/.20/.67,.94/.06/.45}{
    \draw[black!18,thin] (0,0)--(\a,\b);
    \fill[hypblue] ({\t*\a},{\t*\b}) circle (1.7pt);
  }
  \draw[hypblue,thick] (0,1)--(1,0)
     node[pos=.43,sloped,above=4pt] {$H$};
  \node[below left] at (0,0) {$0$};
  \node[below] at (1,0) {$\tfrac12$};
  \node[left] at (0,1) {$\tfrac12$};
  \node[font=\footnotesize] at (.50,-.17) {(a) Ray-by-ray cutoffs};
 \end{scope}
 \draw[->,hyporange,thick] (1.13,.50)--(1.51,.50);
 \node[align=center,font=\footnotesize] at (1.32,.73)
   {reshuffle\\frontier mass};
 \node[font=\footnotesize] at (1.32,.35) {$u|_H=r$};
 \begin{scope}[shift={(1.70,0)}]
  \draw[->,black!70] (0,0)--(1.08,0) node[right] {$\alpha$};
  \draw[->,black!70] (0,0)--(0,1.08) node[above] {$\beta$};
  \foreach \a/\b in {
   .06/.94,.20/.80,.35/.65,.50/.50,.65/.35,.80/.20,.94/.06}{
    \draw[black!18,thin] (0,0)--(\a,\b);
    \fill[hyporange] ({.577350*\a},{.577350*\b}) circle (1.7pt);
  }
  \draw[hypblue,thick] (0,1)--(1,0)
     node[pos=.43,sloped,above=4pt] {$H$};
  \draw[hyporange,thick] (0,.577350)--(.577350,0);
  \node[below left] at (0,0) {$0$};
  \node[below] at (1,0) {$\tfrac12$};
  \node[left] at (0,1) {$\tfrac12$};
  \node[below] at (.577350,0) {$p^*$};
  \node[left] at (0,.577350) {$p^*$};
  \node[font=\footnotesize] at (.50,-.17) {(b) Common cutoff $s=p^*$};
 \end{scope}
\end{tikzpicture}
\caption{Frontier reshuffling.}
\label{fig:hyp-reshuffling}
\end{figure}

Figure~\ref{fig:hyp-reshuffling} illustrates how this freedom
supports a posted price menu. Each ray fixes an error composition
$(\xi,1-\xi)$, with types $x=s(\xi,1-\xi)$ varying in total error $s\in[0,1/2]$.
Ignoring the mutual IC across different rays, the designer only screens total error $s$ in different rays. It is a one-dimensional screening problem and posted price is optimal. However, each ray can have different
optimal price  cutoffs, as in panel~(a). Moreover, the mutual IC constraints may bind, making the analysis of optimal menu more complex.

The redistribution of frontier measure from BM can help align the cutoffs and avoid mutual IC.
Under $g\geq0$, let $p^*$ be an
optimal full-information posted price. Assign each ray a frontier
weight equal to the interior revenue mass above $p^*$ on that
ray. The first-order condition for $p^*$ ensures that these weights
sum to $B$, making this an admissible reshuffling. With the adjusted
weights, each ray's cutoff objective is maximized at $p^*$,
yielding the straight boundary $\alpha+\beta=p^*$ in panel~(b).

\subsection{Interim information sales}
\label{sec:bbs-discussion}

A lender reviewing a loan application may buy a credit report to
learn more about the applicant. It has already formed a view of the
applicant's risk, and that assessment determines what the report is
worth for the decision at hand. The same lender may license a dataset
of past applications and repayment outcomes to improve its scoring
model for future applicants. At purchase, it knows where its current
model tends to make mistakes. The future applications and their
scores are still unknown. The first transaction is interim, as in
\citet{BergemannBonattiSmolin2018}, henceforth BBS. The second is
ex ante, as in our model. BBS screen a realized posterior. We screen
the private dataset that induces the distribution of future posteriors.

\begin{figure}[htbp]
\centering
\resizebox{.8\linewidth}{!}{%
\begin{tikzpicture}[x=8.8cm,y=8.8cm,>=Latex,font=\small,
  line cap=round,line join=round]
\begin{scope}
  \draw[->,black!70] (0,0)--(.56,0) node[right] {$\alpha$};
  \draw[->,black!70] (0,0)--(0,.56) node[above] {$\beta$};
  \draw[hypblue,line width=1.05pt] (0,.5)--(0,0)--(.5,0);
  \draw[hyporange,line width=.9pt,fill=white]
    (0,.5) circle (2.1pt) (.5,0) circle (2.1pt);
  \fill[hypblue] (.24,0) circle (1.9pt)
    node[above=3pt,text=hypblue] {$\theta_0$};
  \node[below left] at (0,0) {$0$};
  \node[below=3pt] at (.5,0) {$\tfrac12$};
  \node[left=3pt] at (0,.5) {$\tfrac12$};
  \node at (.25,-.13) {(a) Interim: realized posterior};
\end{scope}

\begin{scope}[shift={(.79,0)}]
  \coordinate (A0) at (.24,0);
  \coordinate (A1) at (.175,.22);
  \fill[hyplight] (0,0)--(0,.5)--(.5,0)--cycle;
  \draw[->,black!70] (0,0)--(.56,0) node[right] {$\alpha$};
  \draw[->,black!70] (0,0)--(0,.56) node[above] {$\beta$};
  \draw[hyporange,line width=.9pt] (0,.5)--(.5,0)
    node[pos=.52,sloped,above=3pt,text=black!65] {$H$};
  \draw[->,hypgreen,line width=1.05pt,shorten <=3pt,shorten >=4pt]
    (A0)--(A1);
  \fill[hypblue] (A0) circle (1.9pt)
    node[below=3pt,text=hypblue] {$A_0$};
  \fill[hypgreen] (A1) circle (1.9pt)
    node[right=4pt,text=hypgreen] {$A_1$};
  \node[text=hypblue] at (.09,.31) {$D$};
  \node[below left] at (0,0) {$0$};
  \node[below=3pt] at (.5,0) {$\tfrac12$};
  \node[left=3pt] at (0,.5) {$\tfrac12$};
  \node at (.25,-.13) {(b) Ex ante: private dataset};
\end{scope}
\end{tikzpicture}%
}
\caption{Type spaces under interim and ex ante sales.}
\label{fig:bbs-type-spaces}
\end{figure}
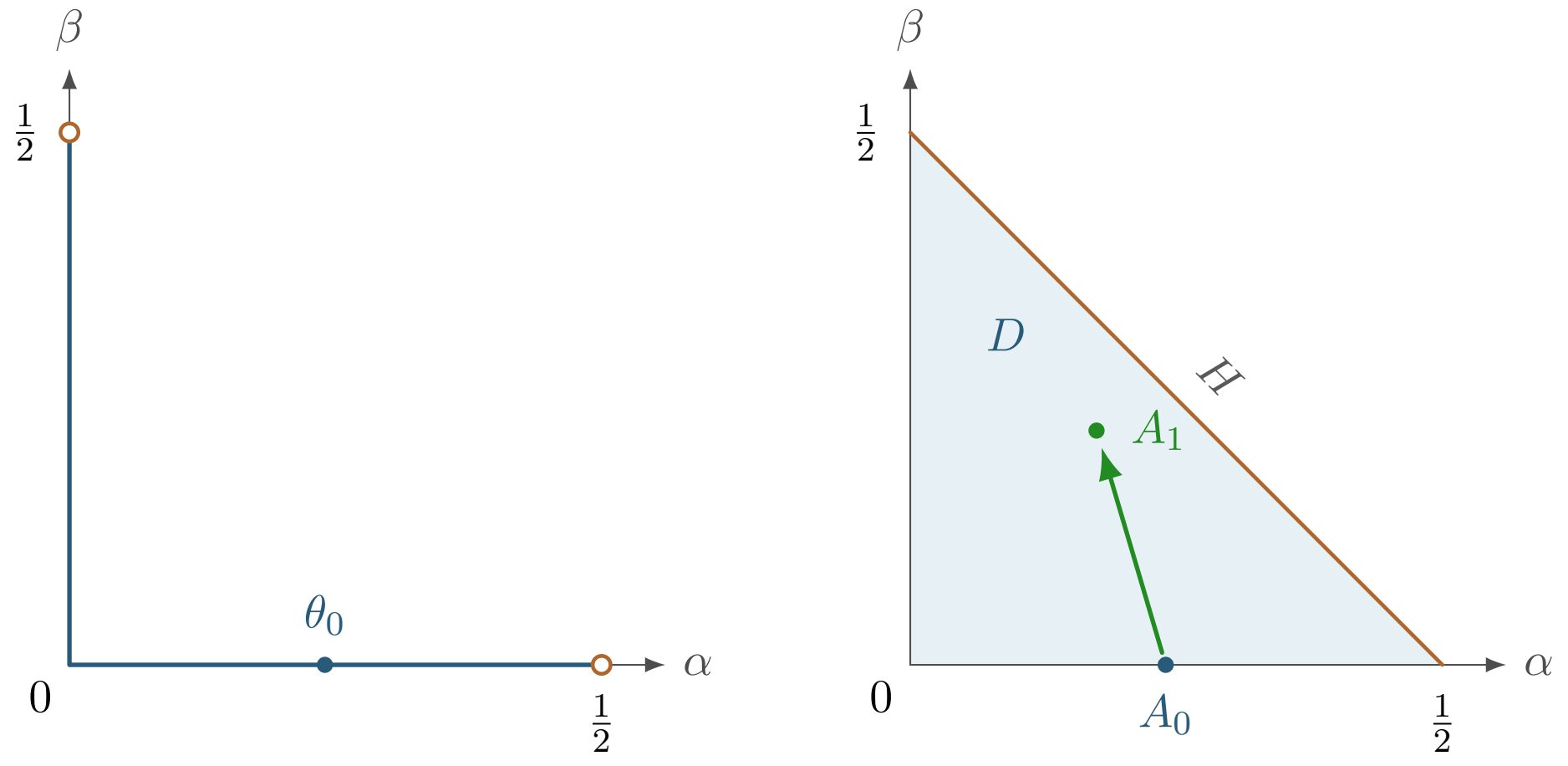

In the hypothesis testing, let $\theta=\Pr(\omega=1)$ be the
buyer's posterior in BBS and $v=\min\{\theta,1-\theta\}$ the
probability of a mistake without supplemental information. If the
buyer predicts state $2$, the error is a false positive, and if he
predicts state $1$, it is a false negative. The posterior interval
therefore folds onto the two error axes:
\begin{equation}
 \theta\leq\tfrac12:\quad (\alpha,\beta)=(v,0),\qquad
 \theta\geq\tfrac12:\quad (\alpha,\beta)=(0,v).
 \label{eq:bbs-folding-main}
\end{equation}
The two endpoints with $v=\tfrac12$ represent the same balanced
belief. An ex ante buyer can make both kinds of mistakes on future
cases, so its private type can lie in the interior of $D$
(Figure~\ref{fig:bbs-type-spaces}).

\paragraph{Information goods and data goods.}
The sales timing and when the buyer observes his prediction change both
what the seller can screen and what the buyer can adjust. In BBS, buyers with the same probability of a mistake can value a partial
experiment differently because they need stronger evidence about
different states to change their decisions. The position of
information therefore gives the seller a way to screen buyers who
value full information equally. Retuning is not feasible for buyers in BBS since a single private signal has been realized.

With ex ante sales, private datasets have richer horizontal differences
since they vary in the composition of prediction errors, giving the seller
more scope to target particular errors. However, the buyer can also retune his
model, as the move from $A_0$ to $A_1$ in
Figure~\ref{fig:bbs-type-spaces}, panel~(b), illustrates. He can then use a
supplement aimed at the resulting mix of errors. Blackwell
monotonicity limits the seller's ability to screen those differences for extraction.

\paragraph{Implementability and optimal design.} We can rephrase the BBS into the error representation.
In the binary recommendation representation, a concentrated experiment in the sense of \citet[p.~8, item~(ii)]{BergemannBonattiSmolin2018} satisfies $\max_i \rho_i=1$, where $\rho_i$ is the probability of recommending the correct action in state $i$.
Write $\chi(\theta)=\rho_1(\theta)-\rho_2(\theta)$ for the difference
of the experiment assigned to posterior $\theta$. The allocations
$x_1(\alpha)=1+\chi(\alpha)$ and $x_2(\beta)=1-\chi(1-\beta)$ are
the slopes of incremental rent along the two error branches.
Rent starts at zero on each branch, and the two envelope formulas
must give the same rent at the balanced belief.
\begin{equation}
 \int_0^{1/2}x_1(\alpha)\,d\alpha
 =\int_0^{1/2}x_2(\beta)\,d\beta.
 \label{eq:bbs-resource-main}
\end{equation}
Condition \eqref{eq:bbs-resource-main}, together with each $x_i$ being nondecreasing and $\bar x_1+\bar x_2\leq 2$,
characterizes implementability, where $\bar x_i$ is the limit of $x_i$ at the balanced belief.

Appendix~\ref{app:bbs-extreme-points} uses the balance condition to
give an extreme-point interpretation of the optimal menu in
\citet{BergemannBonattiSmolin2018}. An optimal menu
offers full information and at most one partial experiment. For example, When both are offered, relabel the states so that buyers with lower
posteriors buy partial information. Let $a$ be the price of full
information, $b$ the posterior at which buyers start buying partial
information, and $c$ the posterior at which they switch to full
information. Buyers with $\theta\in[b,c)$ buy the partial experiment
$(\rho_1,\rho_2)=(\lambda,1)$ at price $\lambda b$, and those with
$\theta\in[c,1-a]$ buy full information at price $a$. Condition \eqref{eq:bbs-resource-main} gives
$a=\lambda b+(1-\lambda)c$, which is $c$'s indifference between partial information and full information. All other buyers
take the outside option. In error coordinates, the assignments are
\begin{equation}
 \begin{aligned}
  x_1(\alpha)
    &=\lambda\mathbf{1}\{\alpha\geq b\}
       +(1-\lambda)\mathbf{1}\{\alpha\geq c\},\\
  x_2(\beta)
    &=\mathbf{1}\{\beta\geq a\}
       +(1-\lambda)\mathbf{1}\{\beta>1-c\}.
 \end{aligned}
 \label{eq:bbs-optimal-cutoffs}
\end{equation}

\clearpage
\appendix
\section{Proof of the revelation principle}\label{app:binary}

We prove Theorem~\ref{thm:dataset-revelation}
in two parts. First, we show that the recommended rows in a
normalized contract can reproduce an optimal continuation after
any report. Second, we show that normalizing a fully IC mechanism
preserves truthful utilities and weakly increases payments at
every report. 

\subsection{Garbling and incentive compatibility}\label{app:garbling}

For a garbling $G\in\mathcal G_n$, $W^o(PG,r)$ is defined by
\eqref{eq:model-prescribed} and \eqref{PG}.

\begin{lemma}\label{lem:imitation}
At report $r$, a type $P$ who draws $s'$ according to $G(\cdot\mid s)$
and follows $\sigma_r(s',y)$ obtains gross payoff $W^o(PG,r)$.
Moreover, $W^o(PG,r)\leq W(P,r)$.
\end{lemma}

\begin{proof}
Conditional independence gives the joint law
\[
 \Pr(\omega,s',y)
 =\sum_s P(\omega,s)G(s'\mid s)E_r(y\mid\omega)
 =(PG)(\omega,s')E_r(y\mid\omega).
\]
Following \(\sigma_r\) under this joint law gives \(W^o(PG,r)\).
To compare it with \(W(P,r)\), fix a pair \((s,y)\) that occurs
with positive probability. The buyer draws \(s'\) using only \(s\),
so this draw leaves his posterior over the state unchanged.
His conditional payoff from garbling and then following the
recommendation is therefore
\[
 \sum_{s'}G(s'\mid s)
 \sum_{\omega}\Pr(\omega\mid s,y)\,
       \upsilon\bigl(\sigma_r(s',y),\omega\bigr)
 \leq
 \max_{a\in A}\sum_{\omega}\Pr(\omega\mid s,y)\,
       \upsilon(a,\omega).
\]
The left-hand side averages the payoffs of the recommended actions
under the same posterior. Averaging the inequality over \((s,y)\)
gives \(W^o(PG,r)\leq W(P,r)\).
\end{proof}

\begin{lemma}[Response encoding]\label{thm:general-encoding}\label{prop:encoding}
Suppose that for every $P,r\in\Theta$ there is $G_{P,r}\in\mathcal G_n$
such that $PG_{P,r}\in\Theta$ and
\begin{equation}\label{eq:appencoding}
 W(P,r)=W^o(PG_{P,r},r).
\end{equation}
Then IC is equivalent to TT and BM, and hence to TT, OB, and BM.
\end{lemma}

\begin{proof}
Suppose the mechanism satisfies IC. For any \(P,r\in\Theta\),
following the recommendation at report \(r\) is a feasible
continuation, so
\[
 W^o(P,r)-t_r\leq W(P,r)-t_r\leq U(P).
\]
This gives TT. At \(r=P\), the left-hand side equals \(U(P)\),
so both inequalities bind and OB follows. If \(Q=PG\in\Theta\),
Lemma~\ref{lem:imitation} gives
\[
 U(P)\geq W(P,Q)-t_Q
       \geq W^o(Q,Q)-t_Q=U(Q).
\]
Thus IC also implies BM.

Conversely, suppose TT and BM hold. Fix \(P,r\in\Theta\) and take
\(Q=PG_{P,r}\in\Theta\) from \eqref{eq:appencoding}. TT for type
\(Q\) compares reporting \(r\) with reporting \(Q\). BM compares
the truthful utilities of \(P\) and \(Q\), since \(P\) can garble
its signal to reproduce \(Q\). Hence
\begin{equation}\label{eq:encodingproof}
 W(P,r)-t_r=W^o(Q,r)-t_r
 \overset{\mathrm{TT}}{\leq}U(Q)
 \overset{\mathrm{BM}}{\leq}U(P).
\end{equation}
For \(r=P\), this inequality gives OB because
\(U(P)=W^o(P,P)-t_P\) and \(W(P,P)\geq W^o(P,P)\).
For every other report, it rules out the report even when the
buyer chooses an optimal continuation. Together these inequalities
give IC.
\end{proof}

\subsection{Encoding optimal response}

Suppose $m=J=2$. Let $s_i^r$ denote the
private label of rank $i$ in the report's ordering. Specifically,
order its nonnull labels by increasing $r(2,s)/[r(1,s)+r(2,s)]$,
place null labels last, and break ties by the original label order.
Let $\tau_r$ denote this permutation, so $s_i^r=\tau_r(i)$. 
\begin{definition}[Normalized binary contracts]\label{def:normalized}
A mechanism belongs to \(\mathcal N_n\) if each contract uses the
outcome labels \(y_0,\ldots,y_n\) and satisfies the following
conditions. At report \(r\), write \(e_{rij}=E_r(y_j\mid i)\).
The likelihood matrix is
\begin{equation}\label{eq:normalizedmatrix}
 E_r=
 \begin{pmatrix}
 e_{r10}&e_{r11}&\cdots&e_{r1,n-1}&0\\
 0&e_{r21}&\cdots&e_{r2,n-1}&e_{r2n}
 \end{pmatrix}.
\end{equation}
For each \(j=1,\ldots,n-1\), the two entries in column \(j\) are
either both positive or both zero. A column with two zero entries
represents an unused outcome. Over the columns with two positive
entries, the likelihood ratios \(e_{r2j}/e_{r1j}\) are weakly
increasing in \(j\).

Order the private labels that occur under \(r\) by increasing
posterior probability of state \(2\), breaking ties by label order.
Place the labels with zero probability last, in label order. Write
\(s_i^r\) for the label of rank \(i\). At outcome \(y_j\), the
contract recommends
\begin{equation}\label{eq:cutoffprofiles}
 c_r^j(s_i^r)=
 \begin{cases}
 a_1,& j\leq n-i,\\
 a_2,& j>n-i,
 \end{cases}
 \qquad j=0,\ldots,n.
\end{equation}
\end{definition}

\begin{proposition}\label{prop:binaryencoding}
On $\Theta_{2,n}$, condition
\eqref{eq:appencoding} holds for every $M\in\mathcal N_n$. 
\end{proposition}
\begin{proof}
The two actions' payoff difference has opposite signs in the two
states. Write
\[
 d_1=\upsilon(a_1,1)-\upsilon(a_2,1)>0,
 \qquad
 d_2=\upsilon(a_2,2)-\upsilon(a_1,2)>0.
\]
Fix a type \(P\), a report \(r\), and an actual private label \(s\).
For each \(j\in\{1,\ldots,n-1\}\) with \(e_{r1j}>0\), the gain
from choosing \(a_2\) rather than \(a_1\) at \((s,y_j)\), weighted
by the probability of signal pair, is
\[
 d_2P(2,s)e_{r2j}-d_1P(1,s)e_{r1j}.
\]
The gain is nonnegative when
\[
 d_2P(2,s)\frac{e_{r2j}}{e_{r1j}}\geq d_1P(1,s).
\]
Since the likelihood ratios are weakly increasing across the
columns with positive entries, this inequality continues to hold
at every later such column once it holds. Let \(k(s)\) be the
first such column, with \(k(s)=n\) if there is none. Choosing
\(a_2\)  when \(j\geq k(s)\) gives an optimal continuation.
At a zero column, either action contributes zero to expected
payoff. At \(y_0\) and \(y_n\), the rule selects the action
optimal in the revealed state.

By \eqref{eq:cutoffprofiles}, this continuation is  the
recommendation row for \(s^r_{n-k(s)+1}\). Define \(G_{P,r}\) to
send each actual label \(s\) to that row's label with probability
one. The cutoff and the label rank are determined by finitely many comparisons. The buyer can therefore choose the appropriate
row using \(s\) alone and follow it at every \(y_j\). Therefore, \(PG_{P,r}\in\Theta_{2,n}\) and
\(
 W(P,r)=W^o(PG_{P,r},r).
\)
\end{proof}
\subsection{Binary normalization}\label{app:normalization}

\begin{proposition}[Normalization]\label{prop:normalization}
For every IC mechanism \(M\) on \(\Theta_{2,n}\), there exists
an IC mechanism \(\widetilde M\in\mathcal N_n\) such that
\begin{align}
 U_{\widetilde M}(P)&=U_M(P) &&(P\in\Theta_{2,n}),
 \label{eq:normalizationU}\\
 \widetilde t_r&\geq t_r &&(r\in\Theta_{2,n}),
 \label{eq:normalizationt}\\
 W_{\widetilde M}(P,r)-\widetilde t_r
 &\leq W_M(P,r)-t_r &&(P,r\in\Theta_{2,n}).
 \label{eq:normalizationW}
\end{align}
\end{proposition}
\begin{proof}
Fix a report \(r\), write \(e_i(y)=E_r(y\mid i)\), and let \(d_i>0\)
be the payoff gain from choosing \(a_i\) over the other action in
state \(i\). For each original supplementary signal \(y\), choose an
optimal recommendation for type \(r\) at every private signal \(s\).
If \(y\) reveals state \(i\), choose \(a_i\) at every \(s\). If both
likelihoods are positive, choose \(a_2\) at a nonnull \(s\) exactly when
\begin{equation}\label{eq:strictbinarychoice}
 d_2r(2,s)e_2(y)>d_1r(1,s)e_1(y).
\end{equation}
Choose \(a_2\) at null private signals in this case, and discard any
\(y\) with zero likelihood in both states. Since the nonnull private
signals are ordered by increasing posterior probability of state \(2\)
and the null signals come last, the resulting profile is \(c_r^{J(y)}\)
for some \(J(y)\in\{0,\ldots,n\}\). Write
\[
 \bar e_{rij}:=\sum_{y:J(y)=j}e_i(y).
\]

To construct the new contract, draw \(Y\) from \(E_r\). If
\(1\le J(Y)\le n-1\), send \(y_{J(Y)}\). Otherwise, send \(y_0\)
in state \(1\) and \(y_n\) in state \(2\). At \(y_j\), recommend
\(c_r^j\). This gives \eqref{eq:normalizedmatrix}. Every active
interior column pools signals with positive likelihood in both states.
Among those original signals, \(J(y)\) is nondecreasing in
\(e_2(y)/e_1(y)\). Each pooled ratio is a state-\(1\)-likelihood-weighted
average of the ratios in its group, so the active interior ratios are
ordered by \(j\).

Fix any type \(P\). Take an optimal rule in the new contract that
chooses \(a_1\) at \(y_0\) and \(a_2\) at \(y_n\), as these signals
reveal the state. In the original contract, the buyer can apply this
rule to \((s,y_{J(Y)})\). Under the common draw \(Y\), the two rules
choose the same action except when state \(1\) is paired with
\(J(Y)=n\) or state \(2\) with \(J(Y)=0\). The expected payoff gain
in the new contract is therefore
\begin{equation}\label{eq:normalizationincrement}
 \Delta_r:=\pi_1d_1\bar e_{r1n}+\pi_2d_2\bar e_{r20}\geq 0.
\end{equation}
This gain is the same for every \(P\) because the prior is common.
Since the original buyer can optimize over all rules,
\[
 W_{\widetilde M}(P,r)-\Delta_r\leq W_M(P,r).
\]
For type \(r\), applying the new recommended rule to
\((s,y_{J(Y)})\) gives the original optimal rule by construction.
Thus following the new recommendations gives \(r\) gross payoff
\(W_M(r,r)+\Delta_r\). 

Set \(\widetilde t_r=t_r+\Delta_r\). At every report, the intended
type's truthful net payoff is unchanged, and every type's optimized
net payoff from that report weakly decreases. These comparisons give
\eqref{eq:normalizationU}--\eqref{eq:normalizationW} and preserve
full IC and IR. 
\end{proof}
\begin{proof}[Proof of Theorem~\ref{thm:dataset-revelation}]
Proposition~\ref{prop:normalization} allows us to restrict attention
to \(\mathcal N_n\) without loss of revenue.
Proposition~\ref{prop:binaryencoding} verifies the response-encoding
condition on this class. By Lemma~\ref{thm:general-encoding}, IC
is therefore equivalent to TT, OB, and BM within \(\mathcal N_n\).
Imposing IR and nonnegative payments on both sides gives
\eqref{eq:encoding-characterization}.
\end{proof}

\subsection{Binary classification as bundling over two nested chains}
\label{app:binary-nested-bundling}

Consider binary classification with \(n\) private labels and unit
payoff for a correct classification. For a normalized contract at
report \(r\), use the report's label order \(s_1^r,\ldots,s_n^r\)
and write \(e_{rij}=E_r(y_j\mid i)\) as in
Definition~\ref{def:normalized}. At \(y_j\), the cutoff rule
\eqref{eq:cutoffprofiles} recommends \(a_1\) on the prefix \(A_j^r\)
and \(a_2\) on the complementary suffix \(B_j^r\), where
\[
 A_j^r=\{s_i^r:i\le n-j\},\qquad
 B_j^r=\{s_i^r:i>n-j\},\qquad j=0,\ldots,n.
\]
As \(j\) increases, \(A_j^r\) shrinks from \(S\) to \(\varnothing\)
and \(B_j^r\) grows from \(\varnothing\) to \(S\). For instance, with
three private labels, the two chains are
\[
\begin{array}{c|cccc}
 j       &0&1&2&3\\ \hline
 A_j^r   &S&\{s_1^r,s_2^r\}&\{s_1^r\}&\varnothing\\
 B_j^r   &\varnothing&\{s_3^r\}&\{s_2^r,s_3^r\}&S .
\end{array}
\]

There are two goods for each label \(s\), one for taking \(a_1\)
at \(s\) in state \(1\) and one for taking \(a_2\) at \(s\) in state
\(2\). Their values to type \(P\) are the joint masses \(P(1,s)\)
and \(P(2,s)\), respectively. Thus the additive bundle values are
\(v_P(A_j^r)=\sum_{s\in A_j^r}P(1,s)\) and
\(v_P(B_j^r)=\sum_{s\in B_j^r}P(2,s)\). By conditional independence,
\begin{equation}
 \label{eq:binary-nested-bundling-payoff}
 W^o(P,r)
 =\sum_{j=0}^n e_{r1j}v_P(A_j^r)
  +\sum_{j=0}^n e_{r2j}v_P(B_j^r).
\end{equation}
The contract therefore specifies two lotteries $e_{r1j},e_{r2j}$ over nested bundles,
one for each state, at a single price \(t_r\). The two lotteries are linked by the common
outcome index \(j\) and the normalization conditions: \(e_{r20}=e_{r1n}=0\),
the interior columns are either positive in both states or zero in
both, and \(e_{r2j}/e_{r1j}\) is weakly increasing across the positive
interior columns. Conversely, these conditions on the two probability
vectors define a normalized supplementary dataset. 

Without supplementary data, \(P\) chooses a single cutoff in its own
order, so
\[
 \ubar(P)=\max_{0\le j\le n}
           \{v_P(A_j^P)+v_P(B_j^P)\}.
\]
Under (TT), the buyer selects a report to maximize the obedient net
payoff \(W^o(P,r)-t_r\). His total utility is
\(U(P)=W^o(P,P)-t_P\), and his incremental rent is \(U(P)-\ubar(P)\).
By Theorem~\ref{thm:dataset-revelation}, on the full domain
\(\Theta_{2,n}\), maximizing revenue over these contracts subject to
(TT), (OB), (BM), and (IR) attains the original IC--IR revenue bound.

For \(n=2\) and the uniform prior, the middle cutoff follows the
ordered private prediction and is optimal without supplementary data.
Writing \(x=(\alpha,\beta)=(P(1,s_2^P),P(2,s_1^P))\) and
\(q=(q_1,q_2)=(e_{P10},e_{P22})\) at the truthful contract,
\eqref{eq:binary-nested-bundling-payoff} then gives
\(U(P)-\ubar(P)=\alpha q_1+\beta q_2-t_P\), the two-good
representation in Appendix~\ref{app:hyp-representation}.

\section{Proof of the hypothesis-testing problem}
\label{app:hyp-representation}

We apply the revelation principle to two states, two actions, and two
private labels. The prior is uniform and the payoff is one for a correct
prediction and zero otherwise. 

\subsection{Selling two goods representation}

Write \(P_x\) for the joint law with error pair \(x=(\alpha,\beta)\),
ordered so that private label \(i\) recommends action \(a_i\):
\[
 P_x=\begin{pmatrix}
        \tfrac12-\alpha&\alpha\\
        \beta&\tfrac12-\beta
      \end{pmatrix}.
\]
The rows are states and the columns are private labels. Put
\[
 D=\{x\in\mathbb R_+^2:s(x):=\alpha+\beta\leq\tfrac12\},
 \qquad H=\{x\in D:s(x)=\tfrac12\},\qquad
 d(x)=\tfrac12-s(x).
\]
The two labels recommend optimal actions precisely when \(x\in D\).
Any dataset in the full domain \(\Theta_{2,2}\) has such a
representative. Order its columns by their posterior probabilities of
state \(2\) when both labels occur, breaking posterior ties by label
order. If one label is null, put it last. We retain both endpoints of $H$ as
fixed-label representations of the same uninformative experiment. At \(P_x\), the outside payoff is
\(\ubar(P_x)=1-s(x)\), and the incremental value of full information is
\(s(x)\).

By Theorem~\ref{thm:dataset-revelation}, it suffices to consider
contracts in \(\mathcal N_2\). Their supplementary datasets have the form
\begin{equation}\label{c:experiment}
 E(q)=
 \begin{pmatrix}q_1&1-q_1&0\\0&1-q_2&q_2\end{pmatrix},
 \qquad q=(q_1,q_2)\in[0,1]^2.
\end{equation}
The first and last outcomes, \(y_0\) and \(y_2\), reveal states
\(1\) and \(2\). At the middle outcome \(y_1\), the normalized
recommendation follows the ordered private label. For \(P_x\) this rule
leaves joint error probabilities \(\alpha(1-q_1)\) and
\(\beta(1-q_2)\), by conditional independence of the two datasets.
Thus the recommended response gives incremental net payoff
\(x\cdot q-t\). 

\begin{lemma}[Obedience]\label{c:optimized}
For every \(x\in D\) and \(q\in[0,1]^2\),
\begin{equation}\label{c:optimized-equation}
 W(x;q)=
 \max\left\{1-s(x)+x\cdot q,\ 
                  \tfrac12(1+q_1),\ \tfrac12(1+q_2)\right\}.
\end{equation}
Obedience  holds if and
only if
\begin{equation}\label{c:obedience}
 (\tfrac12-\alpha)(1-q_1)\geq\beta(1-q_2),\qquad
 (\tfrac12-\beta)(1-q_2)\geq\alpha(1-q_1).
\end{equation}
\end{lemma}
\begin{proof}
At a revealing outcome, the buyer knows the state and chooses the
matching action. At \(y_1\), the recommendation is to choose \(a_i\)
after private label \(i\). Following the recommendation is optimal
exactly when, at each label, the recommended state has at least as
much joint probability as the other state. These two comparisons
are precisely \eqref{c:obedience}.

Break ties in favor of the recommendation. Reversing the
recommendation at both labels would then require both inequalities
in \eqref{c:obedience} to be strictly reversed. This requires
\(q_1,q_2<1\). Multiplying the two strict inequalities and cancelling
\((1-q_1)(1-q_2)\) gives
\[
 \alpha\beta>(\tfrac12-\alpha)(\tfrac12-\beta),
\]
which contradicts \(\alpha+\beta\leq\tfrac12\).

This leaves three possible optimal responses at \(y_1\). Following
the private label leaves errors \(\alpha(1-q_1)\) and
\(\beta(1-q_2)\), giving payoff \(1-s(x)+x\cdot q\).
Choosing \(a_2\) at both labels is correct in state \(2\) at every
outcome and in state \(1\) only when that state is revealed, giving
\(\tfrac12(1+q_1)\). Symmetrically, choosing \(a_1\) at both labels
gives \(\tfrac12(1+q_2)\). Taking the maximum gives
\eqref{c:optimized-equation}.
\end{proof}

\subsection{Characterization of truthtelling and individual rationality}\label{c:utility}

\begin{lemma}[Affine utility representation (\citet{Rochet1987})]\label{c:utility}
Truthtelling for the affine payoff \(x\cdot q-t\)
is equivalent to \(u\) being convex and \(q(x)\) supporting \(u\)
at each \(x\), with
\begin{equation}\label{c:payments}
 t(x)=x\cdot q(x)-u(x).
\end{equation}
\end{lemma}

By \citet{Rochet1987}, 
The restriction \(q(x)\in[0,1]^2\) makes \(u\) coordinatewise
nondecreasing and \(1\)-Lipschitz in the \(\ell^1\) norm.
At the origin $(0,0)$, IR and nonnegative payments imply \(u(0)=0\).
Denote the resulting class of rents $u$ on \(D\) by \(\mathcal U\).

Conversely, every \(u\in\mathcal U\) admits a selection of
supporting slopes \(q(x)\in[0,1]^2\). The supporting inequality
at the origin gives \(t(x)\geq0\), and monotonicity gives \(u\geq0\).
Thus \(\mathcal U\) is the set of implementable rents under
TT and IR with nonnegative payments. At interior differentiability
points, \(q(x)=\nabla u(x)\).

\subsection{Blackwell monotonicity and the frontier}

\begin{lemma}[BM is no distortion on the frontier]\label{c:bm}
For \(u\in\mathcal U\), the total utility 
\(U(x)=1-s(x)+u(x)\) satisfies BM 
if and only if \(u\) is constant on \(H\).
\end{lemma}

\begin{proof}
All datasets on \(H\) are uninformative and hence Blackwell
equivalent. Their outside payoff is \(\tfrac12\), so BM implies
\(u|_H=r\) for some \(r\).

For the converse, take \(u|_H=r\). A garbling with probabilities
\(a\) of sending label \(1\) to label \(2\) and \(b\) of sending
label \(2\) to label \(1\) produces \(P_z\) from \(P_x\)  when
\begin{equation}\label{c:reachable}
 z=(1-a-b)x+a h_1+b h_2,,\qquad
 s(z)=s(x)+(a+b)d(x).
\end{equation}
where \(h_1=(\tfrac12,0)\) and \(h_2=(0,\tfrac12)\). If \(d(x)>0\), membership of \(z\) in \(D\) implies \(a+b\leq1\).
Conversely, every \(a,b\geq0\) with \(a+b\leq1\) defines such a
garbling. Thus the reachable error pairs form
\(\operatorname{co}\{x,(\tfrac12,0),(0,\tfrac12)\}\). This formula
also holds for \(x\in H\), since all points on \(H\) are Blackwell
equivalent and a garbling of an uninformative dataset stays
uninformative. 

Fix \(x\in D\).
The point \(h=x+d(x)(1,0)\) lies on \(H\), and
\(\|h-x\|_1=d(x)\). Since \(u(h)=r\), the Lipschitz bound gives
\[
 u(x)\geq u(h)-\|h-x\|_1=r-d(x).
\]
Using \(d(x)=\tfrac12-s(x)\), we obtain
\[
 U(x)=1-s(x)+u(x)
 \geq 1-s(x)+r-d(x)=\tfrac12+r.
\]
The right-hand side is the total utility of every frontier type.

Now let \(z\) be any type reachable by garbling \(P_x\). By the
reachable-set formula, we can write
\[
 z=(1-a-b)x+a h_1+b h_2,
 \qquad a,b\geq0,\quad a+b\leq1,
\]
Since \(U=1-s+u\) is convex and \(U(h_1)=U(h_2)=\tfrac12+r\),
\[
 U(z)
 \leq (1-a-b)U(x)+aU(h_1)+bU(h_2)
 = (1-a-b)U(x)+(a+b)(\tfrac12+r)
 \leq U(x).
\]
Thus it establishes BM.
\end{proof}

\subsection{The reduced problem}\label{reduce}
\begin{lemma}
For $u\in\mathcal{U}$, the following conditions are equivalent:
OB, BM, IC, and \(u|_H\) is constant.
\end{lemma}

\begin{proof}
By the Rochet characterization in Lemma~\ref{c:utility},
\[
 x\cdot q_z-t_z\leq u(x)\qquad(x,z\in D),
\]
with equality at \(x=z\), and \(u\) is \(1\)-Lipschitz in
\(\ell_1\).

Suppose OB holds. For \(h\in\operatorname{relint}H\),
Lemma~\ref{c:optimized} and \(d(h)=0\) imply
\[
 h\cdot q_h\geq \tfrac12\max\{q_{h,1},q_{h,2}\}.
\]
Since \(2h_1\) and \(2h_2\) are positive weights summing to one,
this inequality requires \(q_{h,1}=q_{h,2}\). Thus
\(h'\cdot q_h-t_h=h\cdot q_h-t_h=u(h)\) for every \(h'\in H\).
TT gives \(u(h')\geq u(h)\). Interchanging \(h\) and \(h'\) when
both lie in \(\operatorname{relint}H\) gives equality. Continuity
extends this equality to the endpoints of \(H\).

Now suppose \(u|_H\equiv r\). Applying the TT constraints at  the two endpoints of \(H\) to any $z\in D$ gives
\(\tfrac12q_{z,i}-t_z\leq r\) for \(i=1,2\) and every \(z\in D\).
The point \(x+d(x)(1,0)\) lies on \(H\), so the Lipschitz bound
gives \(r-d(x)\leq u(x)\). Therefore
\[
 x\cdot q_z-t_z\leq u(x),
 \qquad
 \tfrac12q_{z,i}-t_z-d(x)\leq r-d(x)\leq u(x)
 \quad(i=1,2).
\]
These inequalities bound every term in the optimized payoff from
report \(z\) in Lemma~\ref{c:optimized}. The prescribed response
under the truthful report gives \(u(x)\), so IC holds.

Finally, IC implies OB and BM. Under BM, the types on \(H\) have
the same total utility because they are Blackwell equivalent.
They also have the same outside payoff, so \(u\) is constant on
\(H\).
\end{proof}

Let \(\mathcal U^{\rm BM}
=\{u\in\mathcal U : u|_H \text{ is constant}\}\). Therefore, the implementable set of incremental rent functions under IC and IR
with nonnegative payments is  \(\mathcal U^{\rm BM}\).
\begin{proposition}[Reformulation of the designer's problem]
\label{c:full-implementation}
 Each
\(u\in\mathcal U^{\rm BM}\) is implementable by a mechanism in
\(\mathcal N_2\). If the distribution of error pairs has density
\(f\) on \(D\), then
\begin{equation}\label{c:binary-design}
 \sup_{M\in\mathcal M_{\mathrm{IC,IR}}}\operatorname{Rev}_F(M)
 =
 \sup_{u\in\mathcal U^{\rm BM}}
 \int_D [x\cdot\nabla u(x)-u(x)]f(x)\,dx.
\end{equation}
\end{proposition}

\begin{proof}
\emph{Implementation in \(\mathcal N_2\).}
Fix \(u\in\mathcal U^{\rm BM}\). Lemma~\ref{c:utility} gives
a reduced TT--IR menu \((q_x,t_x)_{x\in D}\) with nonnegative
payments that implements \(u\). The preceding lemma implies
that this menu is IC.

Let \(\rho=u|_H\) and set \(p=\tfrac12-\rho\).
Since \(u(0)=0\) and \(u\) is nondecreasing and \(1\)-Lipschitz,
\(p\in[0,\tfrac12]\). Full information at price \(p\) gives
incremental rent \(s(x)-p=\rho-d(x)\leq u(x)\), with equality
on \(H\). If \(q_{x,i}=1\) for some \(i\), TT at the frontier
endpoint \(\tfrac12 e_i\) gives \(\tfrac12-t_x\leq\rho\), so
\(t_x\geq p\). Hence
\[
 u(x)=x\cdot q_x-t_x\leq s(x)-p\leq u(x).
\]
Assign full information at \(p\) to every type on \(H\) and to
every type whose assigned contract has a coordinate equal to one.
Each type still attains \(u(x)\) at its assigned contract, and
every contract gives optimized rent at most \(u(x)\).
Thus IC and IR are preserved.

Every retained contract has both middle-outcome likelihoods
positive, and full information has both equal to zero. Assign each type the contract for its canonical
error pair and write the recommendation in that type's label order.
This preserves the assigned payoff and every optimized deviation
payoff. The resulting mechanism belongs to \(\mathcal N_2\).

\emph{Revenue.}
By Theorem~\ref{thm:dataset-revelation}, it suffices to consider
normalized IC--IR mechanisms. For such a mechanism, its reduced
rent is
\[
 u(x)=\max\left\{0,\sup_P
 \big[W(x;q_P)-t_P-(1-s(x))\big]\right\}.
\]
By Lemma~\ref{c:optimized}, this is a supremum of affine functions
with slopes in \([0,1]^2\). IR and nonnegative payments give
\(u(0)=0\), and BM makes \(u\) constant on \(H\). Thus
\(u\in\mathcal U^{\rm BM}\).

For a type \(P\) with canonical error pair
\(x\in\operatorname{int}D\), the truthful recommendation follows
the ordered private prediction at the middle outcome. Since
\(P\) and \(P_x\) have the same rent by BM,
\[
 u(x)=x\cdot q_P-t_P,\qquad
 u(z)\geq z\cdot q_P-t_P\quad(z\in D),
\]
where the inequality follows from IC and Lemma~\ref{c:optimized}.
Thus \(q_P\) supports \(u\) at \(x\). Wherever \(u\) is
differentiable in the interior, \(q_P=\nabla u(x)\), so
\(t_P=x\cdot\nabla u(x)-u(x)\). The density assumption makes this
identity hold almost surely. Integrating gives the revenue of
each normalized mechanism, and the implementation above attains
the corresponding revenue.
This proves \eqref{c:binary-design}.
\end{proof}

\subsection{The revenue identity}

For \(u\in\mathcal U\), apply the divergence theorem to the
Lipschitz vector field \(u(x)f(x)x\) on the triangle \(D\).
Since \(\operatorname{div}(f(x)x)=2f(x)+x\cdot\nabla f(x)\),
\[
 \int_D x\cdot\nabla u(x) f(x)\,dx
 =\int_{\partial D}u(x) f(x) x\cdot n(x)\,dS(x)
   -\int_Du(x)[2f(x)+x\cdot\nabla f(x)]\,dx.
\]
Here \(n\) is the outward unit normal and \(dS\) is arc length.
The coordinate edges have zero flux. Parameterize \(H\) by
\(h(a)=(a,\tfrac12-a)\) for \(a\in[0,\tfrac12]\). On that edge
\(x\cdot n=1/(2\sqrt2)\) and \(dS=\sqrt2\,da\). Define the frontier
measure \(\lambda_H\) and its mass \(B\) by
\begin{equation}\label{d:boundarymeasure}
 \lambda_H(L)=
 \int_0^{1/2}\mathbf 1\{h(a)\in L\}\tfrac12 f(h(a))\,da,
 \qquad B=\lambda_H(H)
\end{equation}
for every Borel \(L\subseteq D\). Subtracting \(\int_D u f\,dx\)
from the divergence identity gives
\begin{equation}\label{d:revenue}
 R_f(u):=\int_Hu\,d\lambda_H-\int_D u(x)g(x)\,dx=Br-\int_D u(x)g(x)\,dx.
\end{equation}
The second equality is from \(u\in\mathcal U^{\rm BM}\) with \(u|_H=r\).

\subsection{The posted-price bound}

\begin{theorem}[Posted pricing in hypothesis testing]\label{thm:hyp-posted}
If \(g\geq0\) on \(D\), a single posted price for full information
maximizes revenue under IC and IR. Its revenue is
\begin{equation}\label{eq:hyp-posted-value}
 \sup_{M\in\mathcal M_{\mathrm{IC,IR}}}\operatorname{Rev}_F(M)
 =\max_{0\leq p\leq1/2}p\,\Pr\{s(X)\geq p\}.
\end{equation}
\end{theorem}
\begin{proof}
By Proposition~\ref{c:full-implementation}, take any feasible
reduced rent \(u\in\mathcal U^{\rm BM}\), set \(p=\tfrac12-r\). Nonnegativity and the Lipschitz bound
from \(0\) to \(H\) give \(0\leq r\leq\tfrac12\).
Lemma~\ref{c:obstacle} and IR imply
\begin{equation}\label{d:envelope}
 u(x)\geq \max\{0,r-d(x)\}=(s(x)-p)_+.
\end{equation}
Since \(u_p|_H=r\), the two boundary terms in
\eqref{d:revenue} coincide. Let $u_p(x):=(s(x)-p)_+$ Therefore
\begin{equation}\label{d:dominance}
 R_f(u_p)-R_f(u)=\int_D g(x)[u(x)-u_p(x)]\,dx\geq0.
\end{equation}
The law of \(s(X)\) has no atoms by absolute continuity of
\(X\), so 
\begin{equation}\label{d:pricerevenue}
 R_f(u_p)=p\,\Pr\{s(X)\geq p\}.
\end{equation}
This function is continuous on the compact interval
\([0,\tfrac12]\), so it attains its maximum. The comparison in
\eqref{d:dominance} bounds every feasible revenue by that maximum,
which the posted price attains.
\end{proof}

\section{Proof for the generalized contamination space}
\label{app:contamination}
\subsection{Type space}

Consider multiclass classification with a uniform prior and matching
payoffs. Let $(S_r)_{r=1}^k$ partition $S$ into nonempty groups. The
clean dataset $C\in\Theta_{m,n}$ has pairwise disjoint statewise
supports. A group may have zero probability under $C$.

Let $\ell(\omega)$ denote the group of true class $\omega$. We assume
that all clean labels for class $\omega$ belong to that group, so
\[
 \operatorname{supp} C(\cdot\mid\omega)
 \subseteq S_{\ell(\omega)}
 \qquad\text{for every }\omega\in\Omega.
\]
Several true classes may belong to the same group.

Let $Q$ be the $n\times k$ stochastic matrix that maps each label to
its group, and let $B_h$ be the $k\times n$ stochastic matrix whose
$r$th row draws a label from $h_r$ on $S_r$. Drawing a label and then
reading its group returns the original group, so $B_hQ=I_k$. The
contamination matrix is $\Gamma_h=QRB_h$.

Set $x_j=\varepsilon h_r(j)$ for $j\in S_r$. Since each $h_r$ is a
probability distribution, the coordinates in each group sum to
$\varepsilon$. Define
\[
 L=\left\{x\in\mathbb R^n:
      \sum_{j\in S_r}x_j=\sum_{j\in S_1}x_j
      \text{ for every }r\right\},
 \qquad
 \varepsilon(x)=\sum_{j\in S_1}x_j.
\]
The equalities impose $k-1$ independent linear restrictions, so
$L$ has dimension $d=n-k+1$. The coordinate domain and its frontier are
\[
 \mathcal D=\{x\in L:x\geq0,\ \varepsilon(x)\leq1\},
 \qquad
 \mathcal H=\{x\in\mathcal D:\varepsilon(x)=1\}.
\]
For $x\in\mathcal D\setminus\{0\}$, recover $h$ by setting
$h_r(j)=x_j/\varepsilon(x)$ for $j\in S_r$, and write
$P_x=P_{\varepsilon(x),h}$. Set $P_0=C$. Since
$P_{\varepsilon,h}=(1-\varepsilon)C+\varepsilon CQRB_h$, for
$j\in S_r$ we have
\begin{equation}
 P_x(\omega,j)
  =(1-\varepsilon(x))C(\omega,j)+(CQR)_{\omega r}x_j.
 \label{eq:gc-affine-coordinates}
\end{equation}
Thus $P_x$ is affine in $x$. The expected revenue under a law of $X$
on $\mathcal D$ equals the expected revenue under the induced law of
$P_X$.

\begin{lemma}\label{lem:gc-blackwell-geometry}
Define
\[
 \kappa=\sum_{r=1}^k\max_{\omega\in\Omega}(CQR)_{\omega r},
 \qquad
 \bar v=1-\kappa,
 \qquad
 v(x)=\bar v\,\varepsilon(x).
\]
Then $\ubar(P_x)=1-v(x)$. The datasets on $\mathcal H$ are Blackwell
equivalent, and every $P_x$ Blackwell dominates each of them. For each
fixed $h$, $P_{\varepsilon,h}\succeq_B P_{\varepsilon',h}$ whenever
$0\leq\varepsilon\leq\varepsilon'\leq1$.
\end{lemma}

\begin{proof}
Let $N=R^{-1}$. Since $R\mathbf 1=\mathbf 1$, we also have
$N\mathbf 1=\mathbf 1$. Fix a column $z=Re_r$ of $R$, and let $i$
be any index at which $z$ attains its maximum. The column $z$ is
nonnegative and nonzero, so $z_i>0$. Using the unit row sum of $N$,
we can write
\[
 (Nz)_i
 = z_i + \sum_{j\ne i}(-N_{ij})(z_i-z_j)
 \ge z_i > 0.
\]
Each term in the sum is nonnegative because $N_{ij}\le 0$ and
$z_i\ge z_j$ for $j\ne i$. Since $Nz=e_r$ has only one positive
entry, at index $r$, we must have $i=r$. Thus $R_{rr}$ is strictly
larger than every other entry in column $r$. 

Under the uniform prior, all of class $\omega'$'s clean probability
mass $1/m$ lies in group $\ell(\omega')$. It follows that
\[
 (CQR)_{\omega'r}=\frac{1}{m}R_{\ell(\omega'),r}.
\]
If $j\in S_r$ and $C(\omega,j)>0$, then $\ell(\omega)=r$. The column
maximum property of $R$ therefore gives
\begin{equation}
 \omega\in\arg\max_{\omega'\in\Omega}(CQR)_{\omega'r}.
 \label{eq:gc-common-maximizer}
\end{equation}

Since the clean supports are pairwise disjoint, such a state $\omega$
also maximizes $C(\cdot,j)$. Hence the two terms on the right of
\eqref{eq:gc-affine-coordinates} have a common maximizing state.
If column $j$ of $C$ is zero, any maximizing state for the second
term also maximizes the first. Using $\sum_{j\in S_r}x_j=\varepsilon(x)$
and the fact that clean labels identify the true class, we obtain
\[
 \ubar(P_x)
 =\sum_{j\in S}\max_{\omega\in\Omega}P_x(\omega,j)
 =(1-\varepsilon(x))
   +\varepsilon(x)\sum_{r=1}^k\max_{\omega\in\Omega}(CQR)_{\omega r}
 =1-v(x).
\]

For the Blackwell comparisons, set
\[
 K_\varepsilon=(1-\varepsilon)I_k+\varepsilon R,
 \qquad
 N_\varepsilon=(1-\varepsilon)R^{-1}+\varepsilon I_k.
\]
The matrix $N_\varepsilon$ has unit row sums and nonpositive
off-diagonal entries. It is therefore strictly diagonally dominant
and invertible. To see that its inverse is nonnegative, suppose
$N_\varepsilon z\geq0$ and let $z_i=\min_j z_j$. If $z_i<0$, then
\[
 (N_\varepsilon z)_i
 =z_i+\sum_{j\ne i}(-(N_\varepsilon)_{ij})(z_i-z_j)
 \leq z_i<0,
\]
a contradiction. Thus $N_\varepsilon z\geq0$ implies $z\geq0$, so
$N_\varepsilon^{-1}$ is nonnegative. Its row sums are one, making it
a stochastic matrix.

Using $B_hQ=I_k$, we have $P_{\varepsilon,h}Q=CQK_\varepsilon$.
Since $K_\varepsilon=RN_\varepsilon$, the garbling
$QN_\varepsilon^{-1}B_{h'}$ gives
\begin{equation}
 P_{\varepsilon,h}(QN_\varepsilon^{-1}B_{h'})
 =CQRB_{h'}=P_{1,h'}.
 \label{eq:gc-frontier-garbling}
\end{equation}
This holds for all $h,h'$ and $\varepsilon\in[0,1]$. Taking
$\varepsilon=1$ establishes Blackwell equivalence on the frontier.
The same construction shows that every $P_x$ Blackwell dominates
every frontier dataset.

Finally, fix $h$. For $\varepsilon<\varepsilon'$, set
$a=(1-\varepsilon')/(1-\varepsilon)$. The stochastic matrix
$aI_n+(1-a)QN_\varepsilon^{-1}B_h$ maps $P_{\varepsilon,h}$ to
\[
 aP_{\varepsilon,h}+(1-a)P_{1,h}=P_{\varepsilon',h}.
\]
If $\varepsilon=\varepsilon'$, the two datasets coincide.
\end{proof}
\subsection{Revenue comparison}

\begin{proof}[Proof of Theorem~\ref{thm:generalized-contamination}]
If $\bar v=0$, Lemma~\ref{lem:gc-blackwell-geometry} and IR imply that
all payments are zero. Suppose $\bar v>0$. We consider the revenue of any
mechanism satisfying TT, BM, and IR.

Define $u(x)=U(P_x)-\ubar(P_x)$ and $c_r=1-W^o(C,r)\geq0$ for report
$r\in\Theta$. By \eqref{eq:gc-affine-coordinates} and
Lemma~\ref{lem:gc-blackwell-geometry}, there is a vector $b_r\in L$ such
that
\[
 W^o(P_x,r)-t_r-\ubar(P_x)=x\cdot b_r-c_r-t_r.
\]
The vectors $b_r$ are uniformly bounded. Indeed,
$W^o(P_x,r)$ is linear in the joint law $P_x$ with coefficients
$\sum_y E_r(y\mid\omega)\mathbf1\{\sigma_r(j,y)=\omega\}\in[0,1]$. TT
and IR imply
\begin{equation}
 u(x)=\max\left\{0,\sup_{r\in\Theta}
                       [x\cdot b_r-c_r-t_r]\right\},
 \qquad 0\leq u(x)\leq v(x).
 \label{f:cont-envelope}
\end{equation}
The truthful report attains the supremum, and the upper bound follows
from $W^o(P_x,P_x)\leq1$ and $t_{P_x}\geq0$. Thus $u$ is convex and
Lipschitz continuous on $\mathcal D$. At each interior point of differentiability
relative to $L$, the truthful branch has slope $\nabla_Lu(x)$, which
implies
\begin{equation}
 t_{P_x}=x\cdot\nabla_Lu(x)-u(x)-c_{P_x}.
 \label{f:cont-payment}
\end{equation}

By BM and Lemma~\ref{lem:gc-blackwell-geometry}, $u$ has a common value
$\rho\in[0,\bar v]$ on $\mathcal H$, and
$U(P_x)\geq 1-\bar v+\rho$ on $\mathcal D$. At $p=\bar v-\rho$, IR
and this last inequality yield
\begin{equation}
 u(x)\geq u_p(x):=(v(x)-p)_+,
 \qquad u|_{\mathcal H}=u_p|_{\mathcal H}.
 \label{f:cont-posted-envelope}
\end{equation}
The utility $u_p$ is delivered by the menu with the outside option
and full information at price $p$.  A buyer of type $P_x$ can obtain  $1-p$ from buying and $\ubar(P_x)$ from not buying.

Take $f$ in \eqref{eq:gc-density} to have a $C^1$ extension to a
relative neighborhood of $\mathcal D$ in $L$. Let $dx$ be
$d$-dimensional Lebesgue measure on $L$ and $dS$ be
boundary surface measure relative to $L$. For a Lipschitz function
$w$ on $\mathcal D$, let
\[
 \mathcal L_f(w)=\int_{\mathcal D}[x\cdot\nabla_Lw(x)-w(x)]f(x)\,dx.
\]
The divergence theorem applied to $xfw$ gives
\begin{equation}
 \mathcal L_f(w)=\int_{\mathcal H}wf\,x\cdot\nu\,dS
                  -\int_{\mathcal D}w(x)g(x)\,dx,
 \label{f:cont-revenue-identity}
\end{equation}
where $\nu$ is the outward unit normal relative to $L$ and
$g(x)=(d+1)f(x)+x\cdot\nabla_L f(x)$. Every other facet
lies in a hyperplane through the origin and satisfies $x\cdot\nu=0$.

The law of $X$ has a density and $u$ is Lipschitz, so
\eqref{f:cont-payment} holds almost surely at $X$. It gives
$\operatorname{Rev}_F(M)=\mathcal L_f(u)-\int_{\mathcal D}c_{P_x} f(x)\,dx$.
The function $v$ is linear on $L$, so
$x\cdot\nabla_Lu_p(x)-u_p(x)=p\mathbf1\{v(x)>p\}$ almost everywhere.
The law of $v(X)$ is atomless because $\bar v>0$ and $F$ has a density.
Using  \eqref{f:cont-posted-envelope} yields
\begin{equation}
 p\Pr\{v(X)\geq p\}-\operatorname{Rev}_F(M)
 =\int_{\mathcal D}g(x)[u(x)-u_p(x)]\,dx
                +\int_{\mathcal D}c_{P_x}f(x)\,dx\geq0.
 \label{f:cont-exact-gap}
\end{equation}
The inequality uses \eqref{eq:gc-density} and the nonnegativity of
$c_{P_x}$. The posted-price revenue is continuous on $[0,\bar v]$, so
it has a maximizer. This implementation is feasible under IC and attains the upper
bound on the TT--BM--IR relaxation. Therefore, a posted price is optimal.
\end{proof}

\subsection{Binary classification }
\label{sec:binary-private-datasets}

We now allow the clean dataset to vary across buyers and take the full type
and report space \(\Theta=\Theta_{2,n}\), with the uniform prior and matching
payoffs. The value of full information is,
\[
    v(P)=1-\underline{u}(P)
        =\sum_{j\in S}\min\{P(1,j),P(2,j)\},
    \qquad \bar v=\tfrac12 .
\]
For \(0<v(P)<\bar v\), put \(b_j=\min\{P(1,j),P(2,j)\}\) and define
\[
    h_j=\frac{b_j}{v(P)}, \qquad
    C(\omega,j)=\frac{P(\omega,j)-b_j}{1-2v(P)} .
\]
The two rows of \(C\) have disjoint supports and each sums to \(1/2\).
Thus every such \(P\) lies on a unique path indexed by \(\xi=(C,h)\),
\begin{equation}
    P_{\xi,v}(\omega,j)=(1-2v)C(\omega,j)+vh_j,
    \qquad 0\le v\le\bar v .
    \label{eq:binary-contamination-path}
\end{equation}
Along each path, replacing some private labels with draws from \(h\)
raises \(v\). The endpoints are fully informative at \(v=0\) and
uninformative at \(v=\bar v\). Let \(P\sim F\) and \(V=v(P)\). On the interior population
\(0<V<\bar v\), let \(\Xi=(C,h)\) and write the conditional revenue
from selling full information at price \(p\) as
\begin{equation}
    R_\xi(p)=p\Pr\{V\ge p\mid \Xi=\xi,\ 0<V<\bar v\},
    \qquad 0\le p\le\bar v .
    \label{eq:binary-conditional-price-revenue}
\end{equation}

\paragraph{Differential virtual value.}
Fix a path \(\xi=(C,h)\), and suppose the conditional distribution
\(F_\xi\) of \(V\) has a continuously differentiable density \(f_\xi\).
Differentiating posted-price revenue gives
\[
    -R_\xi'(p)=p f_\xi(p)-\bigl[1-F_\xi(p)\bigr].
\]
The right-hand side is the density-weighted virtual value. Thus
revenue is concave in price exactly when this virtual value is
nondecreasing along the path. 
\begin{theorem}
\label{thm:binary-private-datasets-posted-price}
Suppose \(R_\xi\) is concave in price for almost every \(\xi\) in the
interior population. A posted price for a fully informative dataset is
optimal, with price
\[
    p^\ast\in\arg\max_{p\in[0,\bar v]}p\Pr\{V\ge p\}.
\]
\end{theorem}

Along each path, a mixture of threshold prices reproduces the buyer's
rent and bounds his payment. BM equalizes rents at the uninformative
endpoints, so the mixtures have the same mean price. Concavity then
implies that posting this price on every path weakly increases revenue.

\begin{proof}[Proof of Theorem~\ref{thm:binary-private-datasets-posted-price}]
Fix a mechanism satisfying TT, BM, and IR with nonnegative payments.
Write \(u_\xi(v)=U(P_{\xi,v})-\underline{u}(P_{\xi,v})\).
By \eqref{eq:binary-contamination-path},
\(\underline{u}(P_{\xi,v})=1-v\). For a fixed report \(r\),
\(W^o(P_{\xi,v},r)-t_r-(1-v)\) is affine in \(v\). TT makes \(u_\xi\)
the upper envelope of these affine functions. IR and nonnegative
payments give \(0\le u_\xi(v)\le v\), hence \(u_\xi(0)=0\).
For \(v<w\le\bar v\), replacing the private label by a draw from \(h\)
with probability \((w-v)/(\bar v-v)\) garbles \(P_{\xi,v}\) into
\(P_{\xi,w}\). Thus BM gives \(u_\xi(w)-u_\xi(v)\le w-v\).
Convexity and the minimum at zero give \(u_\xi(w)\ge u_\xi(v)\).

Use the right derivative \(u'_{\xi,+}\) as the distribution function of
\(\Lambda_\xi\) on \([0,\bar v)\), and put any remaining mass at \(\bar v\).
Integrating the derivative gives
\[
    u_\xi(v)=\int_{[0,\bar v]}(v-a)_+\,d\Lambda_\xi(a).
\]
All \(P_{\xi,\bar v}\) are Blackwell equivalent and have outside option
\(1-\bar v\). BM therefore gives \(u_\xi(\bar v)=\rho\) for a common
\(\rho\in[0,\bar v]\). In particular,
\[
    \int_{[0,\bar v]}a\,d\Lambda_\xi(a)=\bar v-\rho=:p_\rho .
\]

At an interior \(v\), the truthful report gives an affine support to
\(u_\xi\) with slope \(q\le u'_{\xi,+}(v)\) and intercept
\(-c-t_{P_{\xi,v}}\), where \(c=1-W^o(P_{\xi,0},P_{\xi,v})\ge0\).
Consequently,
\[
    t_{P_{\xi,v}}=vq-u_\xi(v)-c
        \le v u'_{\xi,+}(v)-u_\xi(v)
        =\int_{[0,v]}a\,d\Lambda_\xi(a).
\]
Conditional on \(\xi\) in the interior population, Tonelli's theorem
and Jensen's inequality give
\[
    \mathbb E[t_P\mid \Xi=\xi,\ 0<V<\bar v]
       \le \int_{[0,\bar v]}R_\xi(a)\,d\Lambda_\xi(a)
       \le R_\xi(p_\rho).
\]
Types with \(V=0\) pay zero by IR. At \(V=\bar v\), payments are at most
\(1-U(P)=p_\rho\). 
A buyer facing full information at price \(p\) buys when \(V\ge p\),
with purchase at indifference. This menu is fully IC and IR.
Its revenue is upper semicontinuous in \(p\) and attains its maximum
on \([0,\bar v]\).
\end{proof}

\section{Beyond regularity: duality and menu coarsening}
\label{app:hyp-posted}

Recall from Appendix~\ref{app:hyp-representation}. \(D,H,s(x)=\alpha+\beta\),  \(d(x)=\tfrac12-s(x)\), \(f\) be a nonnegative
probability density on \(D\) with a \(C^1\) extension to a
neighborhood of \(D\), \(g(x)=3f(x)+x\cdot\nabla f(x)\). \(\mathcal U^{\rm BM}\) consist of the convex, coordinatewise
nondecreasing functions \(u:D\to\mathbb R\) with \(u(0)=0\) and
\(u|_H\equiv r\) for some \(r\in\mathbb R\), satisfying
\[
  |u(h)-u(x)|\leq \|h-x\|_1
  \qquad (x\in D,\ h\in H).
\]

\subsection{Frontier indifference and horizontal incentive compatibility}
\begin{lemma}[An equivalent frontier bound]\label{c:obstacle}
Suppose \(u:D\to\mathbb R\) is convex and coordinatewise
nondecreasing, with \(u(0)=0\) and \(u|_H\equiv r\). Then
\(u\in\mathcal U^{\rm BM}\) if and only if
\begin{equation}\label{c:vertical}
  0\leq r-u(x)\leq d(x)\qquad (x\in D).
\end{equation}
These conditions are also equivalent to \(u\) being
\(1\)-Lipschitz on \(D\) for the \(\ell^1\) norm.
\end{lemma}

\begin{proof}
For each \(x\in D\), the point \(x+d(x)e_1\) lies on \(H\)
and is coordinatewise above \(x\). Monotonicity and the
frontier Lipschitz bound give \eqref{c:vertical}. Conversely,
\(\|h-x\|_1\geq d(x)\) for every \(h\in H\), so
\eqref{c:vertical} implies the frontier Lipschitz bound.

To obtain the global bound, take distinct \(x,z\in D\) with
\(x\leq z\) coordinatewise. The ray from \(x\) through \(z\)
meets \(H\) at a point \(h\) with \(\|h-x\|_1=d(x)\).
Writing \(\lambda=\|z-x\|_1/d(x)\), convexity gives
\[
  0\leq u(z)-u(x)
  \leq \lambda\bigl(u(h)-u(x)\bigr)
  \leq \|z-x\|_1.
\]
For arbitrary \(x,z\in D\), apply this bound from their
coordinatewise minimum \(w=x\wedge z\) to each point. Both
increments are nonnegative, so
\[
  |u(x)-u(z)|
  \leq \max\{\|x-w\|_1,\|z-w\|_1\}
  \leq \|x-z\|_1.
\]
The global Lipschitz bound implies the frontier bound directly.
\end{proof}
\subsection{Strong duality}
\label{app:hyp-obstacle-dual}

\begin{proof}[Proof of Theorem~\ref{thm:hyp-strong-duality}]
Write $C(D)$ for the space of real continuous functions on $D$ with
the uniform norm, and $\mathcal M(D)$ for the space of finite signed
Radon measures on $D$. For $v\in C(D)$ and $\rho\in\mathcal M(D)$,
write $\langle v,\rho\rangle=\int_D v\,d\rho$.

\emph{The primal problem.}
Let $\mathcal F=\{w\in\mathcal K:w\leq d\}$, where the inequality is
pointwise on $D$. For every $x=(\alpha,\beta)\in D$, the point
$(\alpha+d(x),\beta)$ belongs to $H$ and is coordinatewise above $x$.
Since each $w\in\mathcal K$ is coordinatewise nonincreasing and
vanishes on $H$, it follows that $w\geq 0$ on $D$.

For $u\in\mathcal U^{\rm BM}$, let $r$ denote its common value on $H$.
The function $w=r-u$ belongs to $\mathcal K$, and
Lemma~\ref{c:obstacle} gives $0\leq w\leq d$. Conversely, for
$w\in\mathcal F$, the function $u=w(0)-w$ is continuous, convex,
and coordinatewise nondecreasing, with $u(0)=0$ and $u|_H=w(0)$.
Its frontier gap is $w$, so Lemma~\ref{c:obstacle} gives
$u\in\mathcal U^{\rm BM}$. Since $u(0)=0$,
$w(0)=r$.

The divergence theorem under \eqref{d:revenue},
applied to the vector field $x\mapsto f(x)x$, gives
\[
 \int_D g(x)\,dx
 =\int_D f(x)\,dx+\int_{\partial D} f(x)x\cdot n(x)\,dS(x)
 =1+B,
\]
where $n$ is the outward unit normal and $dS$ is arc length. The
coordinate axes have zero flux, so the boundary term equals $B$.
Substituting $u=r-w$ and $r=w(0)$ into \eqref{d:revenue} therefore yields
\begin{equation}
\label{d:gap-revenue}
\begin{aligned}
 R_f(u)
 &=\left(B-\int_D g(x)\,dx\right)r+\int_D w(x)g(x)\,dx\\
 &=\int_D w(x)g(x)\,dx-w(0)
 =\langle w,\nu\rangle.
\end{aligned}
\end{equation}

The set $\mathcal F$ contains zero. Concavity, coordinatewise
monotonicity, vanishing on $H$, and the bound $w\leq d$ are all preserved under
uniform limits, so $\mathcal F$ is closed in $C(D)$. The bound
$0\leq w\leq d$ gives uniform boundedness. For each $w\in\mathcal F$,
the corresponding rent $u$ is $1$-Lipschitz for the $\ell_1$ norm, so
$|w(x)-w(y)|=|u(x)-u(y)|\leq\|x-y\|_1$ for all $x,y\in D$.
Thus $\mathcal F$ is equicontinuous and, by Arzel\`a--Ascoli, compact
in $C(D)$. Because $\nu$ is finite,
$w\mapsto\langle w,\nu\rangle$ is continuous in the uniform norm.
It follows that the primal value
$P^*=\max_{w\in\mathcal F}\langle w,\nu\rangle$ is finite and attained.

\emph{Weak duality.}
Let $w\in\mathcal F$ and let $\lambda\geq 0$ satisfy
$\lambda\succeq_{\mathcal K}\nu$. The order constraint gives the first
inequality below, and $w\leq d$ and $\lambda\geq0$ give the second:
\begin{equation}
\label{d:weak-duality}
 \langle w,\nu\rangle\leq \langle w,\lambda\rangle
 \leq \langle d,\lambda\rangle.
\end{equation}

\emph{Adding penalty.}
Let $\nu=\nu^+-\nu^-$ be the Jordan decomposition of $\nu$, set
$C=\nu^+(D)$, and define
$\Lambda_C=\{\lambda\in\mathcal M(D):\lambda\geq0,\ \lambda(D)\leq C\}$.
For any $w\in\mathcal K$, define $\widehat w=\min\{w,d\}$.
The functions $w$ and $d$ are continuous, concave, coordinatewise
nonincreasing, and vanish on $H$. Their minimum preserves continuity,
monotonicity, and the boundary condition. Its hypograph is the
intersection of the two convex hypographs, so it is concave as well.
Since $\widehat w\leq d$, we have $\widehat w\in\mathcal F$. By
$w-\widehat w=(w-d)_+\geq0$ and the Jordan decomposition,
\[
 \langle w,\nu\rangle
 =\langle\widehat w,\nu\rangle+\langle(w-d)_+,\nu\rangle
 \leq P^*+\langle(w-d)_+,\nu^+\rangle
 \leq P^*+C\|(w-d)_+\|_\infty.
\]
At a maximizer in $\mathcal F$, the penalty term is zero. Hence
\[
 P^*=\sup_{w\in\mathcal K}\bigl\{\langle w,\nu\rangle
 -C\|(w-d)_+\|_\infty\bigr\}.
\]

For fixed $w\in\mathcal K$, let $m=\|(w-d)_+\|_\infty$. Since
$d-w\geq-m$ on $D$, every $\lambda\in\Lambda_C$ satisfies
$\langle d-w,\lambda\rangle\geq-m\lambda(D)\geq-mC$.
If $m>0$, continuity and compactness give a point $x^*\in D$ with
$w(x^*)-d(x^*)=m$, and $C\delta_{x^*}$ attains the lower bound.
If $m=0$, the zero measure attains it. Consequently, for
$L(w,\lambda)=\langle w,\nu\rangle+\langle d-w,\lambda\rangle$,
\[
 P^*=\sup_{w\in\mathcal K}\min_{\lambda\in\Lambda_C}L(w,\lambda).
\]

\emph{Minimax and dual attainment.}
By the Riesz representation theorem, we identify $\mathcal M(D)$ with
$C(D)^*$ under the pairing above. In the weak$^*$ topology, the map
$\lambda\mapsto\langle v,\lambda\rangle$ is continuous for each
$v\in C(D)$. The condition $\lambda\geq 0$ is equivalent to
$\langle v,\lambda\rangle\geq 0$ for every nonnegative $v\in C(D)$,
and the mass bound is $\langle 1,\lambda\rangle\leq C$.
These inequalities define a weak$^*$-closed convex set.

For a nonnegative measure, $\|\lambda\|_{\mathrm{TV}}=\lambda(D)$.
Thus $\Lambda_C$ lies in the closed dual ball of radius $C$.
This ball is weak$^*$ compact by Banach--Alaoglu, so its closed
subset $\Lambda_C$ is weak$^*$ compact as well.

The set $\mathcal K$ is a convex cone in $C(D)$. For each
$\lambda\in\Lambda_C$, $L(\cdot,\lambda)$ is affine and continuous
in the uniform norm, and for each $w\in\mathcal K$,
$L(w,\cdot)$ is affine and weak* continuous. Sion's minimax theorem,
with $\Lambda_C$ as the compact minimization set, gives
\[
 P^*=\inf_{\lambda\in\Lambda_C}\sup_{w\in\mathcal K} L(w,\lambda).
\]
If $\lambda\succeq_{\mathcal K}\nu$, then
$\langle w,\nu-\lambda\rangle\leq0$ for all $w\in\mathcal K$, with
equality at $w=0$. The inner supremum is then $\langle d,\lambda\rangle$.
If the order constraint fails, there is a $w_0\in\mathcal K$ with
$\langle w_0,\nu-\lambda\rangle>0$. The functions $t w_0$ belong to
$\mathcal K$ for all $t\geq0$, and $L(tw_0,\lambda)\to+\infty$ as
$t\to\infty$.

The set $\{\lambda\in\Lambda_C:\lambda\succeq_{\mathcal K}\nu\}$
contains $\nu^+$, since $\nu^+-\nu=\nu^-$ and every $w\in\mathcal K$
is nonnegative. It is weak* compact because each inequality
$\langle w,\lambda\rangle\geq\langle w,\nu\rangle$ defines a weak*
closed half-space. Since $d\in C(D)$, the cost
$\lambda\mapsto\langle d,\lambda\rangle$ is weak* continuous
and attains its minimum on this set at some $\lambda^*$, with cost
$P^*$. The measure $\lambda^*$ is feasible for the unrestricted dual,
and \eqref{d:weak-duality} bounds the cost of every unrestricted dual
feasible measure below by $P^*$. Thus $\lambda^*$ also attains the
unrestricted dual minimum.

\emph{Complementary slackness.}
For any primal-feasible $w$ and dual-feasible $\lambda$,
\[
 \langle d,\lambda\rangle-\langle w,\nu\rangle
 =\langle d-w,\lambda\rangle+\langle w,\lambda-\nu\rangle.
\]
Each term on the right is nonnegative. By strong duality, the pair
is optimal if and only if both terms vanish. Since $d-w$ is continuous
and nonnegative, $\langle d-w,\lambda\rangle=0$ implies
$\operatorname{supp}\lambda\subseteq\{x\in D:w(x)=d(x)\}$.
Indeed, every point with $d(x)-w(x)>0$ has a neighborhood on which
the gap is bounded away from zero, and that neighborhood has zero
$\lambda$-mass.
\end{proof}
\subsection{Certification of posted price}

\begin{proof}[Proof of Proposition~\ref{prop:hyp-grand-bundling-certificate}]
For $p\in[0,\tfrac12]$, the posted-price rent is $u_p=(s-p)_+$.
Its frontier gap is $w_p=\min\{d,\tfrac12-p\}$ and its contact set
with $d$ is $W_p=\{x\in D:s(x)\geq p\}$. For any primal and dual
feasible pair, the difference of their objective values is
\[
 \int_D d\,d\lambda-\int_D w\,d\nu
 =\int_D(d-w)\,d\lambda+\int_D w\,d(\lambda-\nu).
\]
By \eqref{d:weak-duality} both terms are nonnegative. At $w=w_p$,
the first term vanishes exactly when
$\operatorname{supp}\lambda\subseteq W_p$, by continuity of
$d-w_p$. The second vanishes exactly when
$\int_D w_p\,d(\lambda-\nu)=0$. Theorem~\ref{thm:hyp-strong-duality}
gives an attaining dual measure. The equivalence follows.
\end{proof}

\paragraph{Certifying posted price under $g\ge0$.}
Suppose $g\geq0$, and let $p^*$ maximize full-information posted-price
revenue. This revenue is positive at some interior price and
vanishes at $0$ and $\tfrac12$, since $F$ has a density. Let
$h_S(p)=\int_0^p f(a,p-a)\,da$ be the density of $s(X)$.
The first-order condition is
$\Pr\{s(X)\geq p^*\}-p^*h_S(p^*)=0$.
For $W=W_{p^*}$ and $Z=D\setminus W$, use
$g=f+\operatorname{div}(xf)$ and the divergence theorem on $W$.
The flux on $H$ is $B$, the flux on $\{s=p^*\}$ is
$-p^*h_S(p^*)$, and the axes have zero flux. Hence
\begin{equation}
 \int_W g(x)\,dx=B,\qquad \int_Z g(x)\,dx=1,
 \label{d:massbalance}
\end{equation}
Let $\lambda^*=\mathbf 1_Wg(x)\,dx$. For $w\in\mathcal K$,
monotonicity gives $w(x)\leq w(0)$, so
\[
 \int_D w\,d(\lambda^*-\nu)
 =w(0)-\int_Z w(x)g(x)\,dx\geq0 .
\]
Since $w_{p^*}$ is constant on $Z$, equality holds at $w_{p^*}$.
The support of $\lambda^*$ lies in $W$, so it is a certificate.

Set $S_H(x)=(\tfrac12-\beta,\beta)$ and
$\gamma=(S_H,\operatorname{Id})_\#\lambda^*$. Its first marginal
has mass $B$ and is supported on $H$, so it can replace
$\lambda_H$. Its second marginal is $\lambda^*$. Since
$u(S_H(x))-u(x)\leq d(x)$ and $u\geq0$, we have
\[
 R_f(u)
 =\int_{D\times D}[u(h)-u(x)]\,d\gamma(h,x)
     -\int_Z u(x)g(x)\,dx
 \leq\int_D d\,d\lambda^* .
\]
Both inequalities are tight at $u_{p^*}$.

For Figure~\ref{fig:hyp-reshuffling}, use the ray coordinates
$x=s(\xi,1-\xi)$ with Jacobian $s$ and frontier point
$h(\xi)=\tfrac12(\xi,1-\xi)$. Let
$\widetilde b(\xi)=\int_{p^*}^{1/2}s\,g(s(\xi,1-\xi))\,ds$.
By \eqref{d:massbalance}, $\int_0^1\widetilde b(\xi)\,d\xi=B$,
so the pushforward of $\widetilde b(\xi)\,d\xi$ under $h$ is a
valid frontier replacement. Relax the constraints across distinct
rays. Along a ray, rents are nonnegative and $1$-Lipschitz in $s$.
With frontier rent $\tfrac12-c$, the pointwise smallest such rent
is $(s-c)_+$. It is feasible on that ray and maximizes its
revenue contribution at the given frontier rent because $g\geq0$.
The optimized ray contribution at cutoff $c$ is
\[
 L_\xi(c)=\widetilde b(\xi)(\tfrac12-c)
       -\int_c^{1/2}(s-c)s\,g(s(\xi,1-\xi))\,ds .
\]
Its derivative is $\int_c^{p^*}s\,g(s(\xi,1-\xi))\,ds$, which is
nonnegative for $c\leq p^*$ and nonpositive for $c\geq p^*$.
The common cutoff maximizes each ray contribution and is feasible
in the original problem.

\subsection{Menu coarsening}

\begin{proof}[Proof of Proposition~\ref{prop:hyp-boundary-collapse}]
Let \(u\) be the rent generated by the menu, let \(r=u|_H\), and
set \(p=\tfrac12-r\). By Lemma~\ref{c:obstacle}, full information
at price \(p\) gives rent \(s(x)-p=r-d(x)\leq u(x)\) at every type,
with equality on \(H\). Adding this option therefore leaves the rent
envelope unchanged.

Now consider an original contract \((q,t)\) with \(q_i=1\) for
some \(i\). At the frontier endpoint \(\tfrac12 e_i\), where \(e_i\)
is the \(i\)th coordinate vector, this contract gives rent
\(\tfrac12-t\). Since the original menu gives rent \(r\) at that
endpoint, \(t\geq \tfrac12-r=p\). Thus full information at \(p\)
weakly dominates the contract at every type:
\[
 (s(x)-p)-(x\cdot q-t)
   =(\mathbf 1-q)\cdot x+(t-p)\geq 0.
\]
If \(x\) was assigned this contract, then \(x\cdot q-t=u(x)\).
Since \(s(x)-p\leq u(x)\), the inequality is binding and \(t=p\).
We can therefore delete the boundary contracts, assign full
information to their former buyers, and leave all other assignments
in place. Each type keeps both its rent and its payment.

For IC, every retained contract still satisfies
\(\tfrac12 q_i-t\leq r\) for \(i=1,2\). By the continuation-payoff
bound in Proposition~\ref{c:full-implementation}, its optimized net
rent at type \(x\) is at most
\[
 \max\{x\cdot q-t,\ r-d(x)\}\leq u(x).
\]
Full information at \(p\) also gives at most \(u(x)\), and the
assigned option attains \(u(x)\). This proves IC. Since rents
are unchanged, IR is preserved.
\end{proof}

\begin{proof}[Proof of Corollary~\ref{cor:hyp-canonical-bundling}]
Let \(u\) be a feasible optimal rent with one of the stated
allocation forms, and set \(p=\tfrac12-u|_H\). Both forms imply that
at almost every interior type \(x\), the gradient \(q=\nabla u(x)\)
is zero or has a coordinate equal to one. For Pavlov's form this
follows from the boundary-allocation restriction. In the canonical
form, the slope is zero in the exclusion region, one coordinate
is one in each partial-allocation region, and both are one in the
full-allocation region.

IR and Lemma~\ref{c:obstacle} give \(u(x)\geq(s(x)-p)_+\).
To see the reverse inequality, fix an interior differentiability
point with one of these slopes. The supporting contract
\[
 q=\nabla u(x), \qquad t=x\cdot q-u(x)
\]
gives rent \(u(x)\) at \(x\) and lies below \(u\) everywhere.
Evaluating the support at the origin gives \(t\geq0\).
If \(q=0\), this implies \(u(x)=0\). If \(q_i=1\), evaluating
the support at \(\tfrac12 e_i\) gives \(t\geq p\), so
\(u(x)=x\cdot q-t\leq s(x)-p\). Thus \(u=(s-p)_+\) almost
everywhere, and continuity extends the equality to all of \(D\). Posting price \(p\) for full information
preserves the optimal expected revenue.
\end{proof}
\section{Reformulating BBS}
\label{app:bbs-embedding}

We study the binary matching of
\citet{BergemannBonattiSmolin2018}, henceforth BBS. At purchase the
buyer knows the posterior $\theta=\Pr\{\omega=1\}\in[0,1]$.
His outside decision value is
$\ubar^{\mathrm B}(\theta)=\max\{\theta,1-\theta\}$.
We write $u^{\mathrm B}$ for incremental rent and restrict
attention to the responsive concentrated experiments identified
by BBS. Throughout we impose IR and nonnegative payments.

\subsection{Error representation}

Let $\rho_i$ be the probability of recommending the correct action
in state $i$. A concentrated experiment has
$\max\{\rho_1,\rho_2\}=1$. Set $\chi=\rho_1-\rho_2\in[-1,1]$.
The corresponding experiment, with columns recommending actions
$1$ and $2$, is
\begin{equation}
 E_\chi=\begin{pmatrix}\rho_1(\chi)&1-\rho_1(\chi)\\
                         1-\rho_2(\chi)&\rho_2(\chi)\end{pmatrix},
 \qquad
 (\rho_1(\chi),\rho_2(\chi))=
 \begin{cases}
 (1+\chi,1),&\chi\leq0,\\
 (1,1-\chi),&\chi\geq0.
 \end{cases}
 \label{g:bbs-experiment}
\end{equation}
Full information is $\chi=0$. The two endpoint experiments
$\chi=-1,1$ are uninformative.

Following recommendations gives gross value
$\theta\chi+1+\min\{-\chi,0\}$. The remaining decision rules either
ignore or reverse recommendations. Reversal cannot improve on the
outside action: if $\rho_1=1$, its value is
$(1-\theta)(1-\rho_2)\leq1-\theta$; if $\rho_2=1$, its value is
$\theta(1-\rho_1)\leq\theta$. Thus the optimized incremental value is
\begin{equation}
 V^{\mathrm B}(\theta,\chi)
 =\left[\theta\chi+1+\min\{-\chi,0\}-\ubar^{\mathrm B}(\theta)\right]_+.
 \label{g:bbs-value}
\end{equation}
A contract assigned to $\theta$ is responsive when the expression
inside brackets is nonnegative.

Fold the belief interval into two branches:
\begin{equation}
 \theta\leq \tfrac12:\quad (v,0)=(\theta,0),
 \qquad
 \theta\geq \tfrac12:\quad (0,v)=(0,1-\theta).
 \label{g:bbs-fold}
\end{equation}
The coordinate $v=\min\{\theta,1-\theta\}$ is the buyer's baseline
error. The two copies of $v=\tfrac12$ represent the same balanced belief.
The two extreme beliefs at $v=0$ have identical zero incremental
valuations for every experiment; branch labels may be retained when
specifying their assigned contracts.

For an individual experiment define
\begin{equation}
 \widehat x_1(\chi)=1+\chi,\qquad
 \widehat x_2(\chi)=1-\chi,\qquad
 0\leq\widehat x_i\leq2,\quad
 \widehat x_1+\widehat x_2=2,
 \qquad \kappa(z)=(z-1)_+.
 \label{g:bbs-coordinates}
\end{equation}
Substituting \eqref{g:bbs-fold} into \eqref{g:bbs-value} gives
\begin{equation}
 V_i^{\mathrm B}(v,\chi)
   =\left[v\widehat x_i(\chi)
                        -\kappa(\widehat x_i(\chi))\right]_+,
 \qquad i=1,2.
 \label{g:bbs-axis-values}
\end{equation}
On the first branch, $v=\theta$ and the buyer's default action is
action~2. Following the recommendations gives an incremental payoff of
\[
 \theta\chi+1+\min\{-\chi,0\}-(1-\theta)
 =v(1+\chi)-\chi_+.
\]
Here $\chi_+=1-\rho_2$ is the probability that the experiment
recommends action~1 in state~2, where the buyer's default action
is already correct. The second branch is similar.
\subsection{Implementability}

Let $x_i(v)$ denote the valued coordinate of the experiment
assigned on branch $i$, and let $t_i(v)$ be its payment. Thus
$x_1(v)=\widehat x_1(\chi(v))$ and
$x_2(v)=\widehat x_2(\chi(1-v))$. The identity in
\eqref{g:bbs-coordinates} links the two coordinates of a single
experiment. For $v<\tfrac12$, the assignments $x_1(v)$ and
$x_2(v)$ refer to the experiments assigned to two distinct
posteriors, $v$ and $1-v$.

Let $x_1,x_2:[0,\tfrac12)\to[0,2]$ be measurable. When they are
nondecreasing, write $\bar x_i=\lim_{v\uparrow\tfrac12}x_i(v)$.

\begin{proposition}[Error-coordinate implementability]
\label{prop:bbs-folded}
The pair $(x_1,x_2)$ is implementable by a responsive,
incentive-compatible, and individually rational concentrated
menu with nonnegative payments if and only if
\begin{align}
 &x_i\text{ is nondecreasing},\qquad i=1,2,
 \label{g:bbs-monotone}\\
 &\bar x_1+\bar x_2\leq2,
 \label{g:bbs-seam}\\
 &\int_0^{1/2}x_1(v)\,dv=\int_0^{1/2}x_2(v)\,dv.
 \label{g:bbs-resource}
\end{align}
On each branch, rents and payments are
\begin{equation}
 u_i(v)=\int_0^v x_i(w)\,dw,\qquad
 t_i(v)=v x_i(v)-\kappa(x_i(v))-\int_0^v x_i(w)\,dw.
 \label{g:bbs-transfers}
\end{equation}
\end{proposition}

\begin{proof}
In the symmetric matching case, the experiment index in
\citet[Lemma~2]{BergemannBonattiSmolin2018} is
$q=\rho_1-\rho_2=\chi$. Their lemma characterizes responsive
implementation by monotonicity of $\chi$ on $[0,1]$ and the
condition
\begin{equation}
 \int_0^1\chi(\theta)\,d\theta=0.
 \label{g:bbs-original-integral}
\end{equation}
The envelope formula in their proof gives total net utility
$1+I(\theta)$, where $I(\theta)=\int_0^\theta\chi(z)\,dz$.
The constant is one because the buyer at $\theta=0$ knows the
state and pays zero under individual rationality and nonnegative
payments. Subtracting total net utility from the gross payoff
under the assigned experiment gives the transfer in their
equation (A5):
\begin{equation}
 \begin{aligned}
 I(\theta)&=\int_0^\theta\chi(z)\,dz,\\
 t(\theta)&=\theta\chi(\theta)+\min\{-\chi(\theta),0\}-I(\theta),\\
 u^{\mathrm B}(\theta)&=1+I(\theta)-\ubar^{\mathrm B}(\theta).
 \end{aligned}
 \label{g:bbs-scalar-transfer}
\end{equation}
These implementing payments are nonnegative. If $\chi(\theta)\leq0$,
then $t(\theta)=\theta\chi(\theta)-I(\theta)\geq0$ by monotonicity.
If $\chi(\theta)\geq0$, \eqref{g:bbs-original-integral} gives
$t(\theta)=\int_\theta^1[\chi(z)-\chi(\theta)]\,dz\geq0$.

In error coordinates, the scalar assignment is
\begin{equation}
 \chi(\theta)=
 \begin{cases}
  x_1(\theta)-1,&0\leq\theta<\tfrac12,\\
  1-x_2(1-\theta),&\tfrac12<\theta\leq1.
 \end{cases}
 \label{g:bbs-reconstruction}
\end{equation}
Monotonicity on each branch is equivalent to
\eqref{g:bbs-monotone}. A value at $\theta=\tfrac12$ can join the
two branches monotonically if and only if
$\bar x_1-1\leq1-\bar x_2$, which is \eqref{g:bbs-seam}.
Any $\chi(\tfrac12)\in[\bar x_1-1,1-\bar x_2]$ then works.
A change of variables gives
\begin{equation}
 \int_0^1\chi(\theta)\,d\theta
 =\int_0^{1/2}[x_1(v)-1]\,dv
  +\int_0^{1/2}[1-x_2(v)]\,dv
 =\int_0^{1/2}x_1(v)\,dv-\int_0^{1/2}x_2(v)\,dv.
 \label{g:bbs-change-integral}
\end{equation}
Thus \eqref{g:bbs-resource} is equivalent to
\eqref{g:bbs-original-integral}.

The outside payoff on either branch is $1-v$. Hence the rents
at posteriors $v$ and $1-v$ are respectively
\[
 \begin{aligned}
 u_1(v)&=v+I(v)=\int_0^v x_1(w)\,dw,\\
 u_2(v)&=v+I(1-v)
       =v-\int_{1-v}^1\chi(z)\,dz
       =\int_0^v x_2(w)\,dw,
 \end{aligned}
\]
where the second line uses \eqref{g:bbs-original-integral}.
By responsiveness, the assigned experiment gives gross incremental
value $v x_i(v)-\kappa(x_i(v))$ on branch $i$. Subtracting the
rent gives the payment in \eqref{g:bbs-transfers}.
\end{proof}
\subsection{Designer's problem}
Let $F_{\mathrm B}$ be the cdf of the posterior and $f_{\mathrm B}$
its density. For $v\in[0,\tfrac12]$, define the branch densities
and their tails by
\[
 g_1(v)=f_{\mathrm B}(v),\qquad
 g_2(v)=f_{\mathrm B}(1-v),\qquad
 G_i(v)=\int_v^{1/2}g_i(w)\,dw.
\]
Expected revenue is the sum of expected payments on the two
branches. In \eqref{g:bbs-transfers}, the expected rent on branch
$i$ satisfies
\[
 \int_0^{1/2}g_i(v)\left(\int_0^v x_i(w)\,dw\right)dv
 =\int_0^{1/2}x_i(w)G_i(w)\,dw,
\]
where the equality follows from Fubini's theorem. Substituting
into expected revenue gives
\begin{equation}
 \operatorname{Rev}^{\mathrm B}
 =\sum_{i=1}^2\int_0^{1/2}
 \left[\psi_i(v)x_i(v)-g_i(v)\kappa(x_i(v))\right]\,dv,
 \qquad
 \psi_i(v)=v g_i(v)-G_i(v).
 \label{g:bbs-revenue}
\end{equation}
The seller maximizes \eqref{g:bbs-revenue} subject to
\eqref{g:bbs-monotone}--\eqref{g:bbs-resource}.

\subsection{Extreme-point approach}
\label{app:bbs-extreme-points}

By \citet[Proposition~2]{BergemannBonattiSmolin2018}, some optimal
menu assigns full information to a posterior $\theta^I$. For each
candidate $\theta^I\in[0,1]$, we first solve the seller's problem
subject to $\chi(\theta^I)=0$. Maximizing the resulting value over
$\theta^I$ then recovers the unrestricted optimum.
\paragraph{Cutoff representation.}
Let $J_-=[0,\theta^I]$ and $J_+=[\theta^I,1]$. For a nondecreasing
assignment $\chi$ with $\chi(\theta^I)=0$, represent its increases
on the two sides by probability measures $\mu_-$ on $J_-$ and
$\mu_+$ on $J_+$. On $[0,\theta^I)$, the cdf of $\mu_-$ is
$1+\chi(\theta+)$, with any remaining mass placed at $\theta^I$.
On $[\theta^I,1)$, the cdf of $\mu_+$ is $\chi(\theta+)$, with
any remaining mass placed at $1$. Here $\chi(\theta+)$ denotes
the right limit. Extending both measures by zero outside their
respective intervals gives, for almost every $\theta$,
\[
 \chi(\theta)=-1+\mu_-([0,\theta])+\mu_+([0,\theta]).
\]
Conversely, this formula defines a nondecreasing assignment
away from $\theta^I$. It is nonpositive to the left and
nonnegative to the right, so we may set $\chi(\theta^I)=0$.

Integrating the representation gives
\[
 \int_0^1\chi(\theta)\,d\theta
 =1-\int_{J_-}z\,d\mu_-(z)-\int_{J_+}z\,d\mu_+(z).
\]
Hence, by \eqref{g:bbs-original-integral}, the feasible cutoff
pairs are the probability measures on $J_-$ and $J_+$
satisfying
\begin{equation}
 \int_{J_-}z\,d\mu_-(z)+\int_{J_+}z\,d\mu_+(z)=1.
 \label{g:bbs-cutoff-moment}
\end{equation}

\paragraph{Extreme point.}
On the left of $\theta^I$, the transfer in
\eqref{g:bbs-scalar-transfer} equals
$\int_0^\theta[\chi(\theta)-\chi(s)]\,ds$. On the right, using
\eqref{g:bbs-original-integral}, it equals
$\int_\theta^1[\chi(s)-\chi(\theta)]\,ds$. Substituting the
cutoff representation and applying Fubini's theorem yields
\[
 t(\theta)=
 \begin{cases}
  \displaystyle\int_{[0,\theta]}z\,d\mu_-(z),
       & \theta<\theta^I,\\[6pt]
  \displaystyle\int_{(\theta,1]}(1-z)\,d\mu_+(z),
       & \theta>\theta^I,
 \end{cases}
 \qquad\text{for almost every }\theta.
\]
Integrating these payments against the posterior distribution
gives
\[
 \begin{aligned}
 \operatorname{Rev}^{\mathrm B}
 &=\int_{J_-}z\bigl[F_{\mathrm B}(\theta^I)-F_{\mathrm B}(z)\bigr]
       \,d\mu_-(z)\\
 &\quad+\int_{J_+}(1-z)
       \bigl[F_{\mathrm B}(z)-F_{\mathrm B}(\theta^I)\bigr]
       \,d\mu_+(z).
 \end{aligned}
\]
Since $F_{\mathrm B}$ is continuous, revenue is affine and
weakly continuous in $(\mu_-,\mu_+)$.

Identify the pair with the probability measure
$\nu=\tfrac12(\mu_-\oplus\mu_+)$ on the compact disjoint union
$J=J_-\sqcup J_+$. Feasibility is equivalent to
\[
 \nu(J_-)=\frac12,\qquad \int_J z\,d\nu(z)=\frac12,
\]
where $z$ denotes the cutoff coordinate on either side. The two moment equalities define a weakly closed, convex subset
of $\mathcal P(J)$, so the feasible set is weakly compact and
convex. Since revenue is weakly continuous and affine, the set
of maximizers is nonempty, compact, and convex. By the Krein--Milman theorem \citep{KreinMilman1940}, the set of
maximizers has an extreme point. To see that $\nu^*$ is also extreme in the feasible set, suppose
$\nu^*=\tau\nu_1+(1-\tau)\nu_2$ for feasible measures $\nu_1,\nu_2$
and $0<\tau<1$. By affinity, the revenue at $\nu^*$ is the
weighted average of the revenues at $\nu_1$ and $\nu_2$.
Since neither revenue can exceed the maximum, both must attain
it. Extremality of $\nu^*$ among the maximizers then implies
$\nu_1=\nu_2=\nu^*$.  The feasible set is defined by two moment equalities on
$\mathcal P(J)$. Hence \citet[Theorem~2.1]{Winkler1988} implies
that $\nu^*$ is supported on at most three points of $J$.

\subsection{Optimal menu}

Each side has positive mass, so an optimal cutoff pair can be
chosen with one point on each side or two on one side and one
on the other. Relabel the states if necessary so that the
right side has a single cutoff. Both cases can then be written as
\[
 \mu_-=\lambda\delta_b+(1-\lambda)\delta_c,\qquad
 \mu_+=\delta_{1-a},
 \qquad
 0\leq b\leq c\leq\theta^I\leq1-a\leq1,\quad
 \lambda\in[0,1].
\]
The moment condition gives $a=\lambda b+(1-\lambda)c$.
The case with one point on each side is represented by $b=c=a$.

Away from the cutoffs, $\chi$ takes the successive values
$-1$, $\lambda-1$, $0$, and $1$, with transitions at $b$, $c$,
and $1-a$. The payment formula gives the following menu.
Buyers with $b<\theta<c$ receive the experiment
$(\rho_1,\rho_2)=(\lambda,1)$ and pay $\lambda b$. Buyers with
$c<\theta<1-a$ receive full information and pay $a$. For the
latter buyers, the payment equals $\lambda b+(1-\lambda)c$
to the left of $\theta^I$ and $1-(1-a)$ to the right.
The remaining buyers take the outside option. Substituting
the assignment into $x_1(\alpha)=1+\chi(\alpha)$ and
$x_2(\beta)=1-\chi(1-\beta)$ gives
\eqref{eq:bbs-optimal-cutoffs} up to choices at the cutoffs,
which have zero probability.

The expected revenue under the chosen state labeling is
\begin{equation}
 \lambda b\bigl[F_{\mathrm B}(c)-F_{\mathrm B}(b)\bigr]
 +a\bigl[F_{\mathrm B}(1-a)-F_{\mathrm B}(c)\bigr],
 \qquad a=\lambda b+(1-\lambda)c,
 \label{eq:bbs-optimal-design}
\end{equation}
where $F_{\mathrm B}$ is the posterior cdf under that labeling.
Conversely, for any $\lambda\in[0,1]$ and
$0\leq b\leq c\leq1-a\leq1$ satisfying the price relation,
one may choose $\theta^I\in[c,1-a]$. The cutoff pair belongs
to the corresponding intervals and satisfies
\eqref{g:bbs-cutoff-moment}, so it is implementable.

Maximizing \eqref{eq:bbs-optimal-design} over these parameters therefore gives the optimal revenue.
An optimal menu offers full information and at most one
partial experiment, as in
\citet[Proposition~4]{BergemannBonattiSmolin2018}.


\begin{thebibliography}{99}

\bibitem[Adams and Yellen(1976)]{AdamsYellen1976}
Adams, William James, and Janet L. Yellen. 1976.
``Commodity Bundling and the Burden of Monopoly.''
\emph{The Quarterly Journal of Economics} 90(3): 475--498.

\bibitem[Admati and Pfleiderer(1986)]{AdmatiPfleiderer1986}
Admati, Anat R., and Paul Pfleiderer. 1986.
``A Monopolistic Market for Information.''
\emph{Journal of Economic Theory} 39(2): 400--438.

\bibitem[Admati and Pfleiderer(1990)]{AdmatiPfleiderer1990}
Admati, Anat R., and Paul Pfleiderer. 1990.
``Direct and Indirect Sale of Information.''
\emph{Econometrica} 58(4): 901--928.

\bibitem[Ali et~al.(2022)]{AliEtAl2022}
Ali, S. Nageeb, Nima Haghpanah, Xiao Lin, and Ron Siegel. 2022.
``How to Sell Hard Information.''
\emph{Quarterly Journal of Economics} 137(1): 619--678.

\bibitem[Bergemann and Bonatti(2019)]{BergemannBonatti2019}
Bergemann, Dirk, and Alessandro Bonatti. 2019.
``Markets for Information: An Introduction.''
\emph{Annual Review of Economics} 11: 85--107.
\href{https://doi.org/10.1146/annurev-economics-080315-015439}{DOI}.

\bibitem[Bergemann, Bonatti, and Smolin(2018)]{BergemannBonattiSmolin2018}
Bergemann, Dirk, Alessandro Bonatti, and Alex Smolin. 2018.
``The Design and Price of Information.''
\emph{American Economic Review} 108(1): 1--48.
\href{https://doi.org/10.1257/aer.20161079}{DOI}.

\bibitem[Blackwell(1953)]{Blackwell1953}
Blackwell, David. 1953.
``Equivalent Comparisons of Experiments.''
\emph{The Annals of Mathematical Statistics} 24(2): 265--272.
\href{https://doi.org/10.1214/aoms/1177729032}{DOI}.

\bibitem[Bull and Watson(2007)]{BullWatson2007}
Bull, Jesse, and Joel Watson. 2007.
``Hard Evidence and Mechanism Design.''
\emph{Games and Economic Behavior} 58(1): 75--93.

\bibitem[Carroll(2017)]{Carroll2017}
Carroll, Gabriel. 2017.
``Robustness and Separation in Multidimensional Screening.''
\emph{Econometrica} 85(2): 453--488.

\bibitem[Che and Zhong(2025)]{CheZhong2025}
Che, Yeon-Koo, and Weijie Zhong. 2025.
``Robustly Optimal Mechanisms for Selling Multiple Goods.''
\emph{Review of Economic Studies} 92(5): 2923--2951.

\bibitem[Courty and Li(2000)]{CourtyLi2000}
Courty, Pascal, and Hao Li. 2000.
``Sequential Screening.''
\emph{Review of Economic Studies} 67(4): 697--717.

\bibitem[Daskalakis, Deckelbaum, and Tzamos(2017)]{DDT2017}
Daskalakis, Constantinos, Alan Deckelbaum, and Christos Tzamos. 2017.
``Strong Duality for a Multiple-Good Monopolist.''
\emph{Econometrica} 85(3): 735--767.

\bibitem[Doval and Skreta(2026)]{DovalSkreta2026}
Doval, Laura, and Vasiliki Skreta. 2026.
``Purchase History and Product Personalization.''
\emph{RAND Journal of Economics} 57(1): 78--102.

\bibitem[Dworczak et~al.(2026)]{DworczakPeregoVeldkamp2026}
Dworczak, Piotr, Jacopo Perego, and Laura Veldkamp. 2026.
``Privacy for Sale: Anarchy, Ban, or Market?''
Working paper, August 15.

\bibitem[Eliaz and Spiegler(2026)]{EliazSpiegler2026}
Eliaz, Kfir, and Ran Spiegler. 2026.
``Monopolistic Data Dumping.''
\emph{American Economic Journal: Microeconomics} 18(4): 534--559.
\href{https://doi.org/10.1257/mic.20250260}{DOI}.

\bibitem[Es{\H o} and Szentes(2007)]{EsoSzentes2007}
Es{\H o}, P{\'e}ter, and Bal{\'a}zs Szentes. 2007.
``Optimal Information Disclosure in Auctions and the Handicap Auction.''
\emph{Review of Economic Studies} 74(3): 705--731.

\bibitem[Fotakis et~al.(2021)]{FotakisEtAl2021}
Fotakis, Dimitris, Alkis Kalavasis, Vasilis Kontonis,
and Christos Tzamos. 2021.
``Efficient Algorithms for Learning from Coarse Labels.''
\emph{Proceedings of Machine Learning Research} 134: 2060--2079.

\bibitem[Gan and Li(2026)]{GanLi2026}
Gan, Tan, and Yingkai Li. 2026.
``Screening for Choice Sets.''
Working paper, arXiv:2601.15580.

\bibitem[Haghpanah and Hartline(2021)]{HaghpanahHartline2021}
Haghpanah, Nima, and Jason Hartline. 2021.
``When Is Pure Bundling Optimal?''
\emph{Review of Economic Studies} 88(3): 1127--1156.

\bibitem[Haghpanah et~al.(2026)]{HaghpanahLinSiegel2026}
Haghpanah, Nima, Xiao Lin, and Ron Siegel. 2026.
``Selling Evidence Bit by Bit.''
PIER Working Paper No.~26-019.

\bibitem[Inan et al.(2023)]{InanEtAl2023}
Inan, Hakan, Kartikeya Upasani, Jianfeng Chi, Rashi Rungta, Krithika Iyer,
Yuning Mao, Michael Tontchev, Qing Hu, Brian Fuller, Davide Testuggine,
and Madian Khabsa. 2023.
``Llama Guard: LLM-based Input-Output Safeguard for Human-AI Conversations.''
\href{https://arxiv.org/abs/2312.06674}{arXiv:2312.06674}.

\bibitem[Kleiner and Manelli(2019)]{KleinerManelli2019}
Kleiner, Andreas, and Alejandro Manelli. 2019.
``Strong Duality in Monopoly Pricing.''
\emph{Econometrica} 87(4): 1391--1396.
\href{https://doi.org/10.3982/ECTA15735}{DOI}.

\bibitem[Krein and Milman(1940)]{KreinMilman1940}
Krein, M., and D. Milman. 1940.
``On Extreme Points of Regular Convex Sets.''
\emph{Studia Mathematica} 9(1): 133--138.

\bibitem[Lahr and Niemeyer(2025)]{LahrNiemeyer2025}
Lahr, Patrick, and Axel Niemeyer. 2025.
``Extreme Points in Multi-Dimensional Screening.''
CEPS Working Paper No.~2025/16.
\href{https://www.ceps-paris-saclay.fr/fileadmin/user_upload/Lahr_Niemeyer_CEPS_2025_16.pdf}{Working paper}.

\bibitem[Li and Shi(2017)]{LiShi2017}
Li, Hao, and Xianwen Shi. 2017.
``Discriminatory Information Disclosure.''
\emph{American Economic Review} 107(11): 3363--3385.

\bibitem[Markov et al.(2023)]{MarkovEtAl2023}
Markov, Todor, Chong Zhang, Sandhini Agarwal, Florentine Eloundou Nekoul,
Theodore Lee, Steven Adler, Angela Jiang, and Lilian Weng. 2023.
``A Holistic Approach to Undesired Content Detection in the Real World.''
\emph{Proceedings of the AAAI Conference on Artificial Intelligence}
37(12): 15009--18.

\bibitem[McAfee and McMillan(1988)]{McAfeeMcMillan1988}
McAfee, R. Preston, and John McMillan. 1988.
``Multidimensional Incentive Compatibility and Mechanism Design.''
\emph{Journal of Economic Theory} 46(2): 335--354.

\bibitem[McAfee, McMillan, and Whinston(1989)]{McAfeeMcMillanWhinston1989}
McAfee, R. Preston, John McMillan, and Michael D. Whinston. 1989.
``Multiproduct Monopoly, Commodity Bundling, and Correlation of Values.''
\emph{The Quarterly Journal of Economics} 104(2): 371--383.


\bibitem[Myerson(1981)]{Myerson1981}
Myerson, Roger B. 1981. ``Optimal Auction Design.''
\emph{Mathematics of Operations Research} 6(1): 58--73.

\bibitem[Myerson(1982)]{Myerson1982}
Myerson, Roger B. 1982.
``Optimal Coordination Mechanisms in Generalized Principal-Agent Problems.''
\emph{Journal of Mathematical Economics} 10(1): 67--81.

\bibitem[Natarajan et al.(2013)]{NatarajanEtAl2013}
Natarajan, Nagarajan, Inderjit S. Dhillon, Pradeep K. Ravikumar,
and Ambuj Tewari. 2013.
``Learning with Noisy Labels.''
\emph{Advances in Neural Information Processing Systems} 26: 1196--1204.

\bibitem[Patrini et~al.(2017)]{PatriniEtAl2017}
Patrini, Giorgio, Alessandro Rozza, Aditya Krishna Menon,
Richard Nock, and Lizhen Qu. 2017.
``Making Deep Neural Networks Robust to Label Noise:
A Loss Correction Approach.''
\emph{Proceedings of the IEEE Conference on Computer
Vision and Pattern Recognition}, 1944--1952.

\bibitem[Pavlov(2011)]{Pavlov2011}
Pavlov, Gregory. 2011. ``A Property of Solutions to Linear Monopoly Problems.''
\emph{The B.E. Journal of Theoretical Economics} 11(1), Advances, Article 4.
\href{https://doi.org/10.2202/1935-1704.1663}{DOI}.

\bibitem[Raykar et al.(2010)]{RaykarEtAl2010}
Raykar, Vikas C., Shipeng Yu, Linda H. Zhao,
Gerardo Hermosillo Valadez, Charles Florin, Luca Bogoni,
and Linda Moy. 2010.
``Learning From Crowds.''
\emph{Journal of Machine Learning Research} 11: 1297--1322.

\bibitem[Raykar and Yu(2012)]{RaykarYu2012}
Raykar, Vikas C., and Shipeng Yu. 2012.
``Eliminating Spammers and Ranking Annotators for Crowdsourced Labeling Tasks.''
\emph{Journal of Machine Learning Research} 13: 491--518.

\bibitem[Riley and Zeckhauser(1983)]{RileyZeckhauser1983}
Riley, John, and Richard Zeckhauser. 1983.
``Optimal Selling Strategies: When to Haggle, When to Hold Firm.''
\emph{Quarterly Journal of Economics} 98(2): 267--289.
\href{https://doi.org/10.2307/1885625}{DOI}.

\bibitem[Rochet(1987)]{Rochet1987}
Rochet, Jean-Charles. 1987.
``A Necessary and Sufficient Condition for Rationalizability in a Quasi-linear Context.''
\emph{Journal of Mathematical Economics} 16(2): 191--200.

\bibitem[Rochet and Chon{\'e}(1998)]{RochetChone1998}
Rochet, Jean-Charles, and Philippe Chon{\'e}. 1998.
``Ironing, Sweeping, and Multidimensional Screening.''
\emph{Econometrica} 66(4): 783--826.

\bibitem[Segura-Rodr\'iguez(2022)]{SeguraRodriguez2022}
Segura-Rodr\'iguez, Carlos. 2022.
``Selling Data.''
PIER Working Paper No.~19-006, revised August 2022.
\href{https://doi.org/10.2139/ssrn.3385500}{Working paper}.

\bibitem[Sion(1958)]{Sion1958}
Sion, Maurice. 1958.
``On General Minimax Theorems.''
\emph{Pacific Journal of Mathematics} 8(1): 171--176.
\href{https://doi.org/10.2140/pjm.1958.8.171}{DOI}.

\bibitem[Thanassoulis(2004)]{Thanassoulis2004}
Thanassoulis, John. 2004. ``Haggling over Substitutes.''
\emph{Journal of Economic Theory} 117(2): 217--245.

\bibitem[Winkler(1988)]{Winkler1988}
Winkler, Gerhard. 1988.
``Extreme Points of Moment Sets.''
\emph{Mathematics of Operations Research} 13(4): 581--587.

\bibitem[Yang(2025)]{Yang2025}
Yang, Frank. 2025.
``Nested Bundling.''
\emph{American Economic Review} 115(9): 2970--3013.

\bibitem[Yang and Yang(2026)]{YangYang2026}
Yang, Frank, and Kai Hao Yang. 2026.
``Stochastic Optimization and Coupling.''
Cowles Foundation Discussion Paper No. 2506, March 12.

\bibitem[Yang(2022)]{Yang2022}
Yang, Kai Hao. 2022.
``Selling Consumer Data for Profit: Optimal Market-Segmentation Design
and Its Consequences.''
\emph{American Economic Review} 112(4): 1364--1393.
\href{https://doi.org/10.1257/aer.20210616}{DOI}.

\bibitem[Zhong(2026)]{Zhong2026}
Zhong, Weijie. 2026.
``Bundling Complements.''
Working paper, Stanford University.
\href{https://arxiv.org/abs/2607.07982}{arXiv:2607.07982}.

\end{thebibliography}
\end{document}